\documentclass[twocolumn]{aastex631}

\usepackage{amsmath}
\usepackage{natbib}
\usepackage{color,soul}
\usepackage{url}
\usepackage{color}
\usepackage{soul} % use this (many fancier options)

\definecolor{LightCyan}{rgb}{0.88,1,1}

\newcommand{\done}{}

\newcommand{\new}{}

\newcommand{\lenstool}{{\tt{Lenstool}}}

\newcommand{\kms}{km s$^{-1}$}
\newcommand{\msun}{M$_{\odot}$}
\newcommand{\SPTXXIII}{SPT-CL\,J2325$-$4111}
\newcommand{\SPTV}{SPT-CL\,J0512$-$3848}
\newcommand{\SPTZERO}{SPT-CL\,J0049$-$2440}

\newcommand{\GMone}{c1} % arc we called GM1, not lensed

\newcommand{\Amplistrz}{$\mathcal{A}^{lens}_{| \mu | \geq 3 }$} %the A0.5 mu gt3 symbol

\graphicspath{{./}{figures/}}

\shorttitle{Two Powerful Strong lenses from SPT}
\shortauthors{Mahler et al.}

\begin{document}

\title{Strong Lensing analysis of \SPTXXIII\ and \SPTZERO, two Powerful Cosmic Telescopes ($R_E > 40''$) from the SPT Clusters Sample}
%Two Powerful Strong Lensing Clusters from SPT and Prospects for Untapped Potential of Discovery with the JWST. %This is the old title

\correspondingauthor{Guillaume Mahler}
\email{guillaume.mahler@uliege.be}

\author[0000-0003-3266-2001]{Guillaume Mahler}
\affiliation{STAR Institute, Quartier Agora - All\'ee du six Ao\^ut, 19c B-4000 Li\`ege, Belgium}
\affiliation{Centre for Extragalactic Astronomy, Durham University, South Road, Durham DH1 3LE, UK}
\affiliation{Institute for Computational Cosmology, Durham University, South Road, Durham DH1 3LE, UK}

\author[0000-0002-7559-0864]{Keren Sharon}
\affil{Department of Astronomy, University of Michigan, 1085 S. University Ave, Ann Arbor, MI 48109, USA}

\author[0000-0003-1074-4807]{Matthew Bayliss} 
\affiliation{Department of Physics, University of Cincinnati, Cincinnati, OH 45221, USA}

\author[0000-0001-7665-5079]{Lindsey. E. Bleem}
\affiliation{Argonne National Laboratory, High-Energy Physics Division, Argonne, IL 60439}
\affiliation{Kavli Institute for Cosmological Physics, University of Chicago, 5640 South Ellis Avenue, Chicago, IL 60637, USA}
\affiliation{Department of Astronomy and Astrophysics, University of Chicago, 5640 South Ellis Avenue, Chicago, IL 60637, USA}

\author[0000-0002-4208-798X]{Mark Brodwin}
\affiliation{Department of Physics and Astronomy, University of Missouri, 5110 Rockhill Road, Kansas City, MO 64110, USA}

\author[0000-0003-4175-571X]{Benjamin Floyd}
\affiliation{Faculty of Physics and Astronomy, University of Missouri-Kansas City, 5110 Rockhill Road, Kansas City, MO 64110, USA}

\author{Raven Gassis}
\affiliation{Department of Physics, University of Cincinnati}

\author[0000-0003-1370-5010]{Michael D. Gladders}
\affiliation{Kavli Institute for Cosmological Physics, University of Chicago, 5640 South Ellis Avenue, Chicago, IL 60637, USA}
\affiliation{Department of Astronomy and Astrophysics, University of Chicago, 5640 South Ellis Avenue, Chicago, IL 60637, USA}

\author[0000-0002-3475-7648]{Gourav Khullar}
\affiliation{Department of Physics and Astronomy and PITT PACC, University of Pittsburgh, Pittsburgh, PA 15260, USA}

\author[0000-0002-7868-9827]{Juan D. Remolina Gonz\'{a}lez}
\affiliation{Department of Astronomy, University of Michigan, 1085 S. University Ave, Ann Arbor, MI 48109, USA}

\author[0000-0002-5222-1337]{Arnab Sarkar}
\affiliation{MIT Kavli Institute for Astrophysics and Space Research, 70 Vassar St, Cambridge, MA 02139, USA}

%\collaboration{SPT Collaboration}

%% Mark off the abstract in the ``abstract'' environment. 
\begin{abstract} %about 220 words - maxlim at 250
\done{We report  the results from a study of two massive ($M_{500c} > 6.0 \times 10^{14}$\msun) strong lensing clusters selected from the South Pole Telescope cluster survey for their high Einstein radius ($R_E > 40''$), \SPTXXIII\ and \SPTZERO. 
Ground-based and shallow HST imaging indicated extensive strong lensing evidence in these fields, with giant arcs spanning 18\arcsec\ and 31\arcsec, respectively, motivating further space-based imaging followup. Here, we present multiband HST imaging and ground-based Magellan spectroscopy of the fields, from which we compile detailed strong lensing models. The lens models of \SPTXXIII\ and \SPTZERO\ were optimized using 9, and 8 secure multiple-imaged systems with a final image-plane rms of 0\farcs63 and 0\farcs73, respectively. From the lensing analysis, we measure the projected mass density within 500~kpc of $M(<500 ~{\rm kpc}) = 7.30\pm0.07 \times 10^{14}$\msun, and
$M(<500 ~{\rm kpc})=7.12^{+0.16}_{-0.19}\times 10^{14}$\msun\ for these two clusters, and a sub-halos mass ratio of $0.12\pm{0.01}$ and $0.21^{+0.07}_{-0.05}$, respectively. Both clusters produce a large area with high magnification ($\mu\geq 3$) for a source at $z=9$, \Amplistrz$=4.93^{+0.03}_{-0.04}$ arcmin$^2$, and \Amplistrz $= 3.64^{+0.14}_{-0.10}$ arcmin$^2$ respectively, placing them in the top tier of strong lensing clusters. 
%Both clusters exhibit remarkable highly magnified galaxy spanning 18\arcsec\ and 31\arcsec\ on sky. 
We conclude that these clusters are spectacular sightlines for further observations that will reduce the systematic uncertainties due to cosmic variance. This paper provides the community with two additional well-calibrated cosmic telescopes, as strong as the Frontier Fields, suitable for studies of the highly magnified background Universe. }

\end{abstract}

%% Keywords should appear after the \end{abstract} command. 
%% See the online documentation for the full list of available subject
%% keywords and the rules for their use.

\section{Introduction}
\label{sec:intro}
%We found those clusters from the ground base (PISCO) and space base (HST-SNAP) to be one the largest Einstein radius, as suggested by the lensed features. It has been observed within the HST program (HST Cycle 27 proposal 15937: Focusing The Next Generation of Extraordinary Cluster Lenses for JWST.)

%The Hubble Frontier Field program has been multiply observed and represents a meaningful sample. We argue here that some other cluster could heavily contribute to future study and we research the most promising lensing. We publish in this paper the lens modelling and analysis of lensing power for those three clusters as well as the prospective analysis of JWST observations of this cluster. Notably focusing on two giant arcs at redshifts X and Z and on the additional probability of discovering high-z target z>6 would meaningfully contribute to understanding the UV luminosity function and its possible turnover.

Clusters of galaxies are located at the nodes of the cosmic web and represent the largest structures of dark matter. Their mass distribution present a remarkable self-similarity in the outskirts \citep{McDonald2017}. However, the densest region located at the core of the structure deviates from this scale-free distribution. This region is prone to ongoing merging activity, where both luminous and dark matter co-evolve. While luminous matter is commonly observed, dark matter remain elusive, and only probed indirectly. Fortunately, the densest regions of cluster cores produce strong gravitational lensing, offering constraining power to map the underlying matter distribution regardless of its nature and provide a magnified view of the distant Universe.

Past studies have used strong lensing clusters to study a large breadth of topics, often even in one sight-line. Frequently addressed questions focus on: the detailed mapping of the cluster underlying matter distribution, estimating both visible and dark contributions \citep[e.g.,][]{Jauzac2014,Furtak23}; 
%(e.g. Jauzac et al. 2014, Furtak et al. 2022); 
tests of cold dark matter (CDM) alternative candidates \citep{Natarajan2017,Harvey17,Robertson19,Sirks22}; comparisons of galaxy-galaxy lenses occurrences in the dense environments of clusters to simulations \citep{Robertson21, Meneghetti22}; detection of filamentary structure, connecting cluster halos to the cosmic web \citep[e.g.,][]{Jauzac2012,Tam22}; constraints on the galaxy halo/sub-halo mass ratio \citep{Mahler2019} and sub-halo mass function \citep[e.g.,][]{Natarajan2017,Sirks22}; constraining lensing mass clumpiness, producing microlensing from stars and primordial black holes \citep[e.g.,][]{Kelly2018,Diego18, Mahler23};
measuring spatially-resolved properties of giant arcs, such as winds \citep{Fisher2019}, sizes of star forming clumps \citep{Johnson2017, Claeyssens23}, metallicity gradients and kinematics \citep{Patricio2018}, and leaking ionizing photon radiation \citep{Rivera-Thorsen2019} at tens of parsec resolution at $z>1$; using caustic crossing events to probe stars at the dawn of the universe \citep[e.g.,][]{welch2022,Hsiao2023,adamo2024}; detecting magnified high-redshift galaxies, probing the intrinsically fainter galaxies more representative of early populations \citep{Atek24}  and offering constraints on the faint end of the high-z luminosity function \citep{Atek2015,Bouwens17,Livermore17,deLaVieuville2019}; constraining cosmological parameters \new{such as the dark energy equation of state parameter (w) and $\Omega_M$}
%through the w-M relation 
\citep[e.g.,][]{Jullo2010,Acebron2017}, and time-delay measurements of H$_\textrm{0}$ \citep[e.g.,][]{Grillo2018,Napier23}.

%An apparent discrepancy between the observed and expected number of giant arcs, known as the “arc statistics problem”, has motivated studies of the correlation between strong lensing efficiency, cosmology, and cluster properties (see Meneghetti et al. 2013 for a review).
Owing to their diverse core properties \citep{McDonald2017}, clusters with the same total mass may not exhibit similar strong lensing cross section. To identify strong-lensing lines of sight requires selection based 
properties beyond their total mass. 
Two main methods of discovery are employed to find lines of sight of lensing clusters: lensing-selected and non-lensing selected.
Lensing-selected surveys have traditionally searched for substantial lensing evidence, typically in the form of highly magnified giant arcs by %visual or machine-assisted 
inspection of large data sets of shallow ground-based data. Utilising the SDSS data and catalogs resulted with numerous lenses  \citep[e.g.,][]{Bayliss2011apjl,Bayliss2011gmos,Stark13,Dahle2015,Johnson2017,sharon2020} mainly at low arc and lens redshifts. More recently, imaging from the Dark Energy Survey yielded strong lensing samples based on visual inspection \citep[e.g.,][]{Diehl17, Khullar2021}  or machine-assisted identification \citep[e.g.,][]{Huang21}.
%To fully benefit from this lines of sight to analyze the sources and/or lenses requires space-based observations, such as with HST or JWST (REFS). [refs: (Sharon et al. 2020) and the lensed galaxies (Rigby et al. \xxx, Johnson et al. \xxx\ SGAS1110 paper, more recent papers e.g,. from TEMPLATES, COOL-LAMPS).]

Non-lensing based approaches commonly rely on 
deep or high-resolution optical imaging followup of cluster samples that were selected as likely lenses based on other criteria, usually high total mass as indicated from X-ray, sub-mm, or optical mass proxies.
Examples of such surveys include the Hubble Space Telescope (HST) follow-up of the MAssive Cluster Survey \citep[MACS;][]{Ebeling2001}, the optical ground-based and HST followup of the South Pole Telescope cluster sample \citep{Bleem2015,Bleem2020}, the Local Cluster Substructure Survey \citep[LoCuSS;][]{Smith2005} followup of X-ray selected clusters, HST followup of sub-mm selected lens candidates (dusty GEM; \citealt{Canameras2015}), and spectroscopic followup of X-ray selected clusters with VLT/MUSE (e.g. KALEIDOSCOPE cluster survey P.ID: 0102.A-0718(A), PI:
A. Edge; \citealt{Patel2024}). Several treasury programs with HST employed a hybrid selection approach where lensing-evidence from shallower or low resolution imaging were combined with total mass criteria to increase the sample, e.g. CLASH \citep[25 clusters,][]{Postman2012}, and RELICS \citep[41 clusters,][]{Coe2019}.

To fully benefit from the lines of sight of strong gravitational lenses requires space-based observations, such as with HST or JWST, and relatively shallow HST follow-up of the aforementioned samples resulted with numerous discoveries as listed at the beginning of the introduction. 
However, the high investment of space-based telescope time is reserved for a few, extraordinary lenses, carefully picked as the best lenses coming out of the previous surveys. Such are the Frontiers Fields \citep{Lotz2017}, for which the original program used 840 HST orbits to observe six strong lensing clusters. More recently, studies of cluster-lensed galaxies using JWST (e.g., TEMPLATES, \citealt{Rigby2023}; UNCOVER, \citealt{Bezanson2022}; PEARLS, \citealt{Windhorst2023}; SMACSJ0723, \citealt{Mahler2023smacs,Caminha2022,Golubchik2022}; SPT0615, \citealt{adamo2024}; WHL0137, \citealt{welch2022}) have pushed the limits of redshift, luminosity, and resolution. 

While tremendous discoveries were enabled even in single well-studied lines of sight (e.g., UNCOVER \citealt{Bezanson2022}), both the HST and JWST analyses caution that cosmic variance might play an important role in our ability to infer the properties of high redshift populations of galaxies.  
\cite{Salmon2020} identified 322 new candidates at $z>6$ behind the 41 clusters uniformly observed by HST as part of RELICS, reporting a large field-to-field variance, where in some fields no galaxy candidates above z$>$5.5 were found \citep{Mahler2019}.
A recent JWST analysis \citep{Chemerynska24} reported a potential overabundance of UV galaxies behind the lensing cluster Abell 2744. 
Identifying more clusters that are on par with the lensing power of the Frontier-Fields clusters will open up an important discovery space.

This paper presents two new exquisite well-calibrated strong lensing sight-lines with obvious untapped discovery potential coming at the end of a dedicated search based from SPT optical follow-up using the PISCO imager and HST-SNAP follow-up.
These targets, \SPTXXIII\ and \SPTZERO, are two of the most promising strong lensing clusters from the SPT cluster sample. 

The SPT cluster sample is based on the detection and calibration of the Sunyaev Zel'dovich effect observed in the CMB radiation (\citealt{Bleem2015,Bleem2020} and references therein). The entire cluster sample was followed up with multi-band imaging using Magellan/PISCO \citep{Bleem2020,Stalder2014}, facilitating the identification of strong lensing candidates, among other cluster science (e.g. \citealt{Somboonpanyakul2021}), extensive spectroscopic campaigns using Magellan and Gemini \citep{Bayliss2016}. Some clusters were the target of HST programs mainly for weak lensing calibration \citep[e.g.,][]{Schrabback2021} and HST/SNAP programs provided shallow high resolution imaging \citep{SPTsnap}. 

\SPTXXIII\ was reported on by \cite{Bleem2015, Bleem2020} as a $\xi=12.5$ significance SZ detection \footnote{The SZ significance $\xi$ for SPT clusters is defined as the detection signal-to-noise, maximized over the 12 filter that SPT uses for cluster identification, with scales ranging from $0\farcm25 -3\farcm0$; cf.\ \cite{Vanderlinde2010}.}, with ground-based spectroscopic redshift of $z=0.358$ \citep{Ruel2014}. \cite{Bocquet2019} report $M_{500c}=6.70^{+0.95}_{-1.17}\times10^{14} h_{70}^{-1}$\msun\ from the weak-lensing calibrated SZ signal. It was flagged as a strong lensing cluster in Table~4 of \cite{Bleem2015}.

%$M_{500c}=7.52^{+0.78}_{-0.96}
%$M_{500c}=7.55\pm1.2\

\SPTZERO\ was reported on by \cite{Bleem2020} from an analysis of the SPTPol Extended Cluster Survey as a $\xi=7.44$ significance SZ detection, with ground-based spectroscopic redshift of $z=0.527$ \citep{dePropris1999}. \cite{Bleem2020} report $M_{500c}=6.59^{+0.86}_{-0.98}\times10^{14} h_{70}^{-1}$\msun\ from the weak-lensing calibrated SZ signal. The same paper highlights this cluster as a prominent strong lens, first reported on by \cite{dePropris1999}. 

In this paper, we used multiband HST imaging and ground-based spectroscopy to identify lensing evidence, compute lens models, and calculate lens properties such as mass distribution, magnification, and lensing strength, showing that these clusters are excellent new strong lenses. 
This paper is organized as follows. 
We present the data used in this work in \autoref{sec:data}, and the lens modelling analysis in \autoref{sec:modeling}. In \autoref{sec:prop} we discuss the mass distribution and lensing power of those cluster, and highlight prominent lensed sources. We conclude in \autoref{sec:conclusion}. 
\new{We also report in \autoref{sec:appendixB} the HST imaging and spectroscopic followup of a third cluster from the same program, \SPTV, which did not pan out as a similarly powerful cosmic telescope.}

%On top field-to-field variance mention previsouly this is additional evidence for additional observations of robust strong lensing sight-lines. 
%yielded tremendous discovery and they all relied on a robustly calibrated models from high-resolution imaging exemplifying their discovery potential. 

% units and cosmology
We assume flat $\Lambda$CDM cosmology with 
$\Omega_{\Lambda} = 0.7$, $\Omega_{m}=0.3$, and $H_0 = 70$ \kms\ Mpc$^{-1}$.
Magnitudes are reported in the AB system \citep{Oke1974}.

\section{Data and lensing evidence} \label{sec:data}
\subsection{Hubble Space Telescope}

%https://archive.stsci.edu/proposal_search.php?mission=hst&id=15937
%% description of the observations
\done{The fields studied in this paper were observed by HST as part of Cycle 27 GO-15937 (PI: G. Mahler). Each cluster was observed with four filters, using 3 orbits of HST: one orbit with the ACS/F606W filter, one orbit with the ACS/F814W filter, and one orbit
with WFC3-IR split between two filters, F105W and F140W.  We obtained four sub-exposures per filter with small box sub-pixel dithers for point spread function (PSF) reconstruction and
to cover chip gaps and artifacts such as the ``IR Blobs'' and ``Death Star'' \citep[WFC3 Data Handbook;][]{rajan10}. For the WFC3-IR observations we used a sampling interval parameter \texttt{SPARS25}.}

\done{Observations of \SPTXXIII\ took place on Oct 22, 2019 (WFC3) and Oct 25, 2019 (ACS). 
The second visit was affected by a guide-star re-acquisition error, leading to a lost exposure in the F814W band, and was rescheduled by HOPR 91651 to Dec 10, 2019. The resulting imaging data have 1212~s in WFC3/F105W, 1312~s in WFC3/F140W, 2204~s in ACS/F606W,  
%1698.000/4*3+2204.000 =  3477
and 3477~s in ACS/F814W. The failed F814W sub-exposure was not used. 
Observations of \SPTZERO\ took place on Oct 23, 2019 (WFC3) and Oct 31, 2019 (ACS), and consists of 
1212~s each in WFC3-IR/F105W and F140W, 2124~s in ACS/F606W, and 2184~s in ACS/F814W. }

\begin{figure*}
    \centering
    \includegraphics[width=\linewidth]{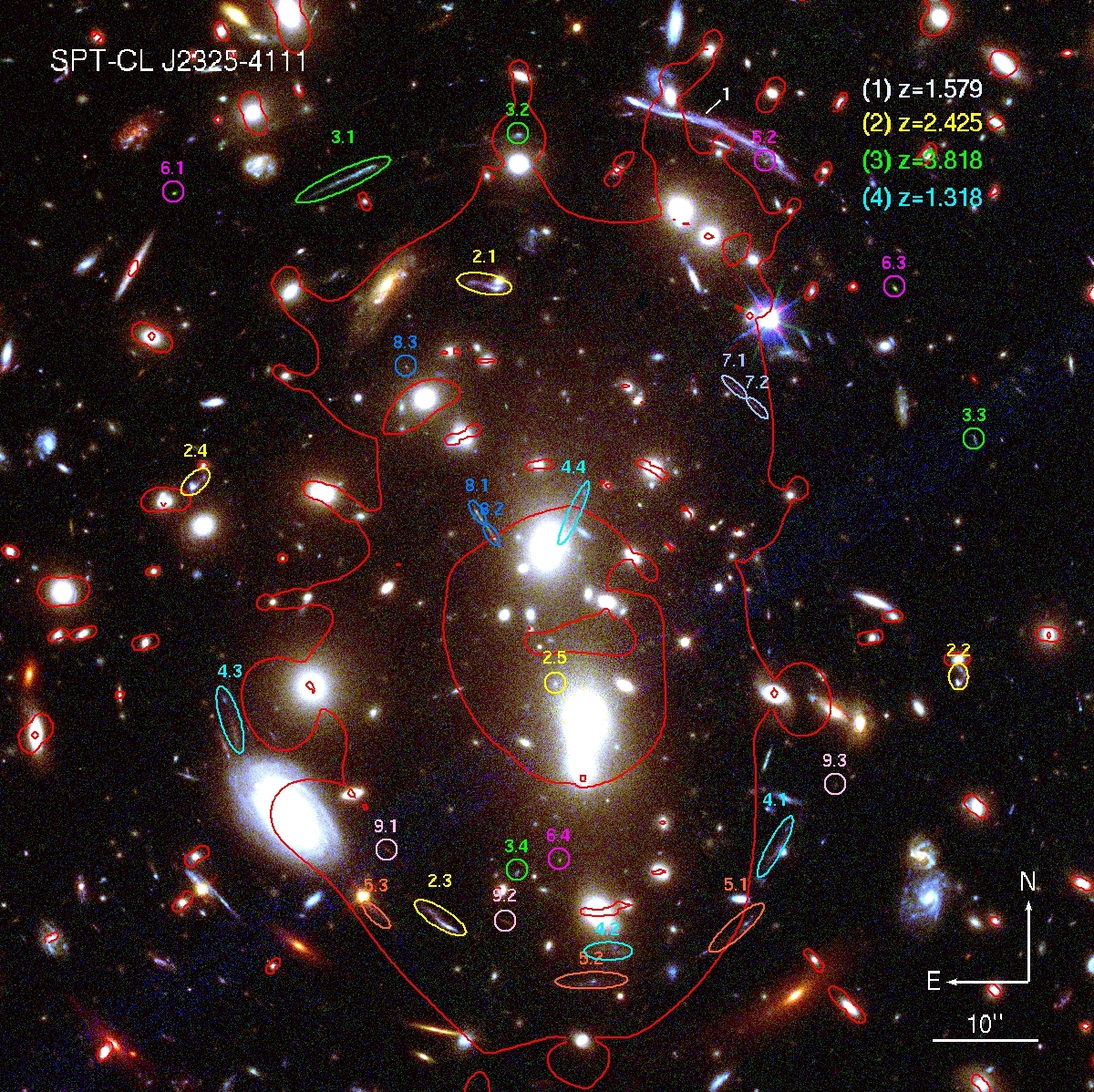}
    \caption{\done{Composite color images of the field of \SPTXXIII, from HST imaging in WFC3IR/F140W (red),  ACS/F814W (green), and ACS/F606W (blue). Multiple images of lensed features are labeled and color-coded by the source ID. The coordinates, redshifts (where available), and references of these strong lensing systems are presented in \autoref{tab:arcstable}. The critical curve for a source plane at $z=2$ is shown in solid red. North is up and East is to the left.}}
    \label{fig:hst2325}
\end{figure*}

\begin{figure*}
    \centering
    \includegraphics[width=\linewidth]{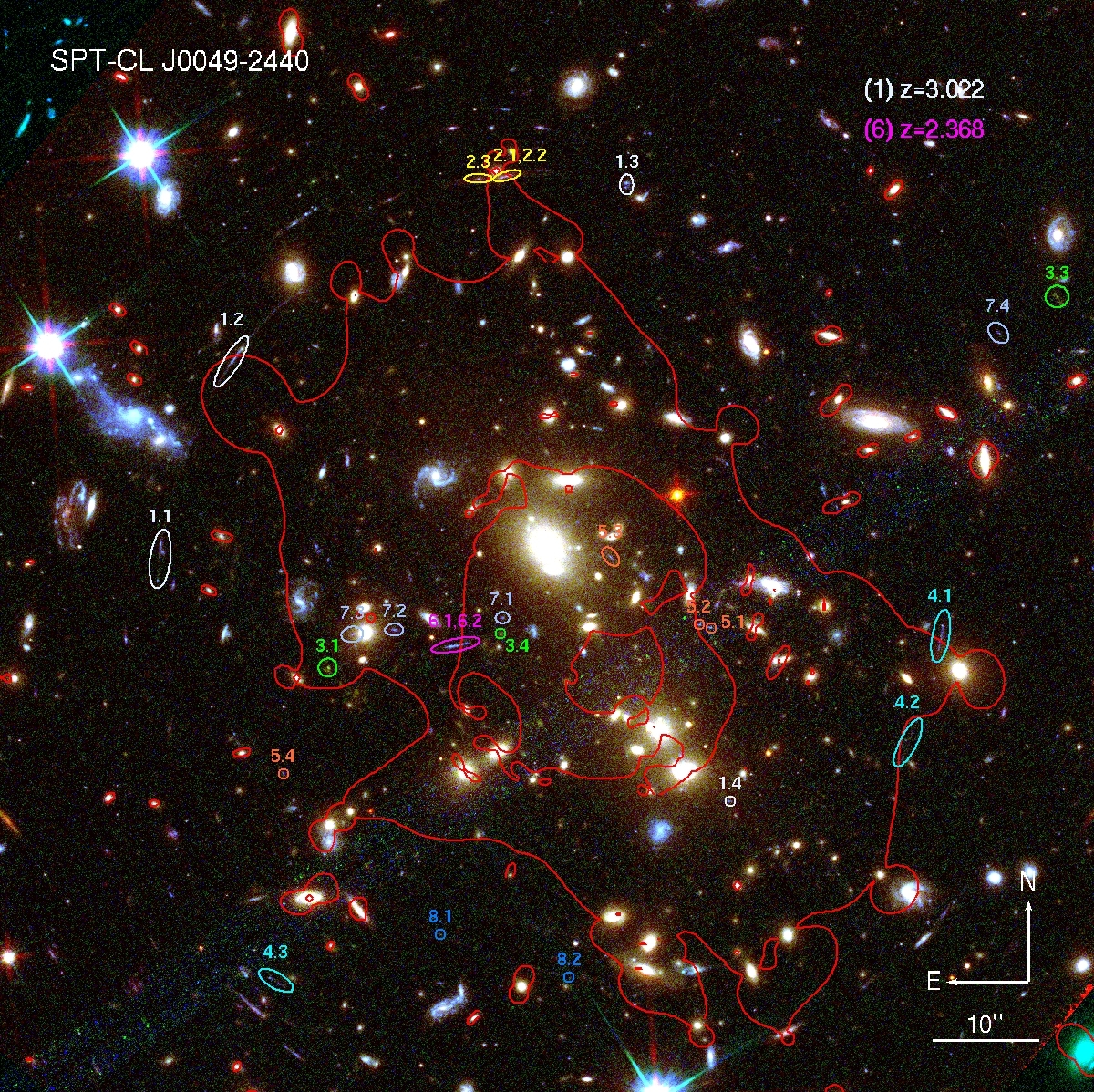}
    \caption{\done{Composite color images of the field of \SPTZERO, from HST imaging in WFC3IR/F140W (red),  ACS/F814W (green), and ACS/F606W (blue). Multiple images of lensed features are labeled and color-coded by the source ID. The coordinates, redshifts (where available), and references of these strong lensing systems are presented in \autoref{tab:arcstable}. The critical curve for a source plane at $z=2$ is shown in solid red. North is up and East is to the left.}}
    \label{fig:hst0049}
\end{figure*}

%% data reduction
\done{Data reduction was done similarly to \citet{sharon2020}.
We combined all the usable sub-exposures of each filter with the AstroDrizzle package \citep{gonzaga12} with a pixel scale of $0\farcs03$ pixel$^{-1}$, and drop size of 1.0 for the ACS images and 0.8 for WFC3. Observations that were executed over multiple visits were aligned to a common world coordinate system (WCS) using \texttt{tweakreg}, and the WCS solutions were applied back to the individual sub-exposures with \texttt{tweakback}. Finally, we drizzled all the images of each field onto the same pixel frame using the same parameters as above. The clusters are presented in \autoref{fig:hst2325} and \autoref{fig:hst0049}}.

\subsection{Lensing Evidence}
Multiple images (arcs) of background galaxies are used as strong lensing evidence to constrain the lens model of each cluster. We identified multiple images of each background source through visual inspection of the HST imaging, based on color and morphology of the images. We have obtained spectroscopic confirmation of a subset of those (see \autoref{sec:specdata} below). Figures \ref{fig:hst2325} and \ref{fig:hst0049} show the identified multiple images in each field over-plotted on the HST data (see also \autoref{fig:stamps} for zoom-ins of all multiply-imaged systems identified). The multiple images of each source are labeled and color-coded by source ID. The coordinates, spectroscopic redshift information and other notes specific to each system are listed in \autoref{tab:arcstable}. In some of these sources we further identified emission clumps or other substructure, that could be matched between images, and used as additional constraints in the lens models. 
The IDs of images of lensed galaxies are labeled as A.B(.C) where A indicates the source ID (or system name); B identifies the lensed image within the multiple images family; C is the ID of emission knot within the image if we used more than one substructure of the images as constraint. For example, ID 1.3.2 would identify the third multiple image of knot \#2 in Source~1.

In \SPTXXIII\ we identified a total of 9 secure strongly-lensed sources with multiple images.  Of these, four strongly-lensed sources are spectroscopically confirmed. 
In the field of \SPTZERO\ we identified a total of 8 secure systems, with spectroscopic redshifts for two. Other arc-like features that were not used in this analysis can be seen in the HST images.
We further discuss the lensing evidence in \autoref{sec:lensing-cstr}

\subsection{Spectroscopy}\label{sec:specdata}
\done{Spectroscopic observations of \SPTXXIII\ and \SPTZERO\ were obtained as part of larger campaigns to follow up SPT-selected clusters of galaxies. We collected all the available spectroscopy for these two clusters for the primary purpose of measuring spectroscopic redshifts of candidate lensed sources and cluster-member galaxies. These observations used \new{the Magellan 6.5-m telescopes at Las Campanas Observatory, with}
the Inamori-Magellan Areal Camera and Spectrograph \citep[IMACS;][]{dressler11} on Magellan I-Baade, the upgraded Low Dispersion Survey Spectrograph (LDSS3-C) on Magellan II-Clay, and the Folded port InfraRed Echellette \citep[FIRE;][]{FIRE} spectrograph on Magellan I-Baade.
We describe the observations and data reduction of Magellan LDSS3, IMACS, and FIRE in the following subsections. Spectroscopy results for specific objects in the following subsection refer to the object IDs in \autoref{fig:hst2325}, \autoref{fig:hst0049}, and \autoref{tab:arcstable}. The compilation of spectroscopic catalogs for these fields is described in \autoref{sec:redshiftcatalog}.}

\subsubsection{Magellan LDSS3 and IMACS MOS sopectroscopy}
%Lindsey: description of LDSS3 observations, masks, spectroscopy data reduction
\done{We obtained spectroscopy of
%used the upgraded LDSS3 spectrograph on the Magellan 6.5-m Clay telescope to observe 
\SPTXXIII\ and \SPTZERO\ with custom designed multi object slit (MOS) masks \citep{remolina_thesis}. Spectroscopic observations that took place before the HST data were in hand used Magellan/PISCO imaging to guide the mask design \citep[see][for description of the PISCO data]{Bleem2020}}.
%Juan's thesis: https://deepblue.lib.umich.edu/handle/2027.42/169844

\done{\SPTXXIII\ was observed with LDSS3 on June 7, 2016, with two multi object slit masks, each observed for $4\times25$ minutes, in clear conditions and seeing of $1\farcs0$. 
% look for mask smf files maybe with the data for the night, on /usr/Cirrus2/data/LDSS3/ut160607
Slits were placed on candidate lensed features as identified from ground-based imaging, including 2.1, 2.3, 3.1, 4.1, 4.3, 
and on cluster-member galaxies selected by color. }
\done{\SPTZERO\ was observed with LDSS3 on Dec 20, 2017, with one multi object slit mask, $2\times20$ min exposure, in clear weather and $0\farcs8$ seeing.
% yay! found the SMF files on SPT Slack. 
Slits were placed on candidate lensing features, including 1.2, 5.1/5.2, 6.1/6.2, and 4.3, and cluster-member galaxies. }

%\subsubsection{Magellan IMACS}
%Someone/Lindsey?: description of IMACS observations, masks, spectroscopy data reduction
\done{\SPTXXIII\ was observed with IMACS on May 27, 2015, in clear weather and seeing of $0\farcs7-0\farcs75$. We observed one multi object slit mask in three subexposures, for a total of 6600 s, using the \texttt{IMACS\_grism\_200} disperser. Only 10 out of the 108 slits were within the HST field of view, targeting arc-like features as identified from ground based data, including the giant arc of source 1 for which a spectroscopic redshift of $z=1.5790$ was previously observed with Gemini/GMOS \citep{Bayliss2016}. The remaining slits in the large field of view of IMACS were placed primarily on cluster-member galaxies, selected by color from ground-based photometry. }

%\red{\textbf{LDSS3, IMACS data reduction} -- spitballing here, based on info from other SPT papers that used LDSS3 (Khullar et al. 2019 \url{https://ui.adsabs.harvard.edu/abs/2019ApJ...870....7K/abstract}) I ran this by Lindsey and she confirms. 4/18/2024}.
\done{We reduced the LDSS3 and IMACS spectra using the Carnegie Observatories System for Multi-Object Spectroscopy \citep[COSMOS\footnote{\url{https://code.obs.carnegiescience.edu/cosmos}};][]{dressler11,oemler17}. The de-biased raw data were flat-fielded using flat field images that were taken immediately before or after each science frame. For wavelength calibration, we used HeNeAr comparison arc frames, also taken immediately before or after each science observation. Sub-exposures of the same mask were coadded, and the off-source area within the same slit was used to subtract the sky. The 1-D spectrum of each targeted object was extracted using custom Python routines following standard methods.}

\done{While yielding ample redshifts for cluster member galaxies and foreground objects, both LDSS3 and IMACS resulted with limited success in securing spectroscopic redshifts of lensed galaxies. This was in part because these observations took place prior to our HST program and many arcs were not identified at the time, and in part due to the wavelength coverage of these instruments. }

\subsubsection{Magellan FIRE spectroscopy of arc candidates}

\done{
\SPTXXIII\ and \SPTZERO\ were spectroscopically observed with 
%the Magellan-I 6.5-m telescope at Last Campanas Observatory using the Folded port InfraRed Echellette (FIRE) 
the FIRE
spectrograph on December 4 and 5 2019, in good conditions, and seeing ranging between $0\farcs55-1\farcs0$. Data reduction used standard \texttt{IRAF} techniques, and the redshifts were measured from lines identified primarily in the 2-D spectra.}

\done{
In the field of \SPTXXIII, we targeted arcs 2.3, 7.1, 8.1, and 3.1 on the first night and arcs 2.2, 4.3, 3.1 and \GMone\ (\new{a blue arc-like feature north of 4.1}) on the second night. Each target was observed with a $1\farcs0$ slit for $2\times602$ sec unless otherwise specified, executing a small A/B dither along the slit between exposures.}

\done{
We measured secure redshifts for three sources with FIRE. Source~2 is at $z=2.4253\pm0.0007$, based on bright [OIII], H$\alpha$, and H$\beta$ lines in emission in the spectrum of Arc~2.3. We identified the same lines in the spectrum of a Arc~2.2, an image of the same source, obtained on the second night. 
The spectrum of Arc~3.1 in night 1 was inconclusive; A deeper observation of $2\times1205$ sec was obtained for Arc~3.1 on the second night, resulting in a secure redshift of $3.8180\pm0.0007$ from two OIII $\lambda\lambda 4959, 5007$\AA\ lines.}

\done{
We identified one line in the spectrum of Arc 4.3, which can be interpreted as either [OIII] $\lambda 5007$\AA\ at $z=2.037$, or H$\alpha$ at $z=1.318$. The lensing analysis strongly favored the $z=1.318$ solution, resulting in a 5$\times$ lower $\chi^2$, and a factor of $10$ reduction in source plane rms. We, therefore, adopt this solution as the redshift of this source.  }

\done{
No emission lines were identified in the spectra of arcs 7.1, 8.1, and \GMone\ at the depth of our data. 
}

\done{
In the field of \SPTZERO, we targeted arcs 1.2, 2.1/2.2, 4.1, and 5.1/5.2, on the first night and arcs 1.2, 1.3, 3.1, and 6.1/6.2 on the second night. We secured a spectroscopic redshift of source 1 at $z=3.022$ from OIII. In arcs 6.1/6.2, we observed a faint line at $16869$\AA, which could be either OIII 5008 at $z=2.368$ or H$\alpha$ at $z=1.570$. We produced lens models for both solutions, and found that the lens model that used the lower redshift as constraint failed to produce the observed radial arc at the observed position, and generated predicted counter images for this arc that are not observed in the data. We therefore proceed with a redshift of $z=2.368$ for this arc.
}

\done{
Spectra of the other targeted arcs in \SPTZERO\ did not yield emission lines. 
Since the spectrum of arc 1.3 did not confirm it as a counter-image of arc 1.2, we did not use it as constraint in the lens model. 
}

%%%%%

%\blue{TODO: Is all the info there}

%%%%%%%%
\subsection{Spectroscopic Redshifts Measurement and Redshift Catalogs}\label{sec:redshiftcatalog}
\done{We extracted redshifts from the observed LDSS3, IMACS, and FIRE spectra as follows. We convolved the reduced 2D spectra with a 1D Gaussian profile along the wavelength axis. The Gaussian parameters were fit to a stack of spectra of a manually-selected clean part of the slit, to increase the signal-to-noise and derive robust values. The spectra are not strictly perpendicular to the slit, ignoring any higher-order distortion, we extracted the 1D spectra along a manually adjusted linear trace on the wavelength axis. The redshift assessment was performed by matching spectral features to the most common emission and absorption lines. }

\done{We assigned a confidence level to each spec-$z$ based on the number and strength of spectral features according to the following rules:
\begin{itemize}
    \item Confidence 3: secure redshift, with several strong spectral features.
    \item Confidence 2: probable redshift, relying on one emission line or several faint absorption features.
    \item Confidence 1: tentative redshift, relying on one spectral feature with low s/n.
\end{itemize}
}

\begin{figure*}
    \centering
    \includegraphics[width=0.45\linewidth]{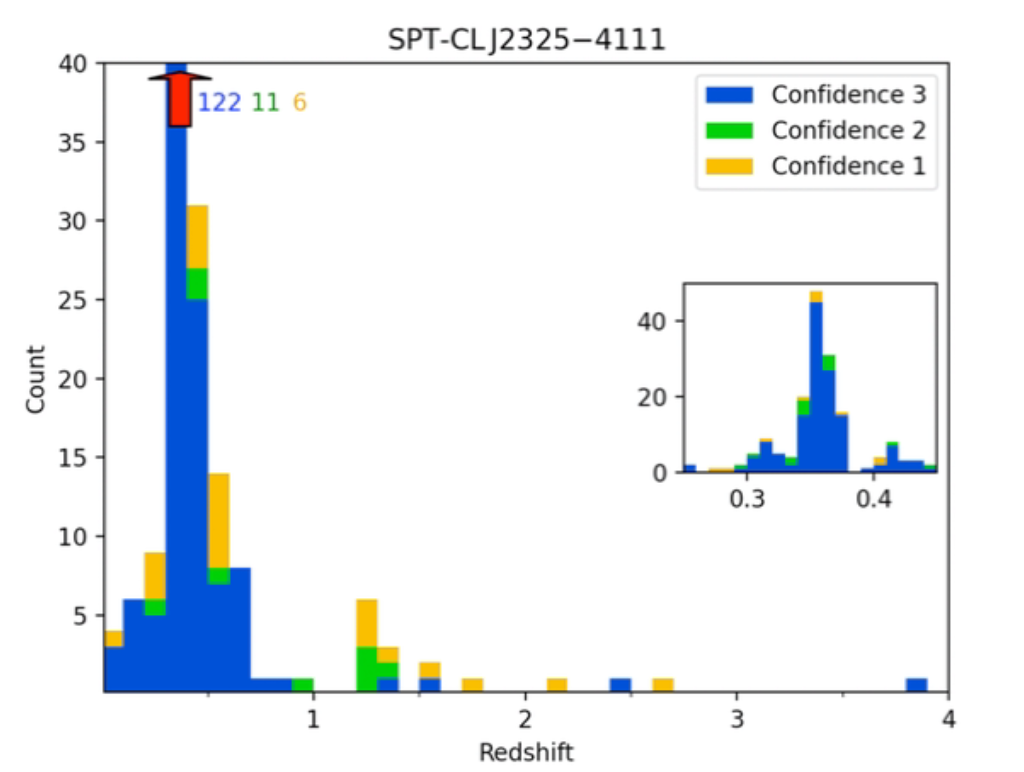}
    \includegraphics[width=0.45\linewidth]{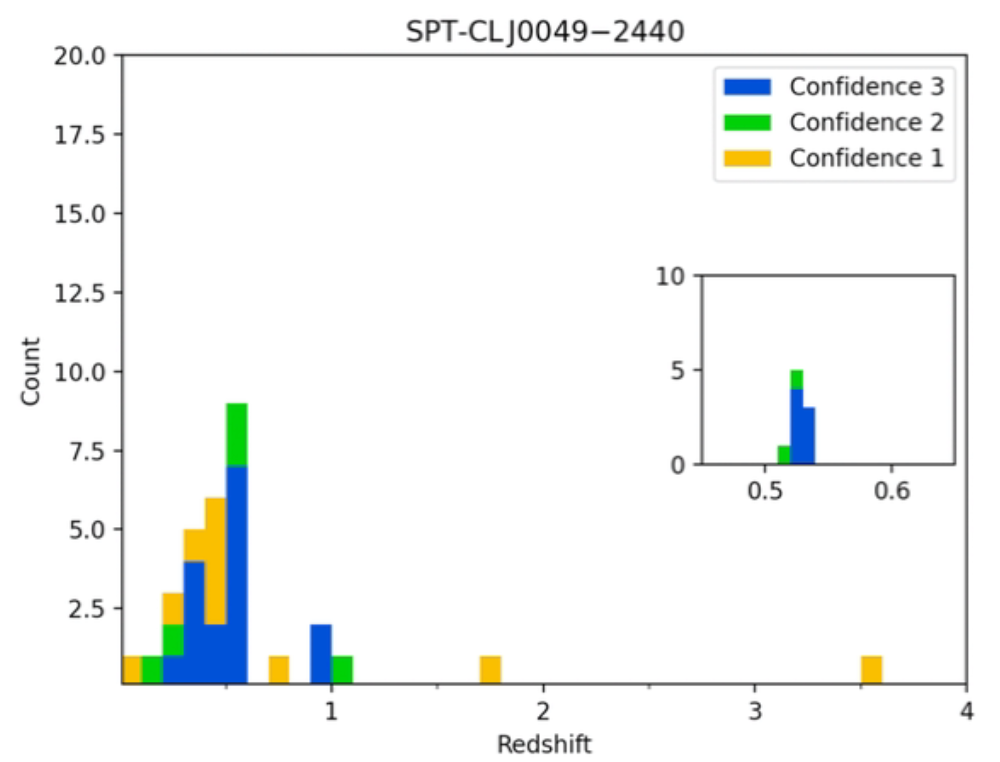}
    \caption{\done{Redshift histogram of all the objects with spectroscopic redshift measurements in the field of view of \SPTXXIII\ (left) and \SPTZERO\ (right). The redshifts are tabulated in \autoref{tab:spt2325allz} and \autoref{tab:spt0049allz}, respectively. The insets show a zoom-in around each cluster redshift, to illustrate the velocity structure in a narrower redshift range. The redshift bin size is $\Delta_z=$0.1 in the main figure and $\Delta_z=$0.01 in the insets. The different colors represent the spec-$z$ confidence level as described in \autoref{sec:redshiftcatalog}.} For better display the left histogram is cut at 40 counts and the red arrow indicates the number of galaxies in the unique redshift bin going beyond 40. }
    \label{fig:histo}
\end{figure*}

\done{We complemented our final catalog with the reported GMOS spectroscopy measurements for \SPTXXIII\ from \cite{Bayliss2016}.
% paper for 2325 looks like it's this one: https://ui.adsabs.harvard.edu/abs/2016ApJS..227....3B/abstract
% the 0049 cluster is not in the tables in Matt's 2016 papar.
For sources that were measured by multiple instruments, we report the highest confidence measurement, where GMOS has the highest confidence, followed by FIRE, LDSS3, and IMACS.
Our final catalog for \SPTXXIII\ contains 230 galaxies, of which 224 are not multiple images of lensed sources, and for \SPTZERO\ we have spectroscopic redshifts for 29 galaxies in the field, and redshifts of two additional lensed sources. The non-arcs catalogs are presented in \autoref{tab:spt2325allz} and \autoref{tab:spt0049allz}, and redshifts of strongly lensed galaxies with multiple images are listed in \autoref{tab:arcstable}.
The redshift histograms for both cluster fields are presented in \autoref{fig:histo}.
}

%\subsubsection{\SPTV}\label{sec:SPT0512}
%\subsubsection{\SPTZERO}\label{sec:SPT0049}
%\subsubsection{\SPTXXIII}\label{sec:SPT2325}

\section{Strong Lensing Analysis} \label{sec:modeling}

\subsection{Lens Modeling}
\done{The strong lens models were constructed using the public lens modeling software \lenstool\ \citep{Jullo2007}. The algorithm uses Markov Chain Monte Carlo (MCMC) sampling to determine the best-fit model, and explores the parameter space to facilitate a statistical assessment of the uncertainties in model parameters and measurements derived from the lens model. The best solution is determined by minimizing the scatter between observed and model-predicted image-plane positions of the lensing constraints, i.e., images of lensed background galaxies. }

\done{As with many other parametric algorithms, \lenstool\ assumes that the mass distribution of the lens can be fairly described by a linear combination of \new{halos described by a parameterized mass distribution}. Several mass density profiles are implemented, e.g., isothermal, or the Navarro–Frenk–White \citep[NFW;][]{Navarro1996} profile. In this work, we use the pseudo-isothermal ellipsoidal mass distribution (dPIE; \citealt{Eliasdottir2007}, also referred to in the literature as PIEMD), which has an elliptical geometry, flattened core, and a truncated isothermal slope of $\rho \propto r^{-2}$. The halo is described by seven parameters: $\alpha$, $\delta$ centroid, ellipticity $e$, position angle $\theta$, core radius $r_{core}$, truncation radius $r_{cut}$, and normalization $\sigma_0$. Note that $\sigma_0$ represents a fiducial central velocity dispersion as defined in the dPIE parameterization, and is not equal to the observed velocity dispersion \citep[ for the parameterization of dPIE, and the relationship between an observed velocity dispersion and $\sigma_0$, we refer the reader to][] {Eliasdottir2007}.
The mass of galaxy clusters is dominated by dark matter, and may best be described by more than one dominant cluster-scale halo. Cluster-member galaxies contain a small fraction of the total cluster mass, but have an important contribution to the complexity of the lensing potential. We model cluster member galaxies as dPIE halos as well, but link their parameters more strongly to their observed stellar mass. In particular, their positional parameters ($\alpha$, $\delta$, $e$, $\theta$) remain fixed to the properties of their light distribution as measured with Source Extractor \citep{SEx}. The slope parameters are linked to the luminosity through scaling relations \citep{Jullo2007} that are optimized in the modeling process for the entire galaxy catalog as a whole.} Some galaxies were optimized separately from the scaling relations, including the BCGs and other galaxies in close proximity to multiple images.
% between $r_{core} \lesssim r \lesssim r_{cut}$.

\done{The modeling process is done iteratively, starting from the most obvious and secure lensing constraints (multiple images and arcs) to inform a preliminary lens model. The model is then used to assist in identifying more images of lensed galaxies that can be used as constraints. When each new set of constraints is added to the analysis, the modeling process is re-initiated in order to not bias the model to fixate on an early solution. 
\autoref{table:parameters} lists the optimized and fixed model parameters for each cluster, with their best-fit solutions and uncertainties as determined from the MCMC analysis. }

\subsection{Identification of lensing constraints}
\label{sec:lensing-cstr}
 We used the locations of multiply-imaged systems as constraints to our modelling. We identified multiple image ``families'' based on the morphology, color and spectroscopic redshift. When spectroscopic redshift was not available we associated the images of the family based on color and morphology only. If the spectroscopic redshift was only measured in a subset of the family we associated the same redshift to all images of the family. 
% %%% moving this to the new subsection before spectroscopy %%%
%The IDs of images of lensed galaxies are labeled as A.B(.C) where A is the number indicating the source ID (or system name); B is the number indicating the ID of lensed image within the multiple images family; C is a number indicating the ID of emission knot within the image if we used more than one substructure of the images as constraints. 
\autoref{tab:arcstable} lists all the identified lensed constraints; {\autoref{fig:hst2325} and \autoref{fig:hst0049} label the multiple images of each source. The coordinates of clumps within images that were used as further constraints are only listed in the table, to avoid over-cluttering the figures.}
We used as constraints in our models only multiple images that were considered as secure. We consider secured candidate when morphology, color and lensing configurations through the iterative process converged toward being images of the same source. Candidate lensed system are suggestive multiply-imaged system or images within a secure system which are often too faint, or contaminated by light from a nearby galaxy, to be securely associated to lensing constraints. Although they can often be geometrically confirmed with by the lens model, we exclude them from the list of constraints in order to avoid confirmation bias (see \autoref{tab:arcstable} for systems marked with a ``$c$'' and considered only as candidate).

 In \SPTXXIII\ we identified nine multiply-imaged systems; one (system 7) is labeled as candidate and eight are considered secure. Image 6.2 is considered a candidate due to contamination from arc~1. 

In \SPTZERO\ we identified nine multiply-imaged systems, all of which are secure. A candidate is identified as image 3 of source~1. This system is further discussed in \autoref{sec:lensed}, and the attempt to spectroscopically confirm the candidate image is discussed in \autoref{sec:specdata}. Image 4.2 is an extended faint arc, likely affected by foreground galaxies. We consider it as a candidate as well.
%Secure system are considered  System 1 is giant arc and an obvious multiply-imaged system. We used several emission knots for constraining the lensing configurations 
%See \autoref{tab:arcstable} for the exact positions (and Appendix).    

\subsubsection{Identification of Cluster-Member Galaxies}\label{sec:clustermembers}
\begin{figure*}
    \centering
    \epsscale{0.55}
    \plotone{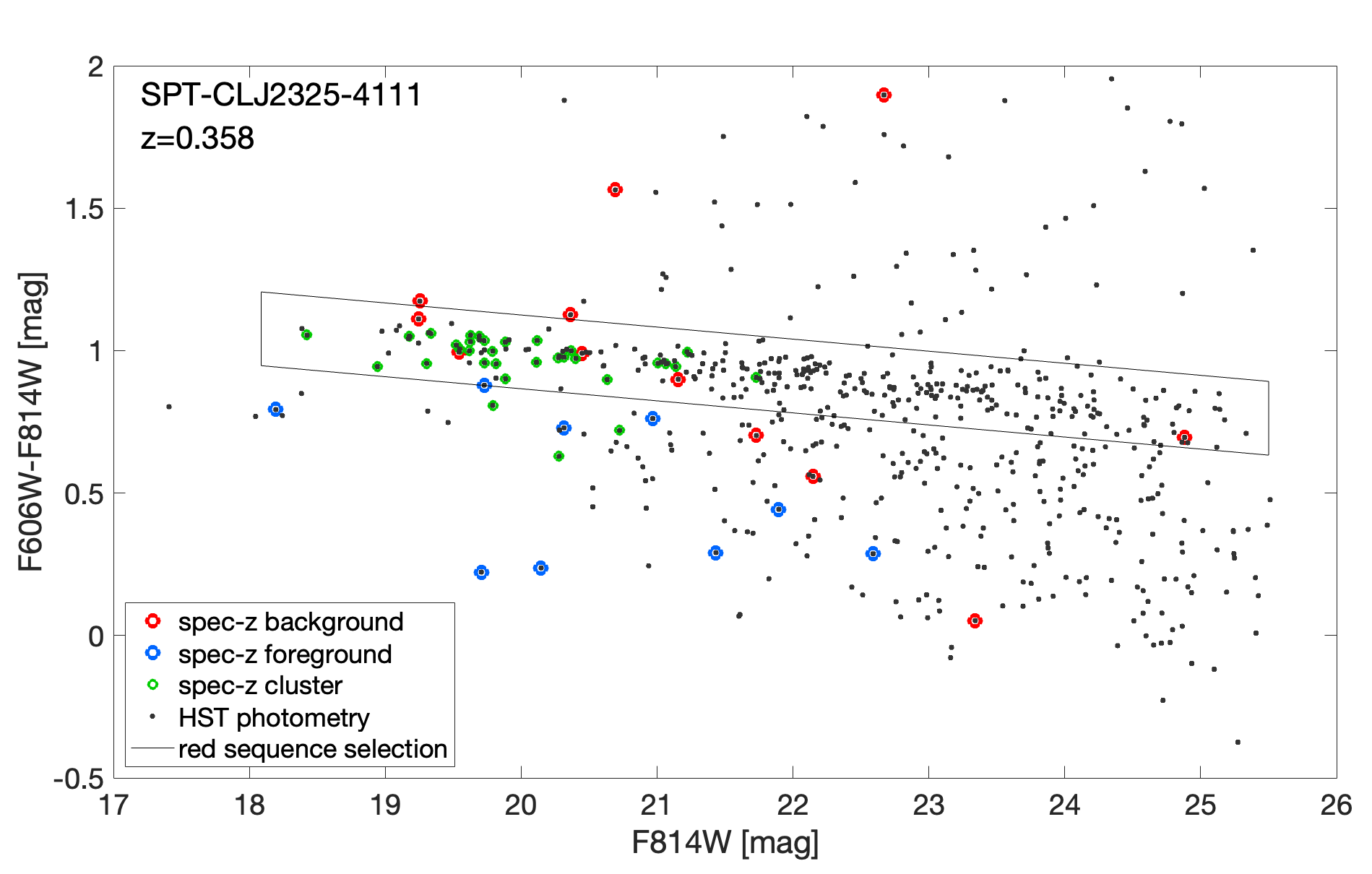}
    \plotone{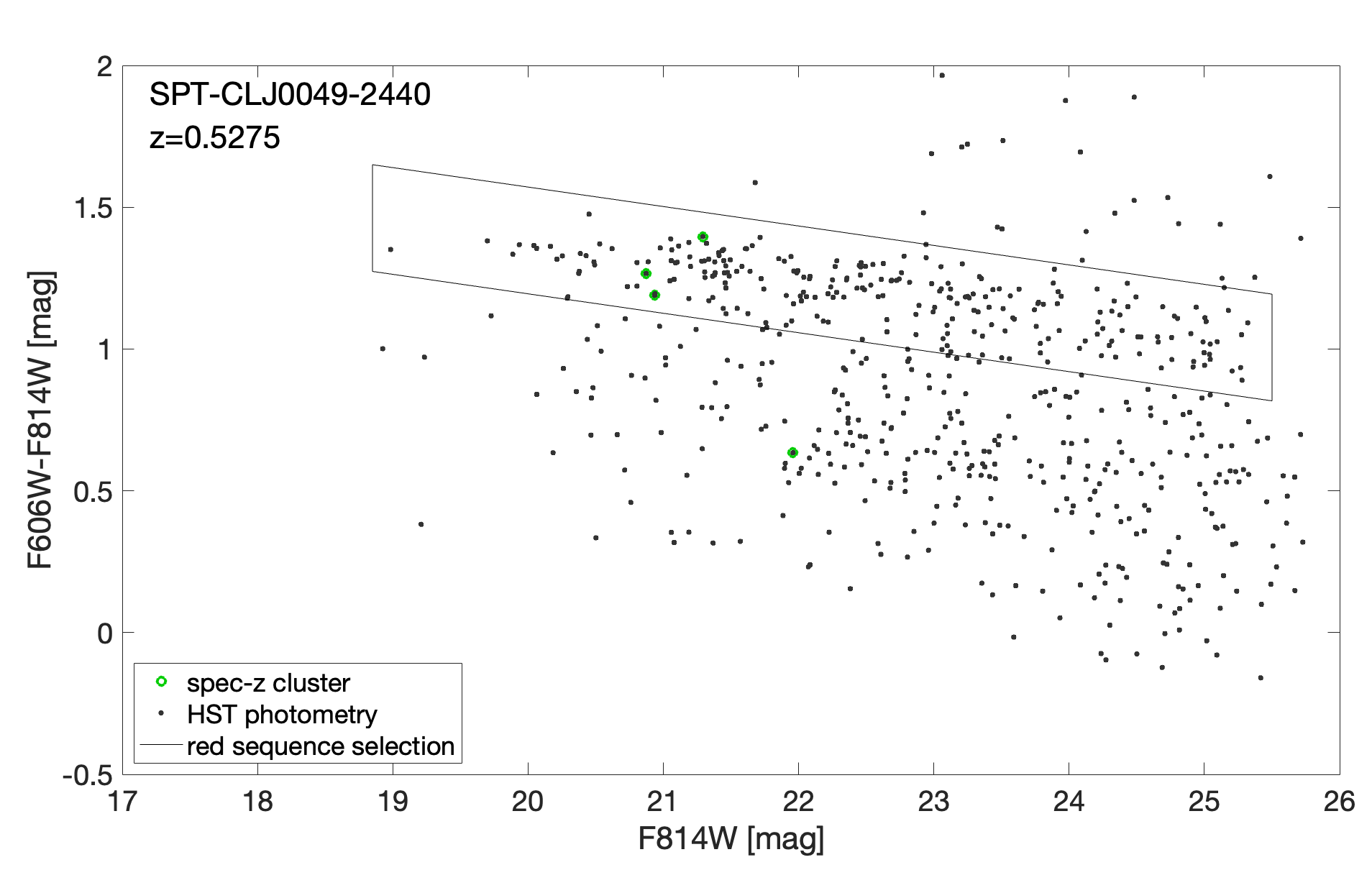}
    \caption{\done{Color magnitude diagrams of \SPTXXIII\ (left) and \SPTZERO\ (right). HST/ACS F606W-F814W color vs F814W magnitude is plotted in black for galaxies within the ACS field of view. Spectroscopically-confirmed galaxies are color-coded by their redshift with respect to the cluster (background, foreground, or at the cluster redshift, see legend). The red-sequence selection is marked with black lines. See \autoref{sec:clustermembers} for more details.}}
    \label{fig:cmd}
\end{figure*}

\done{Cluster-member galaxies were identified based on their color with respect to the cluster red sequence \citep{Gladders2000} in a color-magnitude diagram, using spectroscopic redshift information where available. 
The catalogs were constructed as follows. We started by running Source Extractor \citep{SEx} in dual image mode, using the F814W image for identification and measuring the F814W \texttt{MAG\_AUTO} and F606W-F814W color within the same detection apertures. We flagged and removed stars and artifacts based on their location in a \texttt{MU\_MAX} vs \texttt{MAG\_AUTO} space. We then matched the coordinates of the photometric and spectroscopic catalogs. 
For \SPTXXIII, where numerous objects with spectroscopic redshifts are within the HST field of view, 
we fit a line to the F606W-F814W color vs.\ F814W magnitude of cluster members (those within $\Delta z=0.03$ of the cluster redshift) using iterative 3$\sigma$-clipping, which successfully removes blue cluster-member galaxies from the fit. We set the red sequence selection as $5\sigma$ above and below the fit, with a faint-end limit of 25.5 mag and a bright end limit set by the magnitude of the brightest cluster galaxy (BCG), to account for intrinsic scatter in the red sequence and reduce contamination from faint field galaxies.  The model of \SPTXXIII\ includes 277 cluster member galaxies, of which six were optimized separately from the scaling relations; the model of \SPTZERO\ includes 224 cluster members, two of which were optimized separately, as listed in \autoref{table:parameters}.
The color-magnitude diagrams are shown in \autoref{fig:cmd}. Spectroscopically-confirmed cluster members, foreground, and background galaxies within the ACS field of view, are marked in color.}

\subsubsection{Lens Modelling Results}
\done{The lensing analysis of these lines of sight resulted in two well-calibrated ``cosmic telescopes'', as indicated by the low image plane scatter between observed and predicted images of lensed sources: rms $=0\farcs63$ for \SPTXXIII, and rms $=0\farcs73$ for \SPTZERO. 

%{\bf The image plane rms, $0\farcs63$ and $0\farcs73$ for \SPTXXIII\ and \SPTZERO, respectively, 
The image plane rms reflects the ability of the best-fit lens models to describe the lensing potential and reproduce the lensing evidence. 
A high rms could imply that the underlying mass distribution is more complex than what is permitted by the flexibility of the parameterized modelling. On the other hand, a very low rms could point to overfitting. Our results are in the same range of other clusters with similar richness of lensing evidence in the literature (e.g. \citealt{Richard2011,Zitrin2017,Cerny2018}).
%We note that reported uncertainties are only statistical and subject to systematic uncertainties (see, e.g.,  \citealt{Johnson2016} for a in depth quantification). 
In addition to the overall rms, we report in \autoref{tab:arcstable} an indicator of goodness of fit for each image, in the form of the distance between its observed and predicted position. These values highlight which systems and images perform better and help assess the fidelity of the model.

\autoref{fig:hst2325} and \ref{fig:hst0049} show the critical curves for a source at $z=2$ for the best-fit model of each cluster.
We confirm the initial assessment that both of these clusters have a massive core, as indicated by the large separation between the giant arcs and the BCG. We measured the effective Einstein radius of each cluster, defined as  $\theta_E=\sqrt{A/\pi}$, where $A$ is the area of an ellipse fit to the tangential critical curve. For a source at the redshift of the most prominent giant arc in \SPTXXIII,  we measured $\theta_E(z=1.579)=32$\arcsec, whereas for a generic $z=9$ source plane we measured $\theta_E(z=9)=42$\arcsec. Similarly, for \SPTZERO\ the effective Einstein radii are $\theta_E(z=3.022)=36$\arcsec\ and $\theta_E(z=9)=43$\arcsec.}

\done{The clusters are generally well-represented by one cluster-scale halo ($\sigma_0 >1000$ \kms), with contribution from a high-mass galaxy-scale halo ($\sigma_0 \gtrsim 500$ \kms) near the cluster core, and the rest are more typical cluster-member galaxies.  }
\done{The best fit parameters of the lens models and their 68\%-ile upper and lower limits are tabulated in \autoref{table:parameters}. As part of the lensing analysis, the minimization process also solves for the unknown redshifts of multiply-imaged lensed sources that were used as constraints. These are reported in \autoref{tab:arcstable} as median$\pm$ 68\%-ile, determined directly from the MCMC sampling of the parameter space. }

\done{
We also calculated the lensing magnification and the projected mass density map of each cluster. Uncertainties of these lensing outputs were derived from the MCMC sampling of the parameter space, by selecting 100 steps from the chain at random, and producing the relevant mass and magnification outputs for each of these realizations of the model. 
We note that the statistical uncertainties underestimate the true uncertainty and do not take into account systematic errors (see \citealt{Meneghetti2017} for a detailed discussion, and \citealt{Johnson2016} for a quantitative assessment of systematic uncertainties as related to the number of arcs and spectroscopic redshifts).

\autoref{tab:arcstable} lists the model-predicted magnification at the observed location of each arc. For arcs without a spectroscopic redshift, the predicted magnification of each realization was calculated for the redshift parameter associated with the same step in the MCMC chain. We discuss the results in the following sections. }

\begin{deluxetable*}{l|rcccccccc}
\tablecolumns{9}
\tablecaption{Lens Model Results and Best-Fit Parameters\label{table:parameters} }
\tablehead{ Model name & Component & $\alpha^{\rm ~a}$ & $\delta^{\rm ~a}$ & $e^{\rm ~b}$ & $\theta^{\rm ~c}$ & $\sigma_0^{\rm ~d}$ & r$_{\rm cut}$ & r$_{\rm core}$\\ 
(Fit statistics)$^{\rm e}$ & \nodata & (\arcsec) & (\arcsec) &   & ($\deg$) & (km\ s$^{-1}$) & (kpc) & (kpc)
}
\startdata
\SPTXXIII\ & Halo 1 (cluster) & 7.34$^{+0.28}_{-0.36}$ & 11.99$^{+0.94}_{-1.01}$ & 0.29$^{+0.01}_{-0.01}$ & 73.7$^{+0.6}_{-0.8}$ & 1332.2$^{+14.2}_{-20.1}$ & [1500] & 39.9$^{+0.1}_{-1.7}$\\
rms = $0\farcs63$;  k = 30 & Halo 2 (galaxy) & [$0.0$] & [$0.0$] & [$0.29$] & [$-83.04$] & 619.1$^{+25.8}_{-20.0}$ & 21.8$^{+4.5}_{-4.4}$ & 4.1$^{+0.5}_{-0.3}$\\
 $\chi^2/\nu$ = 9.0; dof = 16 & Halo 3 (galaxy) & [$3.40$] & [$16.21$] & [$0.31$] & [$77.46$] & 407.1$^{+6.0}_{-17.2}$ & 86.0$^{+17.3}_{-10.8}$ & 2.0$^{+0.1}_{-0.4}$\\
$\log$($\mathcal{L}$) = $-54$ & Halo 4 (galaxy)& [$-9.04$] & [$47.91$] & [$0.16$] & [$-43.21$] & 134.8$^{+15.6}_{-14.5}$ & 9.2$^{+10.2}_{-13.8}$ & 1.4$^{+0.2}_{-1.2}$\\
 $\log$($\mathcal{E}$) = $-123$ & Halo 5 (galaxy)& [$14.95$] & [$30.10$] & [$0.13$] & [$22.48$] & 287.4$^{+8.3}_{-7.7}$ & 54.1$^{+6.7}_{-8.3}$ & 1.2$^{+0.2}_{-0.3}$\\
BIC = 233 AICc = 221 & Halo 6 (galaxy)& [$24.61$] & [$21.27$] & [$0.27$] & [$-18.06$] & 100.2$^{+25.5}_{-0.9}$ & 3.4$^{+4.5}_{-0.2}$ & 0.9$^{+1.1}_{-0.3}$\\
 & Halo 7 (galaxy)& [$35.78$] & [$18.15$] & [$0.12$] & [$25.57$] & 19.7$^{+14.9}_{-6.0}$ & 37.8$^{+35.9}_{-0.8}$ & 2.0$^{+0.4}_{-0.7}$\\
 & $L^{*}$ Galaxy & \nodata & \nodata & \nodata & \nodata & 207.6$^{+12.7}_{-6.1}$ & 56.8$^{+13.0}_{-12.4}$ & \nodata \\
\hline
\SPTZERO\ & Halo 1 (cluster)& $-7.95^{+0.87}_{-2.34}$ & $-8.28^{+0.33}_{-2.22}$ & 0.4$^{+0.02}_{-0.09}$ & 136.5$^{+3.7}_{-1.5}$ & 1145.5$^{+49.3}_{-28.9}$ & [$1500$] & 16.9$^{+3.2}_{-1.8}$\\ 
rms = $0\farcs73$;  k = 24 & Halo 2 (galaxy)& 1.42$^{+0.41}_{-0.53}$ & $-1.89^{+0.39}_{-0.5}$ & 0.37$^{+0.03}_{-0.03}$ & 76.1$^{+23.3}_{-11.3}$ & 494.7$^{+155.0}_{-29.0}$ & 117.7$^{+16.3}_{-28.8}$ & 3.7$^{+1.7}_{-0.6}$\\ 
 $\chi^2/\nu$ = 19.0; dof = 8 & Halo 3 (galaxy)& [$25.08$] & [$10.83$] & [$0.42$] & [$-51.14$] & 267.0$^{+13.5}_{-15.2}$ & 10.7$^{+0.7}_{-0.8}$ & 0.8$^{+0.4}_{-0.2}$\\ 
$\log$($\mathcal{L}$) = $-61$ & $L^{*}$ Galaxy & \nodata & \nodata & \nodata & \nodata & 273.5$^{+4.9}_{-19.1}$ & 175.6$^{+20.5}_{-28.3}$ & \nodata \\ 
 $\log$($\mathcal{E}$) = $-323$ & \nodata & \nodata & \nodata & \nodata & \nodata & \nodata & \nodata & \nodata  \\
BIC = 215 AICc = 218 & \nodata & \nodata & \nodata & \nodata & \nodata & \nodata & \nodata & \nodata \\ 
\hline 
\enddata
\tablecomments{\done{$^{\rm a}$ $\alpha$ and $\delta$ are the positions measured in arcseconds relative to the reference coordinate point for \SPTXXIII\ (R.A. $ = 351.298863$, Decl. $ = -41.203566$) and \SPTZERO\ (R.A. $ =12.295750$, Decl. $ = -24.678583$). $^{\rm b}$ Ellipticity ($e$) is defined as $(a^2-b^2) / (a^2+b^2)$, where $a$ and $b$ are the semi-major and semi-minor axes of the ellipse. $^{\rm c}$ $\theta$ is measured north of east. $^{\rm d}$ $\sigma_0$ is the normalization parameter and represents a fiducial central velocity dispersion as defined in the dPIE parameterization. Statistical uncertainties were inferred from the MCMC optimization and correspond to a 68\% confidence interval. Parameters in square brackets were not optimized. The position and the ellipticities of the mass clumps associated with cluster galaxies were kept fixed according to their light distribution, and the other parameters were determined through scaling relations (see text). $^e$ Fit results for each model are given in the left column. rms is the scatter in the image plane; k is the number of free parameters; dof stands for the number of degrees of freedom; BIC is the Bayesian Information criterion and AICc is the corrected Akaike information criterion. }}
\end{deluxetable*}

\section{Discussion} 
\label{sec:prop}

\subsection{Mass profiles and substructures}

Given the large radial extent of the lensing constraints in these lines of sight, we can accurately measure the total enclosed projected mass density out to relatively large radii. The core mass of \SPTXXIII, measured within 500~kpc, is  
$M(<500~{\rm kpc}) = 7.30\pm0.07 \times 10^{14}$\msun, 
%$M(<500 ~{\rm kpc}) = 730.27^{+7.15}_{-7.18} 10^{14}$\msun.
and the core mass of \SPTZERO\ is $M(<500 ~{\rm kpc})=7.12^{+0.16}_{-0.19}\times 10^{14}$\msun. 
%$711.67^{-16.71}_{+19.34}$\msun
%present a median mass and 68\% error within 500~kpc of $730.27^{+7.15}_{-7.18} 10^{14}$\msun.
% the error percentage at percentile at 15.9 and 84.1 which should match 1 sigma if I am not mistaken
%\SPTZERO\ present a median mass and 68\% error within 500~kpc of $711.67^{-16.71}_{+19.34}$\msun. 

\begin{figure*}
    \centering
    \includegraphics[width=\linewidth]{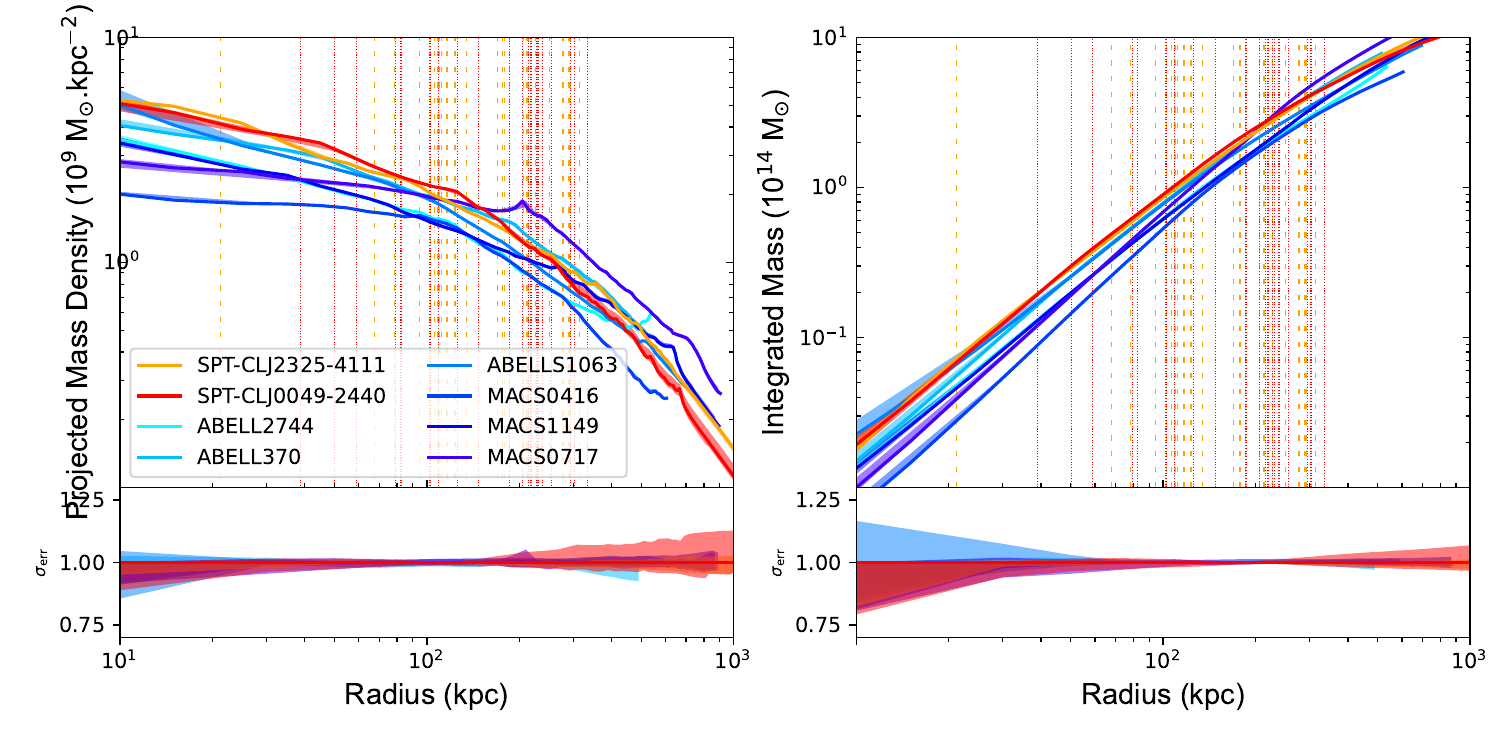}
    \caption{\done{Projected mass density profiles (\textit{left}) and cumulative projected mass profiles (\textit{right}) of \SPTXXIII\ (orange) and \SPTZERO\ (red) plotted against distance from the BCG. Vertical dashed orange and dotted red lines mark the position of multiple images for \SPTXXIII\ and \SPTZERO\ respectively. The six Frontier Fields clusters are plotted for comparison. The mass profiles of the two clusters studied in this work have higher density than the Frontier fields in the innermost $\sim200 {\rm kpc}$, and comparable density and large-scale mass at larger radii.}}
    \label{fig:massprofiles}
\end{figure*}

\autoref{fig:massprofiles} shows the projected mass density profiles of \SPTXXIII\ and \SPTZERO\ in the left panel, and the cumulative enclosed mass as a function of cluster-centric radius in the right panel. In both plots, the distances are measured from the BCG of each cluster. For comparison, we plot on the same figures the density profiles and cumulative mass profiles of the six Frontiers Fields clusters \citep{Lotz2017}, which we derived from the public \lenstool\ models of these clusters (\texttt{Sharon V4}\footnote{\url{https://archive.stsci.edu/prepds/frontier/}}). We find that at projected radii beyond $\sim200$ kpc, the density profiles and cumulative enclosed mass of both clusters are comparable to the average Frontier Fields cluster. However, closer to the cluster cores (R$<$200~kpc), both clusters have higher mass density profiles than the Frontier Fields clusters (left panel of \autoref{fig:massprofiles}). 
%This concentration of projected mass at the center might offer a rather circular shape of this cluster. 
This mass distribution may provide an explanation for the high lensing efficiencey of \SPTXXIII\ and \SPTZERO, consistent with the association of clusters with higher concentration with higher lensing efficiency \citep{Giocoli2012}.

Parametric lens modeling algorithms such as \lenstool\ can separate the contribution to the lensing potential from the different mass components, to calculate the fraction of mass contained in substructures and galaxy scale halos. We estimate the substructure mass by removing Halo~1, which represents the cluster-scale dark matter halo (see \autoref{table:parameters}), and measure the total projected mass density within 500~kpc associated with the remaining mass halos. We find that the substructure mass of \SPTXXIII\ amounts to a fraction of $0.12\pm{0.01}$ of the total mass, and in \SPTZERO, the substructure amounts to $0.21^{+0.07}_{-0.05}$ of the total mass (median and 68\% uncertainty as derived from the MCMC sampling). Previous studies reported amounts as low as 0.01 \citep{Mahler2019} and as high as 0.3 \citep{Mahler2020} of the total mass found in substructures. As shown in \citet{Richard2011}, using a sample of 20 clusters, substructure mass ratios ranges from 0.02 to 0.78 with a median at 0.135. The strong lensing efficiency of sub-halos in the context of predictions from cosmological model (\emph{$\Lambda$-CDM}) has been discussed in previous studies, see e.g., \citet{Grillo2015,Munari2016,Natarajan2017,Meneghetti2020,Meneghetti2022,Meneghetti2023,Bahe2021,Robertson2021,Tokayer2024} for more in-depth discussions of its impact.

%2325 med 0.12312106778315986 lower 1sigma -0.007729899897609374 higher 1 sigma 0.007492383929606786
%0049 med 0.21000676203360025 lower 1sigma -0.054856900506376816 higher 1 sigma 0.07374849452088206

%\begin{figure}
%    \centering
%    \includegraphics[width=\linewidth]{massratio.png}
%    \caption{Have a mass profile, not this ugly plot}
%    \label{fig:massratio}
%\end{figure}

\subsection{Lensing Strength}

\done{\cite{Fox2022} studied the ``lensing strength'' of 74 strong lensing clusters with public lens models and space-based imaging data available at the time. They defined the lensing strength as the total image-plane area in which a source at $z=9$ is magnified by a factor of 3 or above. They found that the lensing strength depends somewhat on the cluster mass, and more strongly on the inner slope of the projected mass density, where shallower inner slope can produce more powerful lenses. They also demonstrated that the Einstein radius and the projected distance between the farthest bright arc and the BCG are good predictors of lensing strength. }

\begin{figure*}
    \centering
    \includegraphics[width=0.45\linewidth]{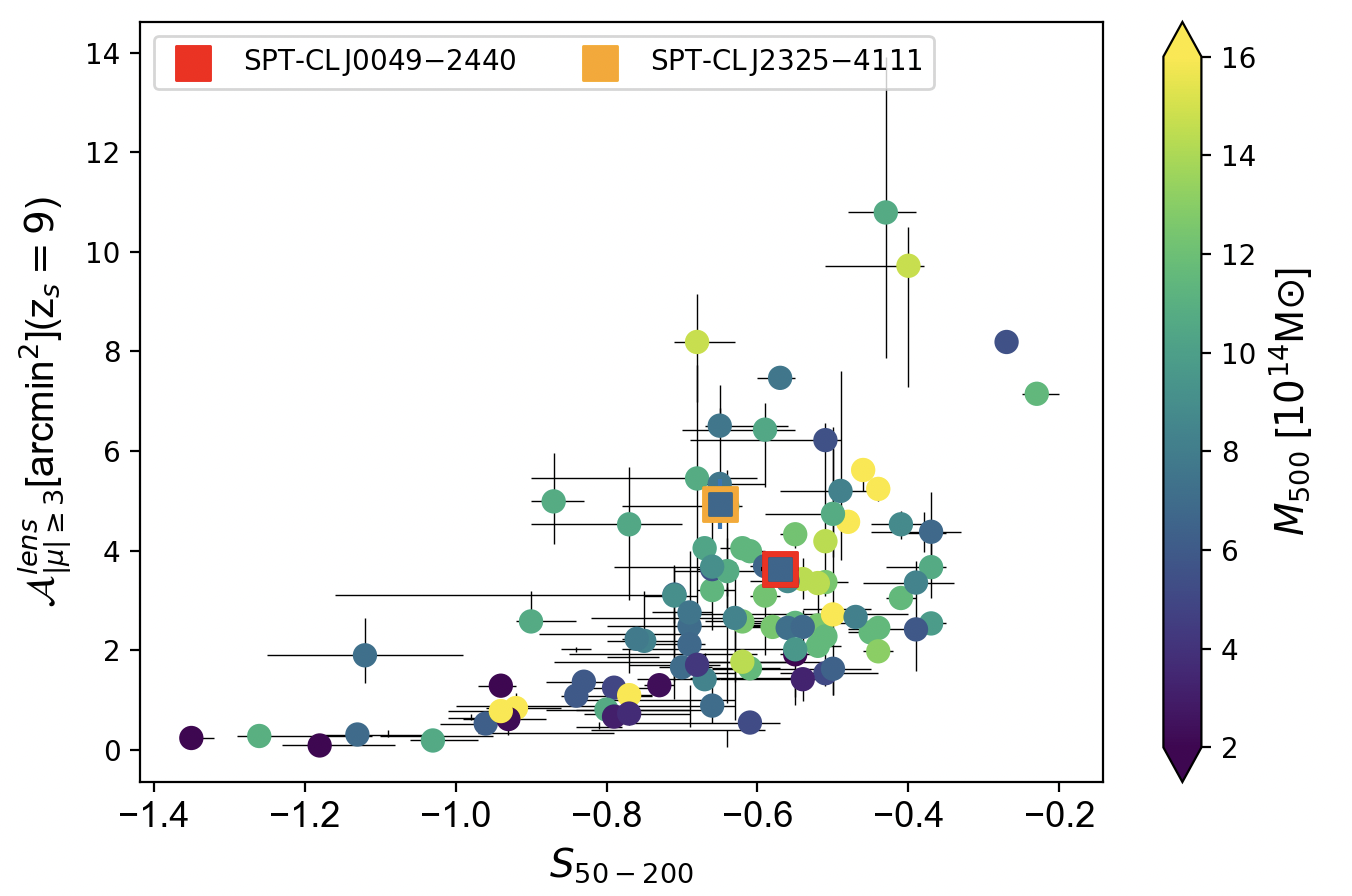}
        \includegraphics[width=0.45\linewidth]{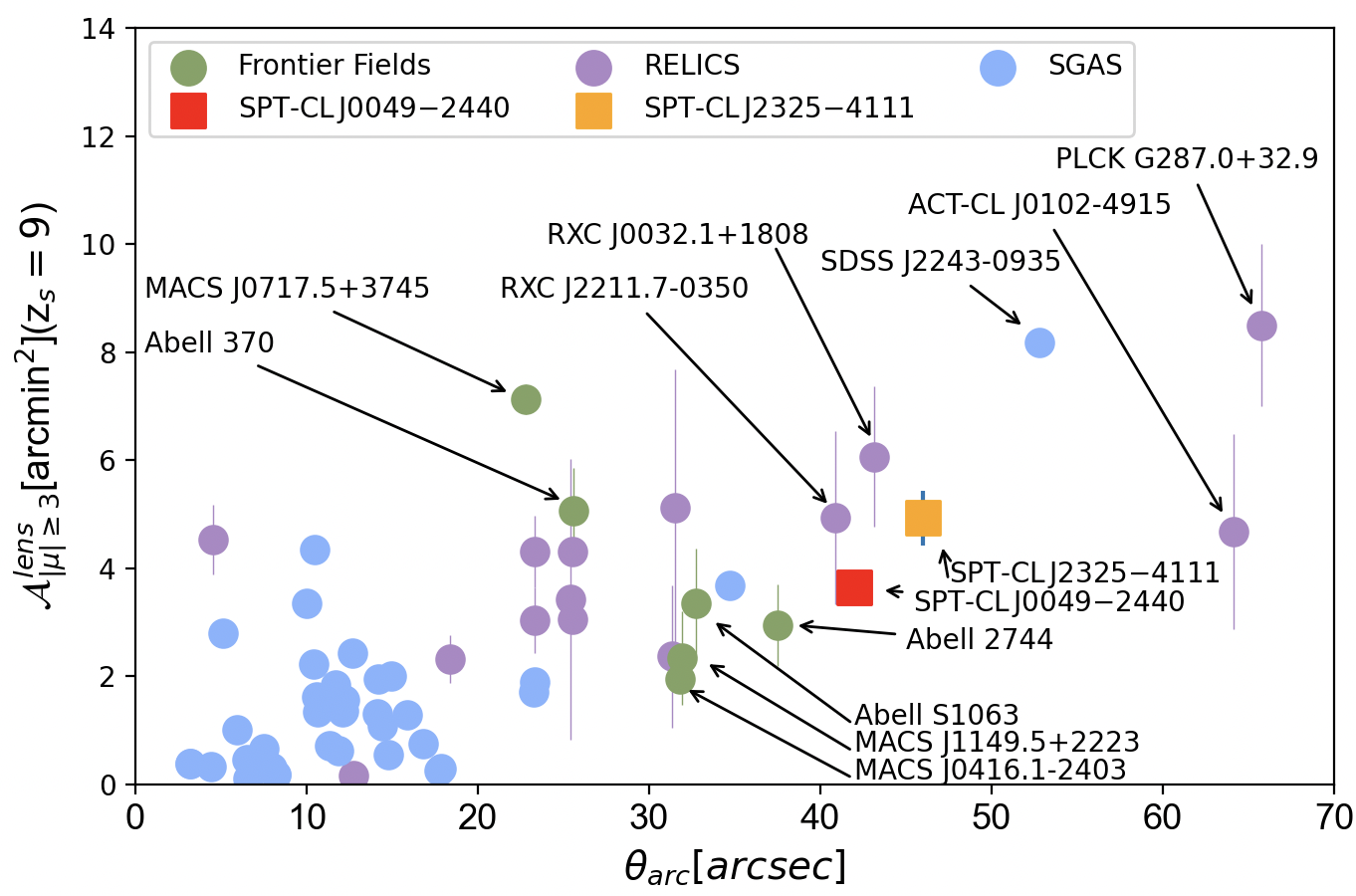}
    \caption{\done{\textit{Left:} Lensing strength $\mathcal{A}^{lens}_{| \mu | \geq 3 }$ plotted against the central slope of the projected mass density profile ($S_{50-200}$). The data points are color-coded by their $M_{500}$ mass. The comparison sample is from \cite{Fox2022} and references therein. The \SPTXXIII\ and \SPTZERO\ datapoints are highlighted with orange and red edges, respectively, with their $M_{500}$ adopted from \cite{Bocquet2019,Bleem2020}.  These two clusters have relatively high lensing strength compared to other clusters with similar properties. \textit{Right:} Lensing strength plotted against the projected separation between the farthest bright arc and the BCG, compared to strong lensing cluster samples. 
    %Error bars reflect systematic uncertainties where multiple lens modeling algorithms. 
    \new{The error bars reflect the systematic uncertainties, determined from the range of measurements obtained by different lens modeling algorithms, where available}.
    Notably, \SPTXXIII\ and \SPTZERO\ have similar lensing strengths to the Frontier Fields. Figures adapted from \cite{Fox2022}.}
    %  2325-4111  M500ch70 6.70+0.95 −1.17
    % 0049-2440   M500ch70 6.59+0.86 −0.98 
    }
    \label{fig:slope}
\end{figure*}

\done{To contextualize the two clusters with respect to the clusters studied by \cite{Fox2022}, we calculated the lensing strength of these clusters, finding \Amplistrz $=4.93^{+0.03}_{-0.04}$ arcmin$^2$ for \SPTXXIII\ (i.e., an area of $\sim4.9$ arcmin$^2$ is magnified by a factor of 3 or higher for a source at $z=9$), and \Amplistrz $= 3.64^{+0.14}_{-0.10}$ arcmin$^2$ for \SPTZERO. 
Following \cite{Fox2022}, we also measured the inner slope of the mass density profile derived from the lens models, $S_{50-200}$, which is defined as the log of the slope of density profile measured between 50 and 200~kpc from the BCG, finding 
 $S_{50-200}=-0.59^{-0.62}_{-0.56}$ for \SPTZERO\ and 
 $S_{50-200}=-0.67^{-0.69}_{-0.66}$ \SPTXXIII. 
In \autoref{fig:slope} we compare the two clusters to a large sample of strong lensing clusters from \cite{Fox2022} spanning a wide range of cluster properties, including clusters from SGAS \citep{sharon2020}, RELICS \citep{Coe2019}, and Frontier Fields \citep{Lotz2017}. The left panel shows the \Amplistrz-$S_{50-200}$ plane, and the right panel shows the lensing strength as a function of separation between the farthest bright arc and the BCG.
We find that \SPTXXIII\ and \SPTZERO\ have a higher lensing strength than most clusters with similar $M_{500}$ or inner slope, consistent with their observed large $\theta_{arc}$ separation; they appear to have a lensing strength comparable to the Frontier Fields.}
% 2325-4111   M500ch70 6.70+0.95 −1.17
% 0049-2440   M500ch70 6.59+0.86 −0.98 
% 
% 

%\subsection{source plane surface and JWST prediction with LF}

%\begin{figure}
%    \centering
%    \includegraphics[width=\linewidth]{ionization.png}
%    \caption{}
%    \label{fig:a}
%\end{figure}
%\begin{figure}
%    \centering
%    \includegraphics[width=\linewidth]{LF-truenumb.png}
%    \caption{}
%    \label{fig:b}
%\end{figure}
%\begin{figure}
%    \centering
%    \includegraphics[width=\linewidth]{numb-12-14.png}
%    \caption{}
%    \label{fig:c}
%\end{figure}

\subsection{Lensed Sources of Interest} 
\label{sec:lensed}
\begin{figure*}
    \centering
    \includegraphics[width=0.45\linewidth]{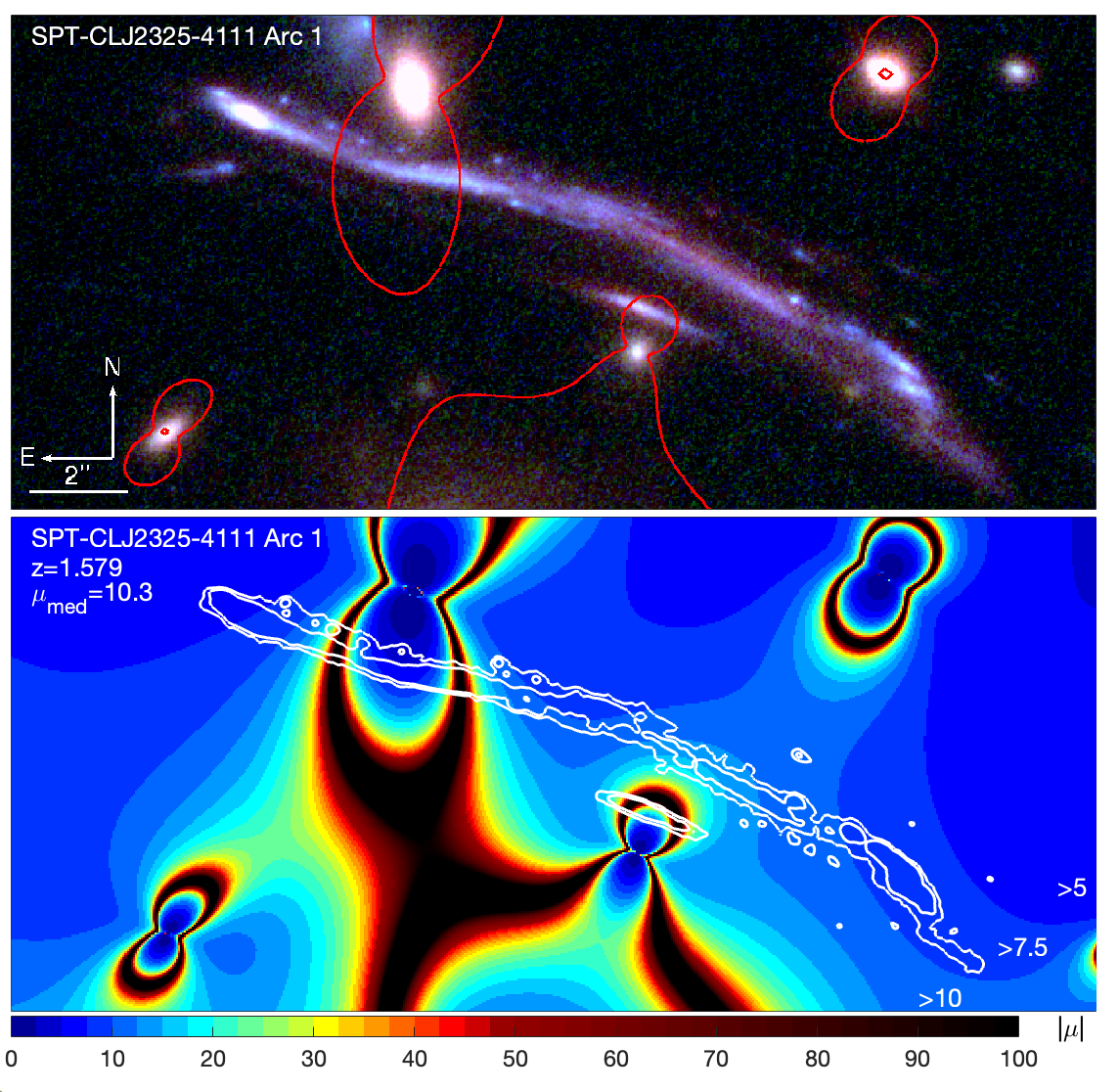}
    \includegraphics[width=0.45\linewidth]{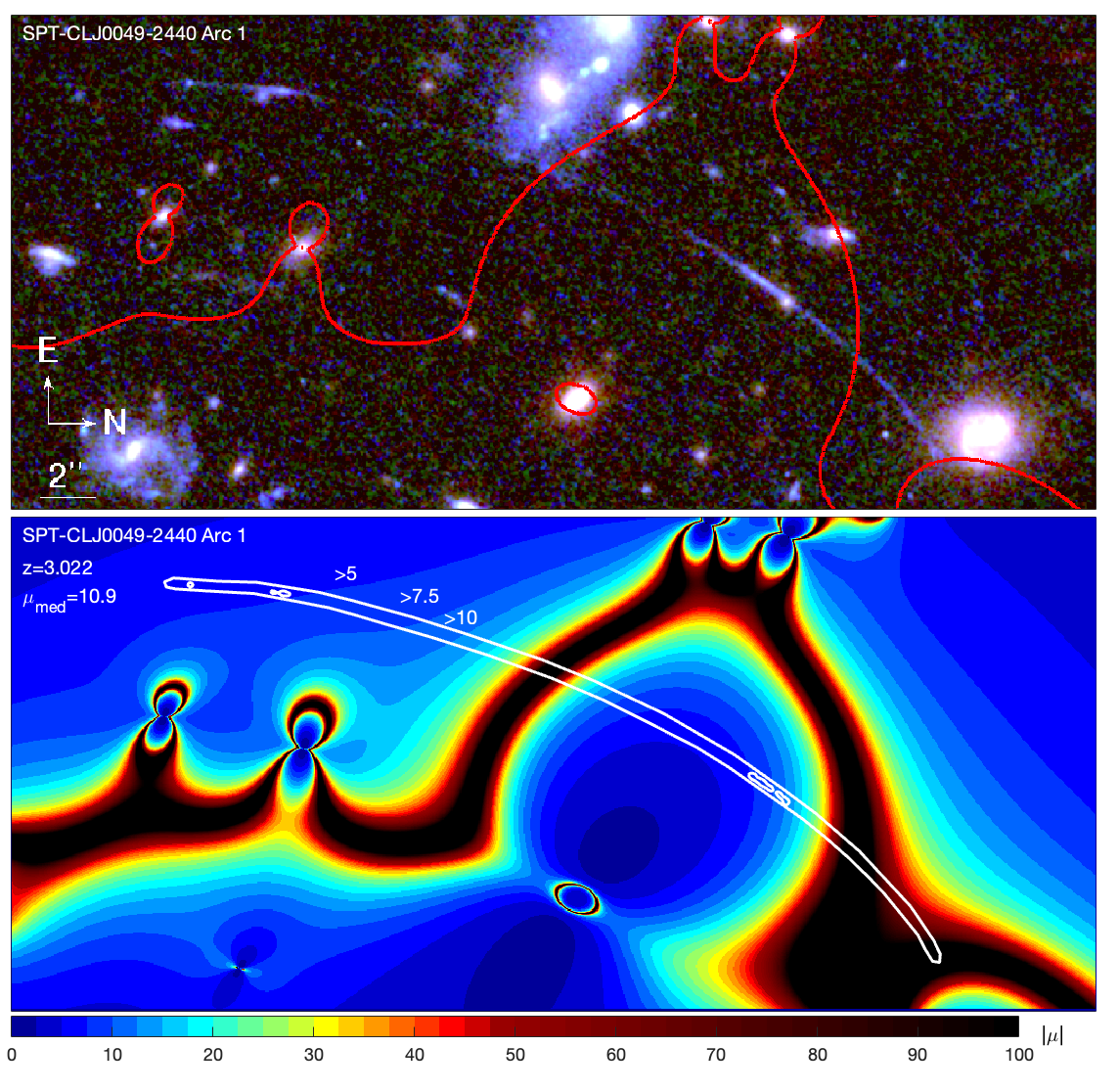}
    \caption{Zoom-in on Arc~1 in \SPTXXIII\ (left) and on Arc~1 in \SPTZERO\ (right). \textit{Top:} color composite (same data as \autoref{fig:hst2325} and \autoref{fig:hst0049}) showing the extent of the giant arc. Critical curves for the arc redshift are overplotted in red. \textit{Bottom:} Map of absolute magnification within the same field of view, computed from the best-fit lens model for the redshift of the arc. The colorbar at the bottom indicates the absolute magnification values. The white contours indicate the location of the giant arcs, which span a wide range of magnifications. The median magnification within the contour is 10.3 and 10.9 for \SPTXXIII\ and \SPTZERO, respectively. 
    Note that the field of \SPTZERO\ is rotated by 90$^\circ$ and less zoomed-in; see compass and scale bar at the bottom left of each imaging panel for reference. Because the arc in this field is faint, the color composite was smoothed with a 3-pixel Gaussian kernel, and the external contours were placed manually to guide the eye.} 
    \label{fig:arc1}
\end{figure*}

\done{Each of the clusters studied in this work lenses numerous sources, which were used to constrain the lens model. While not the focus of this analysis, we highlight two prominent arcs observed in these fields. 
In \SPTXXIII, the image of Source~1, ``J2325 Arc~1'' ($z=1.579$), appears as a bright $18$\arcsec\ long giant arc north of the cluster core, with observed magnitude of m$_{\rm AB}=19.2$ ($19.1$) in the F606W (F814W) band. The multiplicity of the arc was not immediately obvious: it appears that most of the arc is singly-imaged into a high-distortion arc, where only regions next to the nearby cluster member galaxy are multiple images of a small region of the source galaxy. A set of star forming clumps can be mapped with mirror symmetry about the critical curve (\autoref{fig:arc1}). The bright core of the galaxy in the east end of the arc, as well as the long red tail to the west are singly imaged. The bottom panel of \autoref{fig:arc1} shows the magnification map from the best-fit lens model. We estimate a median magnification along the arc of $\mu_{med}=10.3$; in most regions the arc is magnified by at least a factor of 8, with areas very close to the critical curve being magnified by more than 50. 
The brightness, prominent clumps, and indication of more details in the infrared make this arc a promising target for study by JWST.}
% 18 arcsec

\done{The highly extended arc in \SPTZERO, ``J0049 Arc~1'' ($z=3.022$) spans $31$\arcsec\ in the image plane. It is much fainter, and most likely comprised of two or three images along the arc. The most likely counter images (labeled 1.3 and 1.4 in \autoref{fig:hst0049}) have comparable local magnification, but do not suffer from the high distortion of the giant arc. The morphology and image-plane size of the counter images indicates that the source galaxy is quite compact; we measure a FWHM of $0\farcs19$ for image 1.3 using IRAF, which translates to $0.49^{+0.04}_{-0.02}$~kpc in the source plane after dividing by the square root of the lensing magnification. A separate clump, or companion galaxy, is observed nearby ($<0\farcs5$ in the image plane), labeled as source 1.x.2 in \autoref{tab:arcstable}. \autoref{fig:arc1} shows a zoom-in on the faint giant arc and the lensing magnification. We estimate a median magnification of $\mu_{med}=10.9$ along the arc. \autoref{tab:arcstable} lists the measured magnifications and their uncertainties at the positions along the arc and counter images that were used as lensing constraints. Deep, high resolution imaging with JWST could reveal substructures within this galaxy on sub-kpc scale, given its extreme tangential distortion and its large extent in the image plane. }

\startlongtable
\begin{deluxetable*}{lllllll} 
\tablecolumns{7} 
\tablecaption{List of lensing constraints \label{tab:arcstable}} 
\tablehead{\colhead{ID} &
            \colhead{R.A.}    & 
            \colhead{Decl.}    & 
            \colhead{$z_{spec}$}     &  
            \colhead{Dist}     &
            \colhead{$\mu$}     & 
%            \colhead{$z_{spec}$}       & 
            \colhead{Notes}       \\[-8pt]
            \colhead{} &
            \colhead{J2000}     & 
            \colhead{J2000}    & 
            \colhead{or $z_{model}$ }       &
            \colhead{(\arcsec)} &
            \colhead{} &
%            \colhead{Reference}       & 
            \colhead{}             }
\startdata 
\multicolumn{3}{l}{\SPTXXIII\ Source 1}& 1.5790 & && GMOS; Single giant arc; SL by nearby galaxy\\
1.1.1 & 351.296820  & -41.187381 & &0.19&$8.2^{+0.1}_{-0.3}$   & clump 1\\ 
1.2.1 & 351.295943  & -41.187662 & &0.11&$4.4^{+0.4}_{-0.7}$   &  \\ 
1.1.2 & 351.296463 & -41.187529 & &0.16&$24.5^{+3.1}_{-2.3}$  &  clump 2\\ 
1.2.2 & 351.296232 & -41.187615 & &0.09&$12.7^{+3.0}_{-2.0}$  &  \\ 
%11.1 & 351.296463 & -41.187529 & &0.16&$24.5^{+3.1}_{-2.3}$  &  \\ 
%11.2 & 351.296232 & -41.187615 & &0.09&$12.7^{+3.0}_{-2.0}$  &  \\ 
\hline
\multicolumn{3}{l}{\SPTXXIII\ Source 2}& 2.4253 & && FIRE spec of 2.3, from OIII, H$\beta$, H$\alpha$; 2019-12-04\\
2.1.1 & 351.301880 & -41.192335 & &0.51& $2.5^{+0.2}_{-0.2}$ &  clump 1\\ 
2.2.1 & 351.285884 & -41.202655 & &0.31& $5.0^{+0.1}_{-0.3}$ &  \\ 
2.3.1 & 351.303568 & -41.209014 & &0.44& $8.4^{+0.3}_{-0.3}$ &  \\ 
2.4.1 & 351.312449 & -41.197529 & &0.28& $25.1^{+2.7}_{-1.2}$ &  \\ 
2.5.1 & 351.299833 & -41.202644 & &0.28& $1.4^{+0.2}_{-0.1}$ &  \\ 
2.1.2 & 351.302765 & -41.192171 & &1.31& $7.7^{+0.7}_{-0.1}$ &  clump 2\\ 
2.2.2 & 351.285764 & -41.202283 & &0.78& $7.6^{+0.7}_{-0.3}$ &  \\ 
2.3.2 & 351.304206 & -41.208597 & &0.60& $10.2^{+0.5}_{-0.4}$ &  \\ 
2.4.2 & 351.312034 & -41.197256 & &0.30& $13.1^{+0.3}_{-0.7}$ &  \\ 
\hline
\multicolumn{3}{l}{\SPTXXIII\ Source 3}& 3.8180 & && FIRE spec of 3.1, both OIII lines in K-band; 2019-12-05\\
3.1 & 351.306626 & -41.189283 & &0.21&$42.1^{+7.2}_{-2.9}$  &  \\ 
3.2 & 351.301163 & -41.188301 & &0.36& $5.8^{+0.6}_{-0.6}$ &  \\ 
3.3 & 351.285310 & -41.196210 & &0.59& $4.7^{+0.2}_{-0.1}$ &  \\ 
3.4 & 351.301139 & -41.207554 & &0.85& $3.0^{+0.1}_{-0.2}$ &  \\ 
\hline
\multicolumn{3}{l}{\SPTXXIII\ Source 4}& 1.3180 & & & FIRE spec of 4.3, H$\alpha$; 2019-12-05 \\
4.1 & 351.291829 & -41.206459 & &1.03& $7.8^{+0.2}_{-0.2}$&\\ 
4.2 & 351.297771 & -41.209536 & &0.56& $3.5^{+0.2}_{-0.3}$ &  \\ 
4.3 & 351.311327 & -41.203079 & &1.19& $8.6^{+0.3}_{-0.2}$ &  \\ 
4.4 & 351.298830 & -41.197679 & &1.03&  $6.7^{+0.2}_{-0.8}$ &  Radial arc\\ 
\hline
\multicolumn{3}{l}{\SPTXXIII\ Source 5} &  $1.29\pm{0.01}$ & &&\\
5.1 & 351.293221 & -41.208914 & &0.73& $9.7^{+0.3}_{-0.4}$ & \\ 
5.2 & 351.298529 & -41.210439 & &0.27& $8.9^{+0.4}_{-0.6}$ &  \\ 
5.3 & 351.306435 & -41.208372 & &0.82& $13.0^{+0.4}_{-0.7}$ &  \\ 
\hline
\multicolumn{3}{l}{\SPTXXIII\ Source 6} & $7.02\pm{0.20}$ & &&\\
6.1 & 351.313070 & -41.189842 & &1.64 & $4.9^{+0.2}_{-0.2}$ & \\ 
6.2$c$ & 351.292582 & -41.189019 & &\nodata& $14.4^{+1.1}_{-1.1}$ & \\ 
6.3 & 351.288051 & -41.192324 & &  2.80& $12.0^{+0.4}_{-0.5}$ & \\ 
6.4 & 351.299692 & -41.207262 & &0.84& $2.1^{+0.1}_{-0.1}$ & \\ 
\hline
\multicolumn{3}{l}{\SPTXXIII\ Source 7}& \nodata & &&\\
7.1$c$ & 351.293554 & -41.194939 & &&  & Candidate system\\
7.2$c$ & 351.292862 & -41.195442 & &&  & \\
\hline
\multicolumn{3}{l}{\SPTXXIII\ Source 8}& $3.00\pm{0.16}$ & &&\\
8.1 & 351.302596 & -41.198158 & &0.42& $7.7^{+0.6}_{-0.7}$ &  Radial arc \\ 
8.2 & 351.301964 & -41.198866 & &0.48& $22.5^{+7.6}_{-4.4}$ &  Radial arc \\ 
8.3 & 351.305018 & -41.194379 & &0.04& $4.0^{+0.4}_{-1.0}$ &  \\ 
\hline
\multicolumn{3}{l}{\SPTXXIII\ Source 9}& $1.21\pm{0.01}$ & &&\\
9.1 & 351.305663 & -41.207017 & &0.38& $18.5^{+1.0}_{-0.7}$ & \\ 
9.2 & 351.301553 & -41.208875 & &0.49 & $13.4^{+0.8}_{-0.5}$ & \\ 
9.3 & 351.290113 & -41.205325 & &0.42 & $4.9^{+0.1}_{-0.2}$ & \\ 
\hline 
\hline
\multicolumn{3}{l}{\SPTZERO\ Source 1}& 3.0220 & && FIRE spec of 1.2; OIII\\
1.1.1 & 12.306837 & -24.678576 & 3.0220 &0.32& $6.4^{+0.8}_{-0.4}$ &  \\ 
1.2.1 & 12.304789 & -24.673793 & &0.31&$6.8^{+1.1}_{-0.8}$  &  \\ 
1.3.1$c$& 12.293434 & -24.669124 & &\nodata& $8.9^{+1.3}_{-0.6}$ & spectroscopy attempted, not confirmed\\
1.4.1 & 12.290510 & -24.685268 & &0.30& $2.2^{+0.5}_{-0.2}$ &  \\ 
1.1.2 & 12.306927 & -24.679502 & &0.47& $5.7^{+0.9}_{-0.2}$ & clump near Arc~1, assumed to be at the same $z$ \\ 
1.3.2$c$& 12.293478 & -24.669244 & &\nodata & $9.8^{+1.5}_{-0.7}$ &  \\
1.4.2 & 12.290435 & -24.685274 & &0.18& $2.3^{+0.5}_{-0.3}$ &  \\ 
\hline
\multicolumn{3}{l}{\SPTZERO\ Source 2}& $1.52\pm{0.10}$ & &&\\
2.1 & 12.296863 & -24.668930 & &0.15& $19.9^{+4.5}_{-1.3}$ &  \\ 
2.2 & 12.297035 & -24.668981 & &0.14& $4.3^{+1.2}_{-0.7}$ &  \\ 
2.3 & 12.297701 & -24.669017 & &0.22& $5.0^{+0.6}_{-0.4}$ &  \\ 
\hline
\multicolumn{3}{l}{\SPTZERO\ Source 3}& $3.62\pm{0.31}$ & &&\\
3.1 & 12.302022 & -24.681788 & &0.19& $5.5^{+0.7}_{-0.6}$ &  \\ 
3.3 & 12.281064 & -24.672098 & &0.23& $2.5^{+0.2}_{-0.2}$ &  \\ 
3.4 & 12.297052 & -24.680913 & &0.12& $3.7^{+0.6}_{-0.4}$ &  \\ 
\hline
\multicolumn{3}{l}{\SPTZERO\ Source 4}& $3.03\pm{0.16}$ & &&\\
4.1    & 12.284454 & -24.681115 & &0.05& $11.7^{+0.8}_{-0.9}$ &  \\ 
4.2$c$ & 12.285342 & -24.683749 & &\nodata & $13.9^{+      2.9}_{-2.9}$ &\\
4.3 & 12.303653 & -24.689925 & &0.10& $3.9^{+0.2}_{-0.2}$ &  \\ 
\hline
\multicolumn{3}{l}{\SPTZERO\ Source 5}& $1.37\pm{0.03}$ & &&\\
5.1 & 12.291012 & -24.680759 &  &2.43&$12.8^{+0.9}_{-0.9}$ &  Radial arc\\ 
5.2 & 12.291352 & -24.680671 &  &1.98&$12.9^{+1.0}_{-0.8}$ &  Radial arc\\ 
5.3 & 12.293889 & -24.678894 &  &0.40&$9.1^{+0.8}_{-0.3}$ &  \\ 
5.4 & 12.303297 & -24.684562 &  &1.42&$3.2^{+0.2}_{-0.3}$ &  \\ 
\hline
\multicolumn{3}{l}{\SPTZERO\ Source 6}& 2.368 & && FIRE spec of both 6.1 and 6.2; OIII  \\
6.1 & 12.298507 & -24.681234 & 2.368 &0.11& $71^{+70}_{-7}$ &  Radial arc \\ 
6.2 & 12.298314 & -24.681211 & 2.368 &0.20& $860^{+90}_{-12}$&  Radial arc\\ 
\hline
\multicolumn{3}{l}{\SPTZERO\ Source 7}& $2.30\pm{0.10}$ & &&\\
7.1 & 12.296999 & -24.680492 & &0.02& $4.3^{+0.7}_{-0.4}$ &  \\ 
7.2 & 12.300114 & -24.680802 & &0.62& $4.9^{+1.0}_{-0.4}$ &  \\ 
7.3 & 12.301315 & -24.680923 & &0.54& $3.5^{+0.5}_{-0.3}$ &  \\ 
7.4 & 12.282750 & -24.673044 & &0.01& $2.9^{+0.3}_{-0.1}$ &  \\ 
\hline
\multicolumn{3}{l}{\SPTZERO\ Source 8}& $4.96\pm{0.61}$ & &&\\
8.1 & 12.298777 & -24.688750 & &0.10& $13^{+2}_{-1}$ &  \\ 
8.2 & 12.295079 & -24.689882 & &0.08& $56^{+18}_{-11}$ &  \\ 
\enddata 
\tablecomments{The IDs, positions, and redshifts of lensed galaxies that were used as constraints in this work. Where possible, individual emission knots in each image are identified and used as lensing constraints. 
   %The IDs of images of lensed galaxies are labeled as $AB.X$ where $A$ is a number indicating the source ID (or system name);  $B$ is a number indicating the ID of the emission knot within the system; and $X$ is a number indicating the ID of the lensed image within the multiple image family. 
   {\bf The IDs of images of lensed galaxies are labeled as A.B(.C) where A is the number indicating the source ID (or system name); B is the number indicating the ID of lensed image within the multiple images family; C is a number indicating the ID of emission knot within the image if we used more than one substructure of the images as constraints.}
   {\bf Lower-case} $c$ indicates a candidate arc that was not used to constrain the model. {\bf \textit{Dist}, often called individual rmsi, is the distance in the image plane between the observed and model-predicted location of an image. 
   The model-predicted location is calculated as the lensed coordinates of the geometric mean of source positions of all the multiple images of a given source.
   }
   }
\end{deluxetable*}

\section{Summary}
\label{sec:conclusion}

We present strong lensing models of two clusters, 
\SPTXXIII\ and \SPTZERO, based on multi-band HST imaging and spectroscopic redshifts from Magellan/LDSS3, Magellan/FIRE, and Gemini/GMOS. We report the following:

\begin{itemize}
    \item The projected mass of \SPTXXIII\ within 500~kpc from the BCG is  
$M(<500 ~{\rm kpc}) = 7.30\pm0.07 \times 10^{14}$\msun\ (cylindrical mass), with a sub-halos mass ratio of $0.12\pm{0.01}$. 
The projected mass of \SPTZERO\ is
$M(<500 ~{\rm kpc})=7.12^{+0.16}_{-0.19}\times 10^{14}$\msun\ (cylindrical mass), with a sub-halos mass ratio of $0.21^{+0.07}_{-0.05}$. A comparison of the projected mass density profiles of these clusters to those of other strong lenses indicates a higher density within the innermost few hundred kpc than that of the Frontier Fields clusters. 

  \item The Einstein radii for a source at $z=9$  as measured from the lens models are $42$\arcsec\ and $43$\arcsec\ for \SPTXXIII\ and \SPTZERO, respectively.

    \item Following \cite{Fox2022}, we measured the area magnified by a factor of $\mu \geq 3$ for a source at $z=9$ (i.e., lensing strength) of  \Amplistrz$=4.93^{+0.03}_{-0.04}$ arcmin$^2$, and \Amplistrz $= 3.64^{+0.14}_{-0.10}$ arcmin$^2$, for \SPTXXIII\ and \SPTZERO, respectively.

    \item We report on two giant arcs of highly magnified sources in these fields. ``J2325 Arc~1'' ($z=1.5790$) is projected $\theta_{arc}=46$\arcsec\ north of BCG of \SPTXXIII, spanning 18\arcsec\ in the image plane, with a median magnification of $\mu_{\rm med}=10.3$.  ``J0049 Arc~1'' ($z=3.0220$), projected  $\theta_{arc}=42$\arcsec\ northeast of the BCG of \SPTZERO, spanning 31\arcsec\ in the image plane with a median magnification of $\mu_{\rm med}=10.9$. Their high distortions indicate promising resolving capabilities for detailed morphological analysis of galaxies at cosmic noon.
    
\end{itemize}

We conclude that the lensing power measured at these strong lensing sight-lines promotes \SPTXXIII\ and \SPTZERO\ to the top tier of strong lensing clusters known today, on par with well-studied clusters such as the Frontier Fields; these clusters have yet untapped potential for follow-up studies of the high-redshift Universe they magnify. 

\section*{Acknowledgments}
This research is based on observations made with the NASA/ESA Hubble Space Telescope obtained from the Space Telescope Science Institute, which is operated by the Association of Universities for Research in Astronomy, Inc., under NASA contract NAS 5–26555. These observations are associated with program(s) HST-GO-15937.
Support for HST program \#15937 was provided by NASA through a grant from the Space Telescope Science Institute, which is operated by the Association of Universities for Research in Astronomy, Inc., under NASA contract NAS 5-03127. 
The HST data presented in this article were obtained from the Mikulski Archive for Space Telescopes (MAST) at the Space Telescope Science Institute. The specific observations analyzed can be accessed via \dataset[DOI]{https://doi.org/10.17909/5gpg-5575}.
This paper includes data gathered with the 6.5 meter Magellan Telescopes located at Las Campanas Observatory, Chile. NOIRLab IRAF is distributed by the Community Science and Data Center at NSF NOIRLab, which is managed by the Association of Universities for Research in Astronomy (AURA) under a cooperative agreement with the U.S. National Science Foundation. The South Pole Telescope program is supported by
the National Science Foundation (NSF) through award OPP-1852617. Partial support is also provided by the Kavli Institute of Cosmological Physics at the University of Chicago. 
Work at Argonne National Lab is supported by UChicago Argonne LLC, Operator of Argonne National Laboratory (Argonne). Argonne, a U.S. Department of Energy Office of Science Laboratory, is operated under contract no. DE-AC02-06CH11357. 

\software{Source Extractor \citep{SEx}; \lenstool\ \citep{Jullo2007}; NOIRLab IRAF v2.18 \citep{iraf24}; AstroDrizzle \citep{astrodrizzle}; Matlab \citep{MATLAB}; Astropy \citep{Astropy13,Astropy18,Astropy22} ; MATLAB Astronomy and Astrophysics Toolbox \citep[MAAT;][]{Ofek2014}}

\facilities{HST(ACS, WFC3), Magellan(LDSS3, IMACS, PISCO)}

\bibliography{biblio_NEXT-GEN}

\begin{thebibliography}{}
\expandafter\ifx\csname natexlab\endcsname\relax\def\natexlab#1{#1}\fi
\providecommand{\url}[1]{\href{#1}{#1}}

\bibitem[{{Acebron} {et~al.}(2017){Acebron}, {Jullo}, {Limousin}, {Tilquin}, {Giocoli}, {Jauzac}, {Mahler}, \& {Richard}}]{Acebron2017}
{Acebron}, A., {Jullo}, E., {Limousin}, M., {et~al.} 2017, \mnras, 470, 1809

\bibitem[{{Adamo} {et~al.}(2024){Adamo}, {Bradley}, {Vanzella}, {Claeyssens}, {Welch}, {Diego}, {Mahler}, {Oguri}, {Sharon}, {Abdurro'uf}, {Hsiao}, {Messa}, {Zackrisson}, {Brammer}, {Coe}, {Kokorev}, {Ricotti}, {Zitrin}, {Fujimoto}, {Inoue}, {Resseguier}, {Rigby}, {Jim{\'e}nez-Teja}, {Windhorst}, \& {Xu}}]{adamo2024}
{Adamo}, A., {Bradley}, L.~D., {Vanzella}, E., {et~al.} 2024, arXiv e-prints, arXiv:2401.03224

\bibitem[{{Astropy Collaboration} {et~al.}(2013){Astropy Collaboration}, {Robitaille}, {Tollerud}, {Greenfield}, {Droettboom}, {Bray}, {Aldcroft}, {Davis}, {Ginsburg}, {Price-Whelan}, {Kerzendorf}, {Conley}, {Crighton}, {Barbary}, {Muna}, {Ferguson}, {Grollier}, {Parikh}, {Nair}, {Unther}, {Deil}, {Woillez}, {Conseil}, {Kramer}, {Turner}, {Singer}, {Fox}, {Weaver}, {Zabalza}, {Edwards}, {Azalee Bostroem}, {Burke}, {Casey}, {Crawford}, {Dencheva}, {Ely}, {Jenness}, {Labrie}, {Lim}, {Pierfederici}, {Pontzen}, {Ptak}, {Refsdal}, {Servillat}, \& {Streicher}}]{Astropy13}
{Astropy Collaboration}, {Robitaille}, T.~P., {Tollerud}, E.~J., {et~al.} 2013, \aap, 558, A33

\bibitem[{{Astropy Collaboration} {et~al.}(2018){Astropy Collaboration}, {Price-Whelan}, {Sip{\H{o}}cz}, {G{\"u}nther}, {Lim}, {Crawford}, {Conseil}, {Shupe}, {Craig}, {Dencheva}, {Ginsburg}, {VanderPlas}, {Bradley}, {P{\'e}rez-Su{\'a}rez}, {de Val-Borro}, {Aldcroft}, {Cruz}, {Robitaille}, {Tollerud}, {Ardelean}, {Babej}, {Bach}, {Bachetti}, {Bakanov}, {Bamford}, {Barentsen}, {Barmby}, {Baumbach}, {Berry}, {Biscani}, {Boquien}, {Bostroem}, {Bouma}, {Brammer}, {Bray}, {Breytenbach}, {Buddelmeijer}, {Burke}, {Calderone}, {Cano Rodr{\'\i}guez}, {Cara}, {Cardoso}, {Cheedella}, {Copin}, {Corrales}, {Crichton}, {D'Avella}, {Deil}, {Depagne}, {Dietrich}, {Donath}, {Droettboom}, {Earl}, {Erben}, {Fabbro}, {Ferreira}, {Finethy}, {Fox}, {Garrison}, {Gibbons}, {Goldstein}, {Gommers}, {Greco}, {Greenfield}, {Groener}, {Grollier}, {Hagen}, {Hirst}, {Homeier}, {Horton}, {Hosseinzadeh}, {Hu}, {Hunkeler}, {Ivezi{\'c}}, {Jain}, {Jenness}, {Kanarek}, {Kendrew}, {Kern}, {Kerzendorf}, {Khvalko}, {King}, {Kirkby}, {Kulkarni},
  {Kumar}, {Lee}, {Lenz}, {Littlefair}, {Ma}, {Macleod}, {Mastropietro}, {McCully}, {Montagnac}, {Morris}, {Mueller}, {Mumford}, {Muna}, {Murphy}, {Nelson}, {Nguyen}, {Ninan}, {N{\"o}the}, {Ogaz}, {Oh}, {Parejko}, {Parley}, {Pascual}, {Patil}, {Patil}, {Plunkett}, {Prochaska}, {Rastogi}, {Reddy Janga}, {Sabater}, {Sakurikar}, {Seifert}, {Sherbert}, {Sherwood-Taylor}, {Shih}, {Sick}, {Silbiger}, {Singanamalla}, {Singer}, {Sladen}, {Sooley}, {Sornarajah}, {Streicher}, {Teuben}, {Thomas}, {Tremblay}, {Turner}, {Terr{\'o}n}, {van Kerkwijk}, {de la Vega}, {Watkins}, {Weaver}, {Whitmore}, {Woillez}, {Zabalza}, \& {Astropy Contributors}}]{Astropy18}
{Astropy Collaboration}, {Price-Whelan}, A.~M., {Sip{\H{o}}cz}, B.~M., {et~al.} 2018, \aj, 156, 123

\bibitem[{{Astropy Collaboration} {et~al.}(2022){Astropy Collaboration}, {Price-Whelan}, {Lim}, {Earl}, {Starkman}, {Bradley}, {Shupe}, {Patil}, {Corrales}, {Brasseur}, {N{\"o}the}, {Donath}, {Tollerud}, {Morris}, {Ginsburg}, {Vaher}, {Weaver}, {Tocknell}, {Jamieson}, {van Kerkwijk}, {Robitaille}, {Merry}, {Bachetti}, {G{\"u}nther}, {Aldcroft}, {Alvarado-Montes}, {Archibald}, {B{\'o}di}, {Bapat}, {Barentsen}, {Baz{\'a}n}, {Biswas}, {Boquien}, {Burke}, {Cara}, {Cara}, {Conroy}, {Conseil}, {Craig}, {Cross}, {Cruz}, {D'Eugenio}, {Dencheva}, {Devillepoix}, {Dietrich}, {Eigenbrot}, {Erben}, {Ferreira}, {Foreman-Mackey}, {Fox}, {Freij}, {Garg}, {Geda}, {Glattly}, {Gondhalekar}, {Gordon}, {Grant}, {Greenfield}, {Groener}, {Guest}, {Gurovich}, {Handberg}, {Hart}, {Hatfield-Dodds}, {Homeier}, {Hosseinzadeh}, {Jenness}, {Jones}, {Joseph}, {Kalmbach}, {Karamehmetoglu}, {Ka{\l}uszy{\'n}ski}, {Kelley}, {Kern}, {Kerzendorf}, {Koch}, {Kulumani}, {Lee}, {Ly}, {Ma}, {MacBride}, {Maljaars}, {Muna}, {Murphy}, {Norman},
  {O'Steen}, {Oman}, {Pacifici}, {Pascual}, {Pascual-Granado}, {Patil}, {Perren}, {Pickering}, {Rastogi}, {Roulston}, {Ryan}, {Rykoff}, {Sabater}, {Sakurikar}, {Salgado}, {Sanghi}, {Saunders}, {Savchenko}, {Schwardt}, {Seifert-Eckert}, {Shih}, {Jain}, {Shukla}, {Sick}, {Simpson}, {Singanamalla}, {Singer}, {Singhal}, {Sinha}, {Sip{\H{o}}cz}, {Spitler}, {Stansby}, {Streicher}, {{\v{S}}umak}, {Swinbank}, {Taranu}, {Tewary}, {Tremblay}, {de Val-Borro}, {Van Kooten}, {Vasovi{\'c}}, {Verma}, {de Miranda Cardoso}, {Williams}, {Wilson}, {Winkel}, {Wood-Vasey}, {Xue}, {Yoachim}, {Zhang}, {Zonca}, \& {Astropy Project Contributors}}]{Astropy22}
{Astropy Collaboration}, {Price-Whelan}, A.~M., {Lim}, P.~L., {et~al.} 2022, \apj, 935, 167

\bibitem[{{Atek} {et~al.}(2015){Atek}, {Richard}, {Kneib}, {Jauzac}, {Schaerer}, {Clement}, {Limousin}, {Jullo}, {Natarajan}, {Egami}, \& {Ebeling}}]{Atek2015}
{Atek}, H., {Richard}, J., {Kneib}, J.-P., {et~al.} 2015, \apj, 800, 18

\bibitem[{{Atek} {et~al.}(2024){Atek}, {Labb{\'e}}, {Furtak}, {Chemerynska}, {Fujimoto}, {Setton}, {Miller}, {Oesch}, {Bezanson}, {Price}, {Dayal}, {Zitrin}, {Kokorev}, {Weaver}, {Brammer}, {Dokkum}, {Williams}, {Cutler}, {Feldmann}, {Fudamoto}, {Greene}, {Leja}, {Maseda}, {Muzzin}, {Pan}, {Papovich}, {Nelson}, {Nanayakkara}, {Stark}, {Stefanon}, {Suess}, {Wang}, \& {Whitaker}}]{Atek24}
{Atek}, H., {Labb{\'e}}, I., {Furtak}, L.~J., {et~al.} 2024, \nat, 626, 975

\bibitem[{{Bah{\'e}}(2021)}]{Bahe2021}
{Bah{\'e}}, Y.~M. 2021, \mnras, 505, 1458

\bibitem[{{Bayliss} {et~al.}(2011{\natexlab{a}}){Bayliss}, {Gladders}, {Oguri}, {Hennawi}, {Sharon}, {Koester}, \& {Dahle}}]{Bayliss2011apjl}
{Bayliss}, M.~B., {Gladders}, M.~D., {Oguri}, M., {et~al.} 2011{\natexlab{a}}, \apjl, 727, L26

\bibitem[{{Bayliss} {et~al.}(2011{\natexlab{b}}){Bayliss}, {Hennawi}, {Gladders}, {Koester}, {Sharon}, {Dahle}, \& {Oguri}}]{Bayliss2011gmos}
{Bayliss}, M.~B., {Hennawi}, J.~F., {Gladders}, M.~D., {et~al.} 2011{\natexlab{b}}, \apjs, 193, 8

\bibitem[{{Bayliss} {et~al.}(2016){Bayliss}, {Ruel}, {Stubbs}, {Allen}, {Applegate}, {Ashby}, {Bautz}, {Benson}, {Bleem}, {Bocquet}, {Brodwin}, {Capasso}, {Carlstrom}, {Chang}, {Chiu}, {Cho}, {Clocchiatti}, {Crawford}, {Crites}, {de Haan}, {Desai}, {Dietrich}, {Dobbs}, {Doucouliagos}, {Foley}, {Forman}, {Garmire}, {George}, {Gladders}, {Gonzalez}, {Gupta}, {Halverson}, {Hlavacek-Larrondo}, {Hoekstra}, {Holder}, {Holzapfel}, {Hou}, {Hrubes}, {Huang}, {Jones}, {Keisler}, {Knox}, {Lee}, {Leitch}, {von der Linden}, {Luong-Van}, {Mantz}, {Marrone}, {McDonald}, {McMahon}, {Meyer}, {Mocanu}, {Mohr}, {Murray}, {Padin}, {Pryke}, {Rapetti}, {Reichardt}, {Rest}, {Ruhl}, {Saliwanchik}, {Saro}, {Sayre}, {Schaffer}, {Schrabback}, {Shirokoff}, {Song}, {Spieler}, {Stalder}, {Stanford}, {Staniszewski}, {Stark}, {Story}, {Vanderlinde}, {Vieira}, {Vikhlinin}, {Williamson}, \& {Zenteno}}]{Bayliss2016}
{Bayliss}, M.~B., {Ruel}, J., {Stubbs}, C.~W., {et~al.} 2016, \apjs, 227, 3

\bibitem[{{Bertin} \& {Arnouts}(1996)}]{SEx}
{Bertin}, E., \& {Arnouts}, S. 1996, \aaps, 117, 393

\bibitem[{{Bezanson} {et~al.}(2022){Bezanson}, {Labbe}, {Whitaker}, {Leja}, {Price}, {Franx}, {Brammer}, {Marchesini}, {Zitrin}, {Wang}, {Weaver}, {Furtak}, {Atek}, {Coe}, {Cutler}, {Dayal}, {van Dokkum}, {Feldmann}, {Forster Schreiber}, {Fujimoto}, {Geha}, {Glazebrook}, {de Graaff}, {Greene}, {Juneau}, {Kassin}, {Kriek}, {Khullar}, {Maseda}, {Mowla}, {Muzzin}, {Nanayakkara}, {Nelson}, {Oesch}, {Pacifici}, {Pan}, {Papovich}, {Setton}, {Shapley}, {Smit}, {Stefanon}, {Taylor}, \& {Williams}}]{Bezanson2022}
{Bezanson}, R., {Labbe}, I., {Whitaker}, K.~E., {et~al.} 2022, arXiv e-prints, arXiv:2212.04026

\bibitem[{{Bleem} {et~al.}(2015){Bleem}, {Stalder}, {de Haan}, {Aird}, {Allen}, {Applegate}, {Ashby}, {Bautz}, {Bayliss}, {Benson}, {Bocquet}, {Brodwin}, {Carlstrom}, {Chang}, {Chiu}, {Cho}, {Clocchiatti}, {Crawford}, {Crites}, {Desai}, {Dietrich}, {Dobbs}, {Foley}, {Forman}, {George}, {Gladders}, {Gonzalez}, {Halverson}, {Hennig}, {Hoekstra}, {Holder}, {Holzapfel}, {Hrubes}, {Jones}, {Keisler}, {Knox}, {Lee}, {Leitch}, {Liu}, {Lueker}, {Luong-Van}, {Mantz}, {Marrone}, {McDonald}, {McMahon}, {Meyer}, {Mocanu}, {Mohr}, {Murray}, {Padin}, {Pryke}, {Reichardt}, {Rest}, {Ruel}, {Ruhl}, {Saliwanchik}, {Saro}, {Sayre}, {Schaffer}, {Schrabback}, {Shirokoff}, {Song}, {Spieler}, {Stanford}, {Staniszewski}, {Stark}, {Story}, {Stubbs}, {Vanderlinde}, {Vieira}, {Vikhlinin}, {Williamson}, {Zahn}, \& {Zenteno}}]{Bleem2015}
{Bleem}, L.~E., {Stalder}, B., {de Haan}, T., {et~al.} 2015, \apjs, 216, 27

\bibitem[{{Bleem} {et~al.}(2020){Bleem}, {Bocquet}, {Stalder}, {Gladders}, {Ade}, {Allen}, {Anderson}, {Annis}, {Ashby}, {Austermann}, {Avila}, {Avva}, {Bayliss}, {Beall}, {Bechtol}, {Bender}, {Benson}, {Bertin}, {Bianchini}, {Blake}, {Brodwin}, {Brooks}, {Buckley-Geer}, {Burke}, {Carlstrom}, {Rosell}, {Carrasco Kind}, {Carretero}, {Chang}, {Chiang}, {Citron}, {Moran}, {Costanzi}, {Crawford}, {Crites}, {da Costa}, {de Haan}, {De Vicente}, {Desai}, {Diehl}, {Dietrich}, {Dobbs}, {Eifler}, {Everett}, {Flaugher}, {Floyd}, {Frieman}, {Gallicchio}, {Garc{\'\i}a-Bellido}, {George}, {Gerdes}, {Gilbert}, {Gruen}, {Gruendl}, {Gschwend}, {Gupta}, {Gutierrez}, {Halverson}, {Harrington}, {Henning}, {Heymans}, {Holder}, {Hollowood}, {Holzapfel}, {Honscheid}, {Hrubes}, {Huang}, {Hubmayr}, {Irwin}, {James}, {Jeltema}, {Joudaki}, {Khullar}, {Klein}, {Knox}, {Kuropatkin}, {Lee}, {Li}, {Lidman}, {Lowitz}, {MacCrann}, {Mahler}, {Maia}, {Marshall}, {McDonald}, {McMahon}, {Melchior}, {Menanteau}, {Meyer}, {Miquel}, {Mocanu},
  {Mohr}, {Montgomery}, {Nadolski}, {Natoli}, {Nibarger}, {Noble}, {Novosad}, {Padin}, {Palmese}, {Parkinson}, {Patil}, {Paz-Chinch{\'o}n}, {Plazas}, {Pryke}, {Ramachandra}, {Reichardt}, {Remolina Gonz{\'a}lez}, {Romer}, {Roodman}, {Ruhl}, {Rykoff}, {Saliwanchik}, {Sanchez}, {Saro}, {Sayre}, {Schaffer}, {Schrabback}, {Serrano}, {Sharon}, {Sievers}, {Smecher}, {Smith}, {Soares-Santos}, {Stark}, {Story}, {Suchyta}, {Tarle}, {Tucker}, {Vanderlinde}, {Veach}, {Vieira}, {Wang}, {Weller}, {Whitehorn}, {Wu}, {Yefremenko}, \& {Zhang}}]{Bleem2020}
{Bleem}, L.~E., {Bocquet}, S., {Stalder}, B., {et~al.} 2020, \apjs, 247, 25

\bibitem[{{Bocquet} {et~al.}(2019){Bocquet}, {Dietrich}, {Schrabback}, {Bleem}, {Klein}, {Allen}, {Applegate}, {Ashby}, {Bautz}, {Bayliss}, {Benson}, {Brodwin}, {Bulbul}, {Canning}, {Capasso}, {Carlstrom}, {Chang}, {Chiu}, {Cho}, {Clocchiatti}, {Crawford}, {Crites}, {de Haan}, {Desai}, {Dobbs}, {Foley}, {Forman}, {Garmire}, {George}, {Gladders}, {Gonzalez}, {Grandis}, {Gupta}, {Halverson}, {Hlavacek-Larrondo}, {Hoekstra}, {Holder}, {Holzapfel}, {Hou}, {Hrubes}, {Huang}, {Jones}, {Khullar}, {Knox}, {Kraft}, {Lee}, {von der Linden}, {Luong-Van}, {Mantz}, {Marrone}, {McDonald}, {McMahon}, {Meyer}, {Mocanu}, {Mohr}, {Morris}, {Padin}, {Patil}, {Pryke}, {Rapetti}, {Reichardt}, {Rest}, {Ruhl}, {Saliwanchik}, {Saro}, {Sayre}, {Schaffer}, {Shirokoff}, {Stalder}, {Stanford}, {Staniszewski}, {Stark}, {Story}, {Strazzullo}, {Stubbs}, {Vanderlinde}, {Vieira}, {Vikhlinin}, {Williamson}, \& {Zenteno}}]{Bocquet2019}
{Bocquet}, S., {Dietrich}, J.~P., {Schrabback}, T., {et~al.} 2019, \apj, 878, 55

\bibitem[{{Bouwens} {et~al.}(2017){Bouwens}, {Oesch}, {Illingworth}, {Ellis}, \& {Stefanon}}]{Bouwens17}
{Bouwens}, R.~J., {Oesch}, P.~A., {Illingworth}, G.~D., {Ellis}, R.~S., \& {Stefanon}, M. 2017, \apj, 843, 129

\bibitem[{{Ca{\~n}ameras} {et~al.}(2015){Ca{\~n}ameras}, {Nesvadba}, {Guery}, {McKenzie}, {K{\"o}nig}, {Petitpas}, {Dole}, {Frye}, {Flores-Cacho}, {Montier}, {Negrello}, {Beelen}, {Boone}, {Dicken}, {Lagache}, {Le Floc'h}, {Altieri}, {B{\'e}thermin}, {Chary}, {de Zotti}, {Giard}, {Kneissl}, {Krips}, {Malhotra}, {Martinache}, {Omont}, {Pointecouteau}, {Puget}, {Scott}, {Soucail}, {Valtchanov}, {Welikala}, \& {Yan}}]{Canameras2015}
{Ca{\~n}ameras}, R., {Nesvadba}, N.~P.~H., {Guery}, D., {et~al.} 2015, \aap, 581, A105

\bibitem[{{Caminha} {et~al.}(2022){Caminha}, {Suyu}, {Mercurio}, {Brammer}, {Bergamini}, {Acebron}, \& {Vanzella}}]{Caminha2022}
{Caminha}, G.~B., {Suyu}, S.~H., {Mercurio}, A., {et~al.} 2022, \aap, 666, L9

\bibitem[{{Cerny} {et~al.}(2018){Cerny}, {Sharon}, {Andrade-Santos}, {Avila}, {Brada{\v{c}}}, {Bradley}, {Carrasco}, {Coe}, {Czakon}, {Dawson}, {Frye}, {Hoag}, {Huang}, {Johnson}, {Jones}, {Lam}, {Lovisari}, {Mainali}, {Oesch}, {Ogaz}, {Past}, {Paterno-Mahler}, {Peterson}, {Riess}, {Rodney}, {Ryan}, {Salmon}, {Sendra-Server}, {Stark}, {Strolger}, {Trenti}, {Umetsu}, {Vulcani}, \& {Zitrin}}]{Cerny2018}
{Cerny}, C., {Sharon}, K., {Andrade-Santos}, F., {et~al.} 2018, \apj, 859, 159

\bibitem[{{Chemerynska} {et~al.}(2024){Chemerynska}, {Atek}, {Furtak}, {Zitrin}, {Greene}, {Dayal}, {Weibel}, {Fujimoto}, {Kokorev}, {Goulding}, {Williams}, {Nanayakkara}, {Bezanson}, {Brammer}, {Cutler}, {Labbe}, {Leja}, {Pan}, {Price}, {van Dokkum}, {Wang}, {Weaver}, \& {Whitaker}}]{Chemerynska24}
{Chemerynska}, I., {Atek}, H., {Furtak}, L.~J., {et~al.} 2024, \mnras, 531, 2615

\bibitem[{{Claeyssens} {et~al.}(2023){Claeyssens}, {Adamo}, {Richard}, {Mahler}, {Messa}, \& {Dessauges-Zavadsky}}]{Claeyssens23}
{Claeyssens}, A., {Adamo}, A., {Richard}, J., {et~al.} 2023, \mnras, 520, 2180

\bibitem[{{Coe} {et~al.}(2019){Coe}, {Salmon}, {Bradac}, {Bradley}, {Sharon}, {Zitrin}, {Acebron}, {Cerny}, {Cibirka}, {Strait}, {Paterno-Mahler}, {Mahler}, {Avila}, {Ogaz}, {Huang}, {Pelliccia}, {Stark}, {Mainali}, {Oesch}, {Trenti}, {Carrasco}, {Dawson}, {Rodney}, {Strolger}, {Riess}, {Jones}, {Frye}, {Czakon}, {Umetsu}, {Vulcani}, {Graur}, {Jha}, {Graham}, {Molino}, {Nonino}, {Hjorth}, {Selsing}, {Christensen}, {Kikuchihara}, {Ouchi}, {Oguri}, {Welch}, {Lemaux}, {Andrade-Santos}, {Hoag}, {Johnson}, {Peterson}, {Past}, {Fox}, {Agulli}, {Livermore}, {Ryan}, {Lam}, {Sendra-Server}, {Toft}, {Lovisari}, \& {Su}}]{Coe2019}
{Coe}, D., {Salmon}, B., {Bradac}, M., {et~al.} 2019, arXiv e-prints, arXiv:1903.02002

\bibitem[{{Dahle} {et~al.}(2015){Dahle}, {Gladders}, {Sharon}, {Bayliss}, \& {Rigby}}]{Dahle2015}
{Dahle}, H., {Gladders}, M.~D., {Sharon}, K., {Bayliss}, M.~B., \& {Rigby}, J.~R. 2015, \apj, 813, 67

\bibitem[{{de La Vieuville} {et~al.}(2019){de La Vieuville}, {Bina}, {Pello}, {Mahler}, {Richard}, {Drake}, {Herenz}, {Bauer}, {Cl{\'e}ment}, {Lagattuta}, {Laporte}, {Martinez}, {Patr{\'\i}cio}, {Wisotzki}, {Zabl}, {Bouwens}, {Contini}, {Garel}, {Guiderdoni}, {Marino}, {Maseda}, {Matthee}, {Schaye}, \& {Soucail}}]{deLaVieuville2019}
{de La Vieuville}, G., {Bina}, D., {Pello}, R., {et~al.} 2019, \aap, 628, A3

\bibitem[{{de Propris} {et~al.}(1999){de Propris}, {Stanford}, {Eisenhardt}, {Dickinson}, \& {Elston}}]{dePropris1999}
{de Propris}, R., {Stanford}, S.~A., {Eisenhardt}, P.~R., {Dickinson}, M., \& {Elston}, R. 1999, \aj, 118, 719

\bibitem[{{Diego} {et~al.}(2018){Diego}, {Kaiser}, {Broadhurst}, {Kelly}, {Rodney}, {Morishita}, {Oguri}, {Ross}, {Zitrin}, {Jauzac}, {Richard}, {Williams}, {Vega-Ferrero}, {Frye}, \& {Filippenko}}]{Diego18}
{Diego}, J.~M., {Kaiser}, N., {Broadhurst}, T., {et~al.} 2018, \apj, 857, 25

\bibitem[{{Diehl} {et~al.}(2017){Diehl}, {Buckley-Geer}, {Lindgren}, {Nord}, {Gaitsch}, {Gaitsch}, {Lin}, {Allam}, {Collett}, {Furlanetto}, {Gill}, {More}, {Nightingale}, {Odden}, {Pellico}, {Tucker}, {da Costa}, {Fausti Neto}, {Kuropatkin}, {Soares-Santos}, {Welch}, {Zhang}, {Frieman}, {Abdalla}, {Annis}, {Benoit-L{\'e}vy}, {Bertin}, {Brooks}, {Burke}, {Carnero Rosell}, {Carrasco Kind}, {Carretero}, {Cunha}, {D'Andrea}, {Desai}, {Dietrich}, {Drlica-Wagner}, {Evrard}, {Finley}, {Flaugher}, {Garc{\'\i}a-Bellido}, {Gerdes}, {Goldstein}, {Gruen}, {Gruendl}, {Gschwend}, {Gutierrez}, {James}, {Kuehn}, {Kuhlmann}, {Lahav}, {Li}, {Lima}, {Maia}, {Marshall}, {Menanteau}, {Miquel}, {Nichol}, {Nugent}, {Ogando}, {Plazas}, {Reil}, {Romer}, {Sako}, {Sanchez}, {Santiago}, {Scarpine}, {Schindler}, {Schubnell}, {Sevilla-Noarbe}, {Sheldon}, {Smith}, {Sobreira}, {Suchyta}, {Swanson}, {Tarle}, {Thomas}, {Walker}, \& {DES Collaboration}}]{Diehl17}
{Diehl}, H.~T., {Buckley-Geer}, E.~J., {Lindgren}, K.~A., {et~al.} 2017, \apjs, 232, 15

\bibitem[{{Dressler} {et~al.}(2011){Dressler}, {Bigelow}, {Hare}, {Sutin}, {Thompson}, {Burley}, {Epps}, {Oemler}, {Bagish}, {Birk}, {Clardy}, {Gunnels}, {Kelson}, {Shectman}, \& {Osip}}]{dressler11}
{Dressler}, A., {Bigelow}, B., {Hare}, T., {et~al.} 2011, \pasp, 123, 288

\bibitem[{{Ebeling} {et~al.}(2001){Ebeling}, {Edge}, \& {Henry}}]{Ebeling2001}
{Ebeling}, H., {Edge}, A.~C., \& {Henry}, J.~P. 2001, \apj, 553, 668

\bibitem[{{El{\'{\i}}asd{\'o}ttir} {et~al.}(2007){El{\'{\i}}asd{\'o}ttir}, {Limousin}, {Richard}, {Hjorth}, {Kneib}, {Natarajan}, {Pedersen}, {Jullo}, \& {Paraficz}}]{Eliasdottir2007}
{El{\'{\i}}asd{\'o}ttir}, {\'A}., {Limousin}, M., {Richard}, J., {et~al.} 2007, ArXiv e-prints, arXiv:0710.5636

\bibitem[{{Fischer} {et~al.}(2019){Fischer}, {Rigby}, {Mahler}, {Gladders}, {Sharon}, {Florian}, {Kraemer}, {Bayliss}, {Dahle}, {Barrientos}, {Lopez}, {Tejos}, {Johnson}, \& {Wuyts}}]{Fisher2019}
{Fischer}, T.~C., {Rigby}, J.~R., {Mahler}, G., {et~al.} 2019, arXiv e-prints, arXiv:1903.10403

\bibitem[{{Fitzpatrick} {et~al.}(2024){Fitzpatrick}, {Placco}, {Bolton}, {Merino}, {Ridgway}, \& {Stanghellini}}]{iraf24}
{Fitzpatrick}, M., {Placco}, V., {Bolton}, A., {et~al.} 2024, arXiv e-prints, arXiv:2401.01982

\bibitem[{{Fox} {et~al.}(2022){Fox}, {Mahler}, {Sharon}, \& {Remolina Gonz{\'a}lez}}]{Fox2022}
{Fox}, C., {Mahler}, G., {Sharon}, K., \& {Remolina Gonz{\'a}lez}, J.~D. 2022, \apj, 928, 87

\bibitem[{{Fruchter} \& {et al.}(2010)}]{astrodrizzle}
{Fruchter}, A.~S., \& {et al.} 2010, in 2010 Space Telescope Science Institute Calibration Workshop, 382--387

\bibitem[{{Furtak} {et~al.}(2023){Furtak}, {Zitrin}, {Weaver}, {Atek}, {Bezanson}, {Labb{\'e}}, {Whitaker}, {Leja}, {Price}, {Brammer}, {Wang}, {Marchesini}, {Pan}, {Dayal}, {van Dokkum}, {Feldmann}, {Fujimoto}, {Franx}, {Khullar}, {Nelson}, \& {Mowla}}]{Furtak23}
{Furtak}, L.~J., {Zitrin}, A., {Weaver}, J.~R., {et~al.} 2023, \mnras, 523, 4568

\bibitem[{{Giocoli} {et~al.}(2012){Giocoli}, {Meneghetti}, {Ettori}, \& {Moscardini}}]{Giocoli2012}
{Giocoli}, C., {Meneghetti}, M., {Ettori}, S., \& {Moscardini}, L. 2012, \mnras, 426, 1558

\bibitem[{{Gladders} \& {Yee}(2000)}]{Gladders2000}
{Gladders}, M.~D., \& {Yee}, H.~K.~C. 2000, \aj, 120, 2148

\bibitem[{{Gladders} {et~al.}(2019){Gladders}, {Allen}, {Barrientos}, {Bayliss}, {Benson}, {Bleem}, {Brodwin}, {Canning}, {Florian}, {Habib}, {Heitmann}, {Hlavacek-Larrondo}, {Khullar}, {King}, {Li}, {Mantz}, {McDonald}, {Morris}, {Noordeh}, {Sharon}, {Stalder}, {Stark}, {Strazzullo}, \& {von der Linden}}]{SPTsnap}
{Gladders}, M.~D., {Allen}, S.~W., {Barrientos}, L.~F., {et~al.} 2019, {Building the SPT-HST Legacy: Imaging Massive Clusters to z=1.5}, HST Proposal. Cycle 26, ID. \#16017, ,

\bibitem[{{Golubchik} {et~al.}(2022){Golubchik}, {Furtak}, {Meena}, \& {Zitrin}}]{Golubchik2022}
{Golubchik}, M., {Furtak}, L.~J., {Meena}, A.~K., \& {Zitrin}, A. 2022, \apj, 938, 14

\bibitem[{{Gonzaga} {et~al.}(2012){Gonzaga}, {Hack}, {Fruchter}, \& {Mack}}]{gonzaga12}
{Gonzaga}, S., {Hack}, W., {Fruchter}, A., \& {Mack}, J. 2012, {The DrizzlePac Handbook}

\bibitem[{{Grillo} {et~al.}(2015){Grillo}, {Suyu}, {Rosati}, {Mercurio}, {Balestra}, {Munari}, {Nonino}, {Caminha}, {Lombardi}, {De Lucia}, {Borgani}, {Gobat}, {Biviano}, {Girardi}, {Umetsu}, {Coe}, {Koekemoer}, {Postman}, {Zitrin}, {Halkola}, {Broadhurst}, {Sartoris}, {Presotto}, {Annunziatella}, {Maier}, {Fritz}, {Vanzella}, \& {Frye}}]{Grillo2015}
{Grillo}, C., {Suyu}, S.~H., {Rosati}, P., {et~al.} 2015, \apj, 800, 38

\bibitem[{{Grillo} {et~al.}(2018){Grillo}, {Rosati}, {Suyu}, {Balestra}, {Caminha}, {Halkola}, {Kelly}, {Lombardi}, {Mercurio}, {Rodney}, \& {Treu}}]{Grillo2018}
{Grillo}, C., {Rosati}, P., {Suyu}, S.~H., {et~al.} 2018, \apj, 860, 94

\bibitem[{{Harvey} {et~al.}(2017){Harvey}, {Courbin}, {Kneib}, \& {McCarthy}}]{Harvey17}
{Harvey}, D., {Courbin}, F., {Kneib}, J.~P., \& {McCarthy}, I.~G. 2017, \mnras, 472, 1972

\bibitem[{{Hsiao} {et~al.}(2023){Hsiao}, {Coe}, {Abdurro'uf}, {Whitler}, {Jung}, {Khullar}, {Meena}, {Dayal}, {Barrow}, {Santos-Olmsted}, {Casselman}, {Vanzella}, {Nonino}, {Jim{\'e}nez-Teja}, {Oguri}, {Stark}, {Furtak}, {Zitrin}, {Adamo}, {Brammer}, {Bradley}, {Diego}, {Zackrisson}, {Finkelstein}, {Windhorst}, {Bhatawdekar}, {Hutchison}, {Broadhurst}, {Dimauro}, {Andrade-Santos}, {Eldridge}, {Acebron}, {Avila}, {Bayliss}, {Ben{\'\i}tez}, {Binggeli}, {Bolan}, {Brada{\v{c}}}, {Carnall}, {Conselice}, {Donahue}, {Frye}, {Fujimoto}, {Henry}, {James}, {Kassin}, {Kewley}, {Larson}, {Lauer}, {Law}, {Mahler}, {Mainali}, {McCandliss}, {Nicholls}, {Pirzkal}, {Postman}, {Rigby}, {Ryan}, {Senchyna}, {Sharon}, {Shimizu}, {Strait}, {Tang}, {Trenti}, {Vikaeus}, \& {Welch}}]{Hsiao2023}
{Hsiao}, T. Y.-Y., {Coe}, D., {Abdurro'uf}, {et~al.} 2023, \apjl, 949, L34

\bibitem[{{Huang} {et~al.}(2021){Huang}, {Storfer}, {Gu}, {Ravi}, {Pilon}, {Sheu}, {Venguswamy}, {Banka}, {Dey}, {Landriau}, {Lang}, {Meisner}, {Moustakas}, {Myers}, {Sajith}, {Schlafly}, \& {Schlegel}}]{Huang21}
{Huang}, X., {Storfer}, C., {Gu}, A., {et~al.} 2021, \apj, 909, 27

\bibitem[{Inc.(2022)}]{MATLAB}
Inc., T.~M. 2022, MATLAB version: 9.13.0 (R2022b),  Natick, Massachusetts, United States: The MathWorks Inc.
\newblock \url{https://www.mathworks.com}

\bibitem[{{Jauzac} {et~al.}(2012){Jauzac}, {Jullo}, {Kneib}, {Ebeling}, {Leauthaud}, {Ma}, {Limousin}, {Massey}, \& {Richard}}]{Jauzac2012}
{Jauzac}, M., {Jullo}, E., {Kneib}, J.-P., {et~al.} 2012, \mnras, 426, 3369

\bibitem[{{Jauzac} {et~al.}(2014){Jauzac}, {Cl{\'e}ment}, {Limousin}, {Richard}, {Jullo}, {Ebeling}, {Atek}, {Kneib}, {Knowles}, {Natarajan}, {Eckert}, {Egami}, {Massey}, \& {Rexroth}}]{Jauzac2014}
{Jauzac}, M., {Cl{\'e}ment}, B., {Limousin}, M., {et~al.} 2014, \mnras, 443, 1549

\bibitem[{{Johnson} \& {Sharon}(2016)}]{Johnson2016}
{Johnson}, T.~L., \& {Sharon}, K. 2016, ArXiv e-prints, arXiv:1608.08713

\bibitem[{{Johnson} {et~al.}(2017){Johnson}, {Rigby}, {Sharon}, {Gladders}, {Florian}, {Bayliss}, {Wuyts}, {Whitaker}, {Livermore}, \& {Murray}}]{Johnson2017}
{Johnson}, T.~L., {Rigby}, J.~R., {Sharon}, K., {et~al.} 2017, \apjl, 843, L21

\bibitem[{{Jullo} {et~al.}(2007){Jullo}, {Kneib}, {Limousin}, {El{\'{\i}}asd{\'o}ttir}, {Marshall}, \& {Verdugo}}]{Jullo2007}
{Jullo}, E., {Kneib}, J.-P., {Limousin}, M., {et~al.} 2007, New Journal of Physics, 9, 447

\bibitem[{{Jullo} {et~al.}(2010){Jullo}, {Natarajan}, {Kneib}, {D'Aloisio}, {Limousin}, {Richard}, \& {Schimd}}]{Jullo2010}
{Jullo}, E., {Natarajan}, P., {Kneib}, J.-P., {et~al.} 2010, Science, 329, 924

\bibitem[{{Kelly} {et~al.}(2018){Kelly}, {Diego}, {Rodney}, {Kaiser}, {Broadhurst}, {Zitrin}, {Treu}, {P{\'e}rez-Gonz{\'a}lez}, {Morishita}, {Jauzac}, {Selsing}, {Oguri}, {Pueyo}, {Ross}, {Filippenko}, {Smith}, {Hjorth}, {Cenko}, {Wang}, {Howell}, {Richard}, {Frye}, {Jha}, {Foley}, {Norman}, {Bradac}, {Zheng}, {Brammer}, {Benito}, {Cava}, {Christensen}, {de Mink}, {Graur}, {Grillo}, {Kawamata}, {Kneib}, {Matheson}, {McCully}, {Nonino}, {P{\'e}rez-Fournon}, {Riess}, {Rosati}, {Schmidt}, {Sharon}, \& {Weiner}}]{Kelly2018}
{Kelly}, P.~L., {Diego}, J.~M., {Rodney}, S., {et~al.} 2018, Nature Astronomy, 2, 334

\bibitem[{{Khullar} {et~al.}(2021){Khullar}, {Gozman}, {Lin}, {Martinez}, {Matthews Acu{\~n}a}, {Medina}, {Merz}, {Sanchez}, {Sisco}, {Kavin Stein}, {Sukay}, {Tavangar}, {Bayliss}, {Bleem}, {Brownsberger}, {Dahle}, {Florian}, {Gladders}, {Mahler}, {Rigby}, {Sharon}, \& {Stark}}]{Khullar2021}
{Khullar}, G., {Gozman}, K., {Lin}, J.~J., {et~al.} 2021, \apj, 906, 107

\bibitem[{{Livermore} {et~al.}(2017){Livermore}, {Finkelstein}, \& {Lotz}}]{Livermore17}
{Livermore}, R.~C., {Finkelstein}, S.~L., \& {Lotz}, J.~M. 2017, \apj, 835, 113

\bibitem[{{Lotz} {et~al.}(2017){Lotz}, {Koekemoer}, {Coe}, {Grogin}, {Capak}, {Mack}, {Anderson}, {Avila}, {Barker}, {Borncamp}, {Brammer}, {Durbin}, {Gunning}, {Hilbert}, {Jenkner}, {Khandrika}, {Levay}, {Lucas}, {MacKenty}, {Ogaz}, {Porterfield}, {Reid}, {Robberto}, {Royle}, {Smith}, {Storrie-Lombardi}, {Sunnquist}, {Surace}, {Taylor}, {Williams}, {Bullock}, {Dickinson}, {Finkelstein}, {Natarajan}, {Richard}, {Robertson}, {Tumlinson}, {Zitrin}, {Flanagan}, {Sembach}, {Soifer}, \& {Mountain}}]{Lotz2017}
{Lotz}, J.~M., {Koekemoer}, A., {Coe}, D., {et~al.} 2017, \apj, 837, 97

\bibitem[{{Mahler} {et~al.}(2023{\natexlab{a}}){Mahler}, {Natarajan}, {Jauzac}, \& {Richard}}]{Mahler23}
{Mahler}, G., {Natarajan}, P., {Jauzac}, M., \& {Richard}, J. 2023{\natexlab{a}}, \mnras, 518, 54

\bibitem[{{Mahler} {et~al.}(2019){Mahler}, {Sharon}, {Fox}, {Coe}, {Jauzac}, {Strait}, {Edge}, {Acebron}, {Andrade-Santos}, {Avila}, {Brada{\v{c}}}, {Bradley}, {Carrasco}, {Cerny}, {Cibirka}, {Czakon}, {Dawson}, {Frye}, {Hoag}, {Huang}, {Johnson}, {Jones}, {Kikuchihara}, {Lam}, {Livermore}, {Lovisari}, {Mainali}, {Ogaz}, {Ouchi}, {Paterno-Mahler}, {Roederer}, {Ryan}, {Salmon}, {Sendra-Server}, {Stark}, {Toft}, {Trenti}, {Umetsu}, {Vulcani}, \& {Zitrin}}]{Mahler2019}
{Mahler}, G., {Sharon}, K., {Fox}, C., {et~al.} 2019, \apj, 873, 96

\bibitem[{{Mahler} {et~al.}(2020){Mahler}, {Sharon}, {Gladders}, {Bleem}, {Bayliss}, {Calzadilla}, {Floyd}, {Khullar}, {McDonald}, {Remolina Gonz{\'a}lez}, {Schrabback}, {Stark}, \& {van den Busch}}]{Mahler2020}
{Mahler}, G., {Sharon}, K., {Gladders}, M.~D., {et~al.} 2020, \apj, 894, 150

\bibitem[{{Mahler} {et~al.}(2023{\natexlab{b}}){Mahler}, {Jauzac}, {Richard}, {Beauchesne}, {Ebeling}, {Lagattuta}, {Natarajan}, {Sharon}, {Atek}, {Claeyssens}, {Cl{\'e}ment}, {Eckert}, {Edge}, {Kneib}, \& {Niemiec}}]{Mahler2023smacs}
{Mahler}, G., {Jauzac}, M., {Richard}, J., {et~al.} 2023{\natexlab{b}}, \apj, 945, 49

\bibitem[{{McDonald} {et~al.}(2017){McDonald}, {Allen}, {Bayliss}, {Benson}, {Bleem}, {Brodwin}, {Bulbul}, {Carlstrom}, {Forman}, {Hlavacek-Larrondo}, {Garmire}, {Gaspari}, {Gladders}, {Mantz}, \& {Murray}}]{McDonald2017}
{McDonald}, M., {Allen}, S.~W., {Bayliss}, M., {et~al.} 2017, \apj, 843, 28

\bibitem[{{Meneghetti} {et~al.}(2017){Meneghetti}, {Natarajan}, {Coe}, {Contini}, {De Lucia}, {Giocoli}, {Acebron}, {Borgani}, {Bradac}, {Diego}, {Hoag}, {Ishigaki}, {Johnson}, {Jullo}, {Kawamata}, {Lam}, {Limousin}, {Liesenborgs}, {Oguri}, {Sebesta}, {Sharon}, {Williams}, \& {Zitrin}}]{Meneghetti2017}
{Meneghetti}, M., {Natarajan}, P., {Coe}, D., {et~al.} 2017, \mnras, 472, 3177

\bibitem[{{Meneghetti} {et~al.}(2020){Meneghetti}, {Davoli}, {Bergamini}, {Rosati}, {Natarajan}, {Giocoli}, {Caminha}, {Metcalf}, {Rasia}, {Borgani}, {Calura}, {Grillo}, {Mercurio}, \& {Vanzella}}]{Meneghetti2020}
{Meneghetti}, M., {Davoli}, G., {Bergamini}, P., {et~al.} 2020, Science, 369, 1347

\bibitem[{{Meneghetti} {et~al.}(2022{\natexlab{a}}){Meneghetti}, {Ragagnin}, {Borgani}, {Calura}, {Despali}, {Giocoli}, {Granato}, {Grillo}, {Moscardini}, {Rasia}, {Rosati}, {Angora}, {Bassini}, {Bergamini}, {Caminha}, {Granata}, {Mercurio}, {Metcalf}, {Natarajan}, {Nonino}, {Pignataro}, {Ragone-Figueroa}, {Vanzella}, {Acebron}, {Dolag}, {Murante}, {Taffoni}, {Tornatore}, {Tortorelli}, \& {Valentini}}]{Meneghetti22}
{Meneghetti}, M., {Ragagnin}, A., {Borgani}, S., {et~al.} 2022{\natexlab{a}}, \aap, 668, A188

\bibitem[{{Meneghetti} {et~al.}(2022{\natexlab{b}}){Meneghetti}, {Ragagnin}, {Borgani}, {Calura}, {Despali}, {Giocoli}, {Granato}, {Grillo}, {Moscardini}, {Rasia}, {Rosati}, {Angora}, {Bassini}, {Bergamini}, {Caminha}, {Granata}, {Mercurio}, {Metcalf}, {Natarajan}, {Nonino}, {Pignataro}, {Ragone-Figueroa}, {Vanzella}, {Acebron}, {Dolag}, {Murante}, {Taffoni}, {Tornatore}, {Tortorelli}, \& {Valentini}}]{Meneghetti2022}
---. 2022{\natexlab{b}}, \aap, 668, A188

\bibitem[{{Meneghetti} {et~al.}(2023){Meneghetti}, {Cui}, {Rasia}, {Yepes}, {Acebron}, {Angora}, {Bergamini}, {Borgani}, {Calura}, {Despali}, {Giocoli}, {Granata}, {Grillo}, {Knebe}, {Macci{\`o}}, {Mercurio}, {Moscardini}, {Natarajan}, {Ragagnin}, {Rosati}, \& {Vanzella}}]{Meneghetti2023}
{Meneghetti}, M., {Cui}, W., {Rasia}, E., {et~al.} 2023, \aap, 678, L2

\bibitem[{{Munari} {et~al.}(2016){Munari}, {Grillo}, {De Lucia}, {Biviano}, {Annunziatella}, {Borgani}, {Lombardi}, {Mercurio}, \& {Rosati}}]{Munari2016}
{Munari}, E., {Grillo}, C., {De Lucia}, G., {et~al.} 2016, \apjl, 827, L5

\bibitem[{{Napier} {et~al.}(2023){Napier}, {Sharon}, {Dahle}, {Bayliss}, {Gladders}, {Mahler}, {Rigby}, \& {Florian}}]{Napier23}
{Napier}, K., {Sharon}, K., {Dahle}, H., {et~al.} 2023, \apj, 959, 134

\bibitem[{{Natarajan} {et~al.}(2017){Natarajan}, {Chadayammuri}, {Jauzac}, {Richard}, {Kneib}, {Ebeling}, {Jiang}, {van den Bosch}, {Limousin}, {Jullo}, {Atek}, {Pillepich}, {Popa}, {Marinacci}, {Hernquist}, {Meneghetti}, \& {Vogelsberger}}]{Natarajan2017}
{Natarajan}, P., {Chadayammuri}, U., {Jauzac}, M., {et~al.} 2017, \mnras, 468, 1962

\bibitem[{{Navarro} {et~al.}(1996){Navarro}, {Frenk}, \& {White}}]{Navarro1996}
{Navarro}, J.~F., {Frenk}, C.~S., \& {White}, S. D.~M. 1996, \apj, 462, 563

\bibitem[{{Oemler} {et~al.}(2017){Oemler}, {Clardy}, {Kelson}, {Walth}, \& {Villanueva}}]{oemler17}
{Oemler}, A., {Clardy}, K., {Kelson}, D., {Walth}, G., \& {Villanueva}, E. 2017, {COSMOS: Carnegie Observatories System for MultiObject Spectroscopy}, Astrophysics Source Code Library, record ascl:1705.001, ,

\bibitem[{{Ofek}(2014)}]{Ofek2014}
{Ofek}, E.~O. 2014, {MAAT: MATLAB Astronomy and Astrophysics Toolbox}, Astrophysics Source Code Library, record ascl:1407.005, , , ascl:1407.005

\bibitem[{{Oke}(1974)}]{Oke1974}
{Oke}, J.~B. 1974, \apjs, 27, 21

\bibitem[{{Patel} {et~al.}(2024){Patel}, {Jauzac}, {Niemiec}, {Lagattuta}, {Mahler}, {Beauchesne}, {Edge}, {Ebeling}, \& {Limousin}}]{Patel2024}
{Patel}, N.~R., {Jauzac}, M., {Niemiec}, A., {et~al.} 2024, arXiv e-prints, arXiv:2405.04577

\bibitem[{{Patr{\'\i}cio} {et~al.}(2018){Patr{\'\i}cio}, {Richard}, {Carton}, {Contini}, {Epinat}, {Brinchmann}, {Schmidt}, {Krajnovi{\'c}}, {Bouch{\'e}}, {Weilbacher}, {Pell{\'o}}, {Caruana}, {Maseda}, {Finley}, {Bauer}, {Martinez}, {Mahler}, {Lagattuta}, {Cl{\'e}ment}, {Soucail}, \& {Wisotzki}}]{Patricio2018}
{Patr{\'\i}cio}, V., {Richard}, J., {Carton}, D., {et~al.} 2018, \mnras, 477, 18

\bibitem[{{Postman} {et~al.}(2012){Postman}, {Coe}, {Ben{\'\i}tez}, {Bradley}, {Broadhurst}, {Donahue}, {Ford}, {Graur}, {Graves}, {Jouvel}, {Koekemoer}, {Lemze}, {Medezinski}, {Molino}, {Moustakas}, {Ogaz}, {Riess}, {Rodney}, {Rosati}, {Umetsu}, {Zheng}, {Zitrin}, {Bartelmann}, {Bouwens}, {Czakon}, {Golwala}, {Host}, {Infante}, {Jha}, {Jimenez-Teja}, {Kelson}, {Lahav}, {Lazkoz}, {Maoz}, {McCully}, {Melchior}, {Meneghetti}, {Merten}, {Moustakas}, {Nonino}, {Patel}, {Reg{\"o}s}, {Sayers}, {Seitz}, \& {Van der Wel}}]{Postman2012}
{Postman}, M., {Coe}, D., {Ben{\'\i}tez}, N., {et~al.} 2012, The Astrophysical Journal Supplement Series, 199, 25

\bibitem[{{Rajan}(2010)}]{rajan10}
{Rajan}, A. 2010, in WFC3 Data Handbook v. 2, Vol.~2, 2

\bibitem[{{Remolina Gonzàlez}(2021)}]{remolina_thesis}
{Remolina Gonzàlez}, J.~D. 2021, {The Concentration-Mass Relation Across Cosmic Time of Strong Lensing Galaxy Clusters}, v1.0,  Deep Blue, doi:https://dx.doi.org/10.7302/2889.
\newblock \url{https://deepblue.lib.umich.edu/handle/2027.42/169844}

\bibitem[{{Richard} {et~al.}(2011){Richard}, {Kneib}, {Ebeling}, {Stark}, {Egami}, \& {Fiedler}}]{Richard2011}
{Richard}, J., {Kneib}, J.-P., {Ebeling}, H., {et~al.} 2011, \mnras, 414, L31

\bibitem[{{Rigby} {et~al.}(2023){Rigby}, {Vieira}, {Phadke}, {Hutchison}, {Welch}, {Cathey}, {Spilker}, {Gonzalez}, {Adhikari}, {Aravena}, {Bayliss}, {Birkin}, {Bursk}, {Chapman}, {Dahle}, {Elicker}, {Fischer}, {Florian}, {Gladders}, {Hayward}, {Hewald}, {Kettler}, {Khullar}, {Kim}, {Law}, {Mahler}, {Malhotra}, {Murphy}, {Narayanan}, {Olivier}, {Rhoads}, {Sharon}, {Solimano}, {Thiruvengadam}, {Vizgan}, \& {Younker}}]{Rigby2023}
{Rigby}, J.~R., {Vieira}, J.~D., {Phadke}, K.~A., {et~al.} 2023, arXiv e-prints, arXiv:2312.10465

\bibitem[{{Rivera-Thorsen} {et~al.}(2019){Rivera-Thorsen}, {Dahle}, {Chisholm}, {Florian}, {Gronke}, {Rigby}, {Gladders}, {Mahler}, {Sharon}, \& {Bayliss}}]{Rivera-Thorsen2019}
{Rivera-Thorsen}, T.~E., {Dahle}, H., {Chisholm}, J., {et~al.} 2019, Science, 366, 738

\bibitem[{{Robertson}(2021{\natexlab{a}})}]{Robertson21}
{Robertson}, A. 2021{\natexlab{a}}, \mnras, 504, L7

\bibitem[{{Robertson}(2021{\natexlab{b}})}]{Robertson2021}
---. 2021{\natexlab{b}}, \mnras, 504, L7

\bibitem[{{Robertson} {et~al.}(2019){Robertson}, {Harvey}, {Massey}, {Eke}, {McCarthy}, {Jauzac}, {Li}, \& {Schaye}}]{Robertson19}
{Robertson}, A., {Harvey}, D., {Massey}, R., {et~al.} 2019, \mnras, 488, 3646

\bibitem[{{Ruel} {et~al.}(2014){Ruel}, {Bazin}, {Bayliss}, {Brodwin}, {Foley}, {Stalder}, {Aird}, {Armstrong}, {Ashby}, {Bautz}, {Benson}, {Bleem}, {Bocquet}, {Carlstrom}, {Chang}, {Chapman}, {Cho}, {Clocchiatti}, {Crawford}, {Crites}, {de Haan}, {Desai}, {Dobbs}, {Dudley}, {Forman}, {George}, {Gladders}, {Gonzalez}, {Halverson}, {Harrington}, {High}, {Holder}, {Holzapfel}, {Hrubes}, {Jones}, {Joy}, {Keisler}, {Knox}, {Lee}, {Leitch}, {Liu}, {Lueker}, {Luong-Van}, {Mantz}, {Marrone}, {McDonald}, {McMahon}, {Mehl}, {Meyer}, {Mocanu}, {Mohr}, {Montroy}, {Murray}, {Natoli}, {Nurgaliev}, {Padin}, {Plagge}, {Pryke}, {Reichardt}, {Rest}, {Ruhl}, {Saliwanchik}, {Saro}, {Sayre}, {Schaffer}, {Shaw}, {Shirokoff}, {Song}, {{\v{S}}uhada}, {Spieler}, {Stanford}, {Staniszewski}, {Starsk}, {Story}, {Stubbs}, {van Engelen}, {Vanderlinde}, {Vieira}, {Vikhlinin}, {Williamson}, {Zahn}, \& {Zenteno}}]{Ruel2014}
{Ruel}, J., {Bazin}, G., {Bayliss}, M., {et~al.} 2014, \apj, 792, 45

\bibitem[{{Salmon} {et~al.}(2020){Salmon}, {Coe}, {Bradley}, {Bouwens}, {Brada{\v{c}}}, {Huang}, {Oesch}, {Stark}, {Sharon}, {Trenti}, {Avila}, {Ogaz}, {Andrade-Santos}, {Carrasco}, {Cerny}, {Dawson}, {Frye}, {Hoag}, {Johnson}, {Jones}, {Lam}, {Lovisari}, {Mainali}, {Past}, {Paterno-Mahler}, {Peterson}, {Riess}, {Rodney}, {Ryan}, {Sendra-Server}, {Strait}, {Strolger}, {Umetsu}, {Vulcani}, \& {Zitrin}}]{Salmon2020}
{Salmon}, B., {Coe}, D., {Bradley}, L., {et~al.} 2020, \apj, 889, 189

\bibitem[{{Schrabback} {et~al.}(2021){Schrabback}, {Bocquet}, {Sommer}, {Zohren}, {van den Busch}, {Hern{\'a}ndez-Mart{\'\i}n}, {Hoekstra}, {Raihan}, {Schirmer}, {Applegate}, {Bayliss}, {Benson}, {Bleem}, {Dietrich}, {Floyd}, {Hilbert}, {Hlavacek-Larrondo}, {McDonald}, {Saro}, {Stark}, \& {Weissgerber}}]{Schrabback2021}
{Schrabback}, T., {Bocquet}, S., {Sommer}, M., {et~al.} 2021, \mnras, 505, 3923

\bibitem[{{Sharon} {et~al.}(2020){Sharon}, {Bayliss}, {Dahle}, {Dunham}, {Florian}, {Gladders}, {Johnson}, {Mahler}, {Paterno-Mahler}, {Rigby}, {Whitaker}, {Akhshik}, {Koester}, {Murray}, {Remolina Gonz{\'a}lez}, \& {Wuyts}}]{sharon2020}
{Sharon}, K., {Bayliss}, M.~B., {Dahle}, H., {et~al.} 2020, \apjs, 247, 12

\bibitem[{{Simcoe} {et~al.}(2013){Simcoe}, {Burgasser}, {Schechter}, {Fishner}, {Bernstein}, {Bigelow}, {Pipher}, {Forrest}, {McMurtry}, {Smith}, \& {Bochanski}}]{FIRE}
{Simcoe}, R.~A., {Burgasser}, A.~J., {Schechter}, P.~L., {et~al.} 2013, \pasp, 125, 270

\bibitem[{{Sirks} {et~al.}(2022){Sirks}, {Oman}, {Robertson}, {Massey}, \& {Frenk}}]{Sirks22}
{Sirks}, E.~L., {Oman}, K.~A., {Robertson}, A., {Massey}, R., \& {Frenk}, C. 2022, \mnras, 511, 5927

\bibitem[{{Smith} {et~al.}(2005){Smith}, {Kneib}, {Smail}, {Mazzotta}, {Ebeling}, \& {Czoske}}]{Smith2005}
{Smith}, G.~P., {Kneib}, J.-P., {Smail}, I., {et~al.} 2005, \mnras, 359, 417

\bibitem[{{Somboonpanyakul} {et~al.}(2021){Somboonpanyakul}, {McDonald}, {Gaspari}, {Stalder}, \& {Stark}}]{Somboonpanyakul2021}
{Somboonpanyakul}, T., {McDonald}, M., {Gaspari}, M., {Stalder}, B., \& {Stark}, A.~A. 2021, \apj, 910, 60

\bibitem[{{Stalder} {et~al.}(2014){Stalder}, {Stark}, {Amato}, {Geary}, {Shectman}, {Stubbs}, \& {Szentgyorgyi}}]{Stalder2014}
{Stalder}, B., {Stark}, A.~A., {Amato}, S.~M., {et~al.} 2014, in \procspie, Vol. 9147, Ground-based and Airborne Instrumentation for Astronomy V, 91473Y

\bibitem[{{Stark} {et~al.}(2013){Stark}, {Auger}, {Belokurov}, {Jones}, {Robertson}, {Ellis}, {Sand}, {Moiseev}, {Eagle}, \& {Myers}}]{Stark13}
{Stark}, D.~P., {Auger}, M., {Belokurov}, V., {et~al.} 2013, \mnras, 436, 1040

\bibitem[{{Tam} {et~al.}(2022){Tam}, {Umetsu}, \& {Amara}}]{Tam22}
{Tam}, S.-I., {Umetsu}, K., \& {Amara}, A. 2022, \apj, 925, 145

\bibitem[{{Tokayer} {et~al.}(2024){Tokayer}, {Dutra}, {Natarajan}, {Mahler}, {Jauzac}, \& {Meneghetti}}]{Tokayer2024}
{Tokayer}, Y.~M., {Dutra}, I., {Natarajan}, P., {et~al.} 2024, \apj, 970, 143

\bibitem[{{Vanderlinde} {et~al.}(2010){Vanderlinde}, {Crawford}, {de Haan}, {Dudley}, {Shaw}, {Ade}, {Aird}, {Benson}, {Bleem}, {Brodwin}, {Carlstrom}, {Chang}, {Crites}, {Desai}, {Dobbs}, {Foley}, {George}, {Gladders}, {Hall}, {Halverson}, {High}, {Holder}, {Holzapfel}, {Hrubes}, {Joy}, {Keisler}, {Knox}, {Lee}, {Leitch}, {Loehr}, {Lueker}, {Marrone}, {McMahon}, {Mehl}, {Meyer}, {Mohr}, {Montroy}, {Ngeow}, {Padin}, {Plagge}, {Pryke}, {Reichardt}, {Rest}, {Ruel}, {Ruhl}, {Schaffer}, {Shirokoff}, {Song}, {Spieler}, {Stalder}, {Staniszewski}, {Stark}, {Stubbs}, {van Engelen}, {Vieira}, {Williamson}, {Yang}, {Zahn}, \& {Zenteno}}]{Vanderlinde2010}
{Vanderlinde}, K., {Crawford}, T.~M., {de Haan}, T., {et~al.} 2010, \apj, 722, 1180

\bibitem[{{Welch} {et~al.}(2022){Welch}, {Coe}, {Zackrisson}, {de Mink}, {Ravindranath}, {Anderson}, {Brammer}, {Bradley}, {Yoon}, {Kelly}, {Diego}, {Windhorst}, {Zitrin}, {Dimauro}, {Jim{\'e}nez-Teja}, {Abdurro'uf}, {Nonino}, {Acebron}, {Andrade-Santos}, {Avila}, {Bayliss}, {Ben{\'\i}tez}, {Broadhurst}, {Bhatawdekar}, {Brada{\v{c}}}, {Caminha}, {Chen}, {Eldridge}, {Farag}, {Florian}, {Frye}, {Fujimoto}, {Gomez}, {Henry}, {Hsiao}, {Hutchison}, {James}, {Joyce}, {Jung}, {Khullar}, {Larson}, {Mahler}, {Mandelker}, {McCandliss}, {Morishita}, {Newshore}, {Norman}, {O'Connor}, {Oesch}, {Oguri}, {Ouchi}, {Postman}, {Rigby}, {Ryan}, {Sharma}, {Sharon}, {Strait}, {Strolger}, {Timmes}, {Toft}, {Trenti}, {Vanzella}, \& {Vikaeus}}]{welch2022}
{Welch}, B., {Coe}, D., {Zackrisson}, E., {et~al.} 2022, \apjl, 940, L1

\bibitem[{{Windhorst} {et~al.}(2023){Windhorst}, {Cohen}, {Jansen}, {Summers}, {Tompkins}, {Conselice}, {Driver}, {Yan}, {Coe}, {Frye}, {Grogin}, {Koekemoer}, {Marshall}, {O'Brien}, {Pirzkal}, {Robotham}, {Ryan}, {Willmer}, {Carleton}, {Diego}, {Keel}, {Porto}, {Redshaw}, {Scheller}, {Wilkins}, {Willner}, {Zitrin}, {Adams}, {Austin}, {Arendt}, {Beacom}, {Bhatawdekar}, {Bradley}, {Broadhurst}, {Cheng}, {Civano}, {Dai}, {Dole}, {D'Silva}, {Duncan}, {Fazio}, {Ferrami}, {Ferreira}, {Finkelstein}, {Furtak}, {Gim}, {Griffiths}, {Hammel}, {Harrington}, {Hathi}, {Holwerda}, {Honor}, {Huang}, {Hyun}, {Im}, {Joshi}, {Kamieneski}, {Kelly}, {Larson}, {Li}, {Lim}, {Ma}, {Maksym}, {Manzoni}, {Meena}, {Milam}, {Nonino}, {Pascale}, {Petric}, {Pierel}, {Polletta}, {R{\"o}ttgering}, {Rutkowski}, {Smail}, {Straughn}, {Strolger}, {Swirbul}, {Trussler}, {Wang}, {Welch}, {B. Wyithe}, {Yun}, {Zackrisson}, {Zhang}, \& {Zhao}}]{Windhorst2023}
{Windhorst}, R.~A., {Cohen}, S.~H., {Jansen}, R.~A., {et~al.} 2023, \aj, 165, 13

\bibitem[{{Zitrin} {et~al.}(2017){Zitrin}, {Seitz}, {Monna}, {Koekemoer}, {Nonino}, {Gruen}, {Balestra}, {Girardi}, {Koppenhoefer}, \& {Mercurio}}]{Zitrin2017}
{Zitrin}, A., {Seitz}, S., {Monna}, A., {et~al.} 2017, \apjl, 839, L11

\end{thebibliography}

\appendix
%\autoref{tab:arcstable}.
\section{Spectroscopic catalogs}\label{sec:appendixA}
We provide in this section spectroscopic catalogs of all the objects in \SPTXXIII\ and \SPTZERO\ for which spectroscopic redshifts were measured, as described in \autoref{sec:specdata}. For strongly lensed galaxies, see \autoref{tab:arcstable}.
\startlongtable
\begin{deluxetable*}{lllllll} 
\tablecolumns{7} 
\tablecaption{List of spectroscopically identified objects in \SPTXXIII\ field of view \label{tab:spt2325allz}  } 
\tablehead{\colhead{ID} &
            \colhead{R.A.}    & 
            \colhead{Decl.}    & 
            \colhead{$z_{spec}$}     & 
            \colhead{conf.}       & 
            \colhead{Instruments}      &
            \colhead{}       \\[-8pt]
            \colhead{} &
            \colhead{J2000}     & 
            \colhead{J2000}    & 
            \colhead{}       & 
            \colhead{}       & 
            \colhead{}       & 
            \colhead{}             }
\startdata 
01401 & 351.3441640 & -41.2209030 & 0.348900 & 3 & GMOS & \\
01413 & 351.3497470 & -41.1972200 & 0.160000 & 3 & GMOS & \\
01390 & 351.3407030 & -41.1987940 & 0.361300 & 3 & GMOS & \\
01368 & 351.3332750 & -41.1982170 & 0.363900 & 3 & GMOS & \\
01265 & 351.3085740 & -41.2348110 & 0.363800 & 3 & GMOS & \\
01336 & 351.3243450 & -41.2229620 & 0.352100 & 3 & GMOS & \\
01218 & 351.2987820 & -41.2038280 & 0.362400 & 3 & GMOS & \\
01201 & 351.2953760 & -41.1707570 & 0.192000 & 3 & GMOS & \\
01318 & 351.3188660 & -41.1687860 & 0.357900 & 3 & GMOS & \\
01241 & 351.3030290 & -41.1961540 & 0.356700 & 3 & GMOS & \\
01183 & 351.2921760 & -41.2028950 & 0.350500 & 3 & GMOS & \\
01294 & 351.3145180 & -41.1759540 & 0.362100 & 3 & GMOS & \\
01348 & 351.3276450 & -41.2226060 & 0.226000 & 3 & GMOS & \\
01137 & 351.2802040 & -41.1862050 & 0.489700 & 3 & GMOS & \\
01154 & 351.2865250 & -41.1763750 & 0.356900 & 3 & GMOS & \\
01121 & 351.2765190 & -41.1656310 & 0.374400 & 3 & GMOS & \\
01103 & 351.2699750 & -41.1702150 & 0.356200 & 3 & GMOS & \\
03971 & 351.2665630 & -41.1772930 & 0.432500 & 3 & GMOS,LDSS3 & \\
01057 & 351.2577900 & -41.2204410 & 0.375700 & 3 & GMOS & \\
01045 & 351.2523090 & -41.2005090 & 0.367500 & 3 & GMOS & \\
01013 & 351.2375390 & -41.1811040 & 0.351900 & 3 & GMOS & \\
01027 & 351.2445480 & -41.2360100 & 0.357100 & 3 & GMOS & \\
01020 & 351.2410960 & -41.2258420 & 0.432500 & 3 & GMOS & \\
01035 & 351.2487790 & -41.1939490 & 0.349900 & 3 & GMOS & \\
01394 & 351.3416100 & -41.1752660 & 0.321000 & 3 & GMOS & \\
01404 & 351.3447660 & -41.2287250 & 0.356900 & 3 & GMOS & \\
01411 & 351.3487220 & -41.1709890 & 0.347200 & 3 & GMOS & \\
01372 & 351.3346820 & -41.1703420 & 0.362000 & 3 & GMOS & \\
01270 & 351.3091410 & -41.2274320 & 0.355300 & 3 & GMOS & \\
01340 & 351.3247990 & -41.2043660 & 0.362200 & 3 & GMOS & \\
01197 & 351.2954430 & -41.1902700 & 0.491300 & 3 & GMOS & \\
01312 & 351.3177880 & -41.2039130 & 0.356600 & 3 & GMOS,LDSS3 & \\
01225 & 351.3000300 & -41.1990380 & 0.353900 & 3 & GMOS & \\
01245 & 351.3043020 & -41.1951820 & 0.350500 & 3 & GMOS & \\
01290 & 351.3142180 & -41.2351900 & 0.312900 & 3 & GMOS & \\
01349 & 351.3278620 & -41.2207260 & 0.348400 & 3 & GMOS & \\
01329 & 351.3214490 & -41.1846350 & 0.313200 & 3 & GMOS & \\
01148 & 351.2842540 & -41.1864490 & 0.366400 & 3 & GMOS & \\
01135 & 351.2801310 & -41.2032250 & 0.367700 & 3 & GMOS & \\
01094 & 351.2683210 & -41.1910510 & 0.321900 & 3 & GMOS,IMACS & \\
01164 & 351.2882280 & -41.2252490 & 0.348700 & 3 & GMOS & \\
01107 & 351.2721890 & -41.1771660 & 0.321400 & 3 & GMOS & \\
01070 & 351.2609010 & -41.1530830 & 0.359500 & 3 & GMOS & \\
01054 & 351.2563260 & -41.2219040 & 0.374600 & 3 & GMOS & \\
01046 & 351.2526510 & -41.1843500 & 0.335100 & 3 & GMOS & \\
04869 & 351.2485840 & -41.1571180 & 0.702000 & 3 & GMOS & \\
01011 & 351.2369250 & -41.1530090 & 0.220300 & 3 & GMOS & \\
1000000 & 351.3035706 & -41.2089879 & 0.5977 & 3 & LDSS3 & \\
1000002 & 351.3087817 & -41.2094812 & 1.2530 & 3 & LDSS3 & \\
1000004 & 351.2916434 & -41.1847361 & 0.4605 & 3 & LDSS3 & \\
1000006 & 351.3056850 & -41.1922042 & 0.6584 & 3 & LDSS3 & \\
1000007 & 351.3188893 & -41.2330244 & 0.3609 & 3 & LDSS3 & \\
1000008 & 351.3084807 & -41.1828173 & 1.7742 & 3 & LDSS3 & \\
1000009 & 351.3062829 & -41.1834300 & 1.2476 & 3 & LDSS3 & \\
1000010 & 351.3108177 & -41.1830207 & 0.4077 & 3 & LDSS3 & \\
1000011 & 351.3292737 & -41.1694328 & 0.3128 & 3 & LDSS3 & \\
1000012 & 351.3298597 & -41.1746109 & 0.3548 & 3 & LDSS3 & \\
1000013 & 351.3440608 & -41.1963206 & 0.3616 & 3 & LDSS3,IMACS & \\
1000014 & 351.3106927 & -41.1703992 & 0.3591 & 3 & LDSS3 & \\
1000015 & 351.3129579 & -41.1807395 & 0.3034 & 3 & LDSS3 & \\
1000016 & 351.3359554 & -41.1846179 & 0.3488 & 3 & LDSS3 & \\
1000017 & 351.3143582 & -41.1692656 & 0.3522 & 3 & LDSS3 & \\
1000018 & 351.3115126 & -41.1750614 & 0.1728 & 3 & LDSS3 & \\
1000019 & 351.3026622 & -41.2254834 & 0.3493 & 3 & LDSS3 & \\
1000020 & 351.2815989 & -41.2078578 & 0.3585 & 3 & LDSS3 & \\
1000021 & 351.2850689 & -41.2131094 & 0.3587 & 3 & LDSS3 & \\
1000022 & 351.2928224 & -41.2258305 & 0.4028 & 3 & LDSS3 & \\
1000023 & 351.2831678 & -41.2190378 & 0.3634 & 3 & LDSS3 & \\
1000024 & 351.2918273 & -41.2324123 & 0.4881 & 3 & LDSS3 & \\
1000025 & 351.2907580 & -41.2348826 & 0.3635 & 3 & LDSS3 & \\
1000026 & 351.2868297 & -41.2463011 & 0.3603 & 3 & LDSS3,IMACS & \\
1000027 & 351.2827486 & -41.2013488 & 0.3552 & 3 & LDSS3 & \\
1000028 & 351.2940010 & -41.2501492 & 0.3383 & 3 & LDSS3 & \\
1000029 & 351.2756630 & -41.2396156 & 0.6814 & 3 & LDSS3 & \\
1000030 & 351.2530784 & -41.2350867 & 0.5790 & 3 & LDSS3 & \\
1000031 & 351.2689117 & -41.2478494 & 0.4635 & 3 & LDSS3 & \\
1000032 & 351.2955916 & -41.2465336 & 0.3617 & 3 & LDSS3 & \\
1000033 & 351.2653446 & -41.2353025 & 0.3683 & 3 & LDSS3 & \\
1000034 & 351.3481379 & -41.1745702 & 0.3616 & 3 & LDSS3 & \\
1000035 & 351.3354588 & -41.1733367 & 0.3398 & 3 & LDSS3 & \\
1000036 & 351.3690074 & -41.1931554 & 0.3566 & 3 & LDSS3,IMACS & \\
1000037 & 351.3040458 & -41.2116733 & 1.3211 & 3 & LDSS3 & \\
1000037 & 351.3030531 & -41.2119615 & 0.9083 & 3 & LDSS3 & \\
1000000 & 351.2934811 & -41.2082503 & 0.2542 & 3 & LDSS3 & \\
1000001 & 351.2919703 & -41.2070080 & 0.2544 & 3 & LDSS3 & \\
1000002 & 351.3118027 & -41.2034595 & 2.1890 & 3 & LDSS3 & \\
1000003 & 351.3108949 & -41.2000035 & 0.5609 & 3 & LDSS3 & \\
1000004 & 351.2958046 & -41.2101419 & 0.1145 & 3 & LDSS3 & \\
1000005 & 351.3077958 & -41.2157158 & 1.2870 & 3 & LDSS3 & \\
1000006 & 351.3188007 & -41.2043415 & 1.5811 & 3 & LDSS3 & \\
1000006 & 351.3196598 & -41.2045284 & 1.2548 & 3 & LDSS3 & \\
1000008 & 351.2966835 & -41.1869120 & 0.5790 & 3 & LDSS3 & \\
1000009 & 351.2997517 & -41.2165675 & 0.8437 & 3 & LDSS3 & \\
1000010 & 351.3251136 & -41.2092826 & 0.3551 & 3 & LDSS3 & \\
1000011 & 351.3402885 & -41.2132521 & 0.3481 & 3 & LDSS3 & \\
1000012 & 351.3207483 & -41.2384890 & 0.3563 & 3 & LDSS3 & \\
1000013 & 351.3245493 & -41.2030424 & 0.3429 & 3 & LDSS3 & \\
1000014 & 351.3172530 & -41.2319393 & 0.3671 & 3 & LDSS3 & \\
1000015 & 351.3417345 & -41.2322558 & 0.3668 & 3 & LDSS3 & \\
1000016 & 351.3478817 & -41.2339794 & 0.3597 & 3 & LDSS3 & \\
1000017 & 351.3395725 & -41.2305706 & 0.3503 & 3 & LDSS3 & \\
1000018 & 351.3634325 & -41.2211889 & 0.3598 & 3 & LDSS3 & \\
1000019 & 351.3607681 & -41.2200194 & 0.3522 & 3 & LDSS3 & \\
1000020 & 351.3655344 & -41.2040346 & 0.3549 & 3 & LDSS3 & \\
1000021 & 351.2783992 & -41.1973868 & 0.3464 & 3 & LDSS3 & \\
1000022 & 351.2794197 & -41.2123900 & 0.3588 & 3 & LDSS3 & \\
1000023 & 351.2778861 & -41.1858110 & 0.4889 & 3 & LDSS3 & \\
1000024 & 351.2758972 & -41.1709807 & 0.3597 & 3 & LDSS3 & \\
1000025 & 351.2617899 & -41.1796658 & 0.3457 & 3 & LDSS3 & \\
1000026 & 351.2620483 & -41.1927381 & 0.3488 & 3 & LDSS3 & \\
1000027 & 351.2634913 & -41.2052930 & 0.3748 & 3 & LDSS3 & \\
1000028 & 351.2786270 & -41.1887002 & 0.3434 & 3 & LDSS3 & \\
1000029 & 351.3550048 & -41.2530103 & 0.3568 & 3 & LDSS3 & \\
1000030 & 351.3597963 & -41.2380938 & 0.5543 & 3 & LDSS3 & \\
1000031 & 351.3571694 & -41.2087321 & 0.1367 & 3 & LDSS3 & \\
1000032 & 351.3201482 & -41.2467900 & 0.3544 & 3 & LDSS3 & \\
1000033 & 351.2657143 & -41.1523966 & 0.3470 & 3 & LDSS3,IMACS & \\
1000034 & 351.2628524 & -41.1636893 & 0.6594 & 3 & LDSS3 & \\
1000035 & 351.2716906 & -41.1533293 & 0.4840 & 3 & LDSS3 & \\
1000037 & 351.2543306 & -41.1665713 & 0.3458 & 3 & LDSS3 & \\
1000038 & 351.2598936 & -41.1506836 & 0.5102 & 3 & LDSS3 & \\
1000039 & 351.2699605 & -41.2269533 & 0.3551 & 3 & LDSS3 & \\
1000040 & 351.3223366 & -41.2291830 & 0.3668 & 3 & LDSS3 & \\
1000042 & 351.2971077 & -41.2142311 & 1.2487 & 2 & LDSS3,IMACS & \\
1000043 & 351.3197344 & -41.1970816 & 0.3488 & 3 & LDSS3 & \\
1 & 351.3307531 & -41.1975129 & 0.1875 & 3 & IMACS & \\
3 & 351.3087164 & -41.1913390 & 1.3428 & 2 & IMACS & \\
4 & 351.3200478 & -41.1920123 & 1.2885 & 1 & IMACS & \\
6 & 351.3161614 & -41.1916622 & 0.4895 & 3 & IMACS & \\
6 & 351.3145867 & -41.1918808 & 0.3907 & 3 & IMACS & \\
7 & 351.3387738 & -41.1982031 & 0.0472 & 3 & IMACS & \\
8 & 351.0908000 & -41.3856000 & 0.4153 & 2 & IMACS & \\
9 & 351.1680792 & -41.3749842 & 0.5778 & 3 & IMACS & \\
10 & 351.0430226 & -41.3757393 & 0.5786 & 3 & IMACS & \\
11 & 351.2371158 & -41.3606047 & 0.4233 & 3 & IMACS & \\
12 & 351.3294233 & -41.3514408 & 0.3733 & 3 & IMACS & \\
13 & 351.1325777 & -41.3470125 & 0.6027 & 3 & IMACS & \\
14 & 351.2191606 & -41.3444107 & 0.5488 & 1 & IMACS & \\
15 & 350.9874171 & -41.3408072 & 0.6919 & 3 & IMACS & \\
16 & 351.0768385 & -41.3258616 & 0.3180 & 1 & IMACS & \\
17 & 351.2389775 & -41.3234121 & 0.3647 & 2 & IMACS & \\
18 & 351.2486345 & -41.3151546 & 0.3594 & 3 & IMACS & \\
19 & 351.2426654 & -41.3158980 & 0.4471 & 2 & IMACS & \\
20 & 350.9797792 & -41.3149806 & 0.2994 & 3 & IMACS & \\
21 & 351.1704507 & -41.3095811 & 0.6399 & 3 & IMACS & \\
22 & 351.1442806 & -41.3089506 & 0.5796 & 3 & IMACS & \\
23 & 351.4607480 & -41.3064505 & 0.3736 & 3 & IMACS & \\
24 & 351.2438396 & -41.2984098 & 0.3648 & 3 & IMACS & \\
25 & 351.5309839 & -41.2948694 & 0.3603 & 3 & IMACS & \\
26 & 350.9971620 & -41.2896735 & 0.3008 & 3 & IMACS & \\
27 & 351.5024523 & -41.2944313 & 0.5597 & 3 & IMACS & \\
28 & 351.3999741 & -41.2879101 & 0.3593 & 3 & IMACS & \\
29 & 350.9866289 & -41.2834291 & 0.3169 & 3 & IMACS & \\
30 & 351.1946338 & -41.2852857 & 0.3603 & 3 & IMACS & \\
31 & 351.5030315 & -41.2845684 & 0.5808 & 1 & IMACS & \\
32 & 351.1798852 & -41.2834147 & 0.3247 & 3 & IMACS & \\
33 & 351.2698396 & -41.2808630 & 0.3561 & 3 & IMACS & \\
34 & 351.2997460 & -41.2781853 & 0.3486 & 3 & IMACS & \\
35 & 351.0889352 & -41.2772592 & 0.4247 & 3 & IMACS & \\
36 & 351.4587165 & -41.2757527 & 0.3574 & 3 & IMACS & \\
37 & 351.4353233 & -41.2757739 & 0.3539 & 3 & IMACS & \\
38 & 350.9299940 & -41.2689700 & 0.3143 & 3 & IMACS & \\
39 & 351.2705655 & -41.2716035 & 0.3097 & 3 & IMACS & \\
40 & 351.5135874 & -41.2694621 & 0.5815 & 3 & IMACS & \\
41 & 351.4631121 & -41.2631035 & 0.3757 & 3 & IMACS & \\
42 & 351.4658950 & -41.2616775 & 0.3586 & 3 & IMACS & \\
43 & 351.3125910 & -41.2569106 & 0.3655 & 3 & IMACS & \\
44 & 351.1596829 & -41.2537828 & 0.4932 & 3 & IMACS & \\
45 & 351.5377888 & -41.2518169 & 0.4201 & 3 & IMACS & \\
46 & 351.4336385 & -41.2458311 & 0.3479 & 3 & IMACS & \\
46 & 351.4343508 & -41.2456887 & 0.3492 & 2 & IMACS & \\
47 & 351.4390198 & -41.2482455 & 0.3727 & 3 & IMACS & \\
48 & 350.9284946 & -41.2431137 & 0.6768 & 3 & IMACS & \\
49 & 351.2477438 & -41.2423766 & 0.3553 & 3 & IMACS & \\
50 & 351.5111342 & -41.2420319 & 0.0 & 3 & IMACS & \\
50 & 351.5119582 & -41.2417901 & 0.2711 & 1 & IMACS & \\
51 & 351.2195562 & -41.2384889 & 0.3649 & 3 & IMACS & \\
52 & 351.1296429 & -41.2374986 & 0.3747 & 3 & IMACS & \\
54 & 351.5199697 & -41.2356103 & 0.3756 & 3 & IMACS & \\
55 & 351.3677532 & -41.2340216 & 0.3531 & 3 & IMACS & \\
56 & 351.0694769 & -41.2290305 & 0.3001 & 2 & IMACS & \\
57 & 350.9897473 & -41.2280795 & 0.4462 & 3 & IMACS & \\
58 & 351.3798623 & -41.2272595 & 0.3522 & 3 & IMACS & \\
59 & 351.0441194 & -41.2239418 & 0.3178 & 3 & IMACS & \\
60 & 351.1176601 & -41.2235394 & 0.4068 & 3 & IMACS & \\
61 & 351.4847129 & -41.2227832 & 0.3208 & 3 & IMACS & \\
62 & 351.0360547 & -41.2219622 & 0.0 & 3 & IMACS & \\
63 & 351.5055554 & -41.2184035 & 0.3752 & 3 & IMACS & \\
64 & 351.2240318 & -41.2007740 & 0.3305 & 2 & IMACS & \\
67 & 351.1644843 & -41.1949432 & 0.6588 & 3 & IMACS & \\
68 & 350.9375484 & -41.1945333 & 0.4137 & 3 & IMACS & \\
70 & 351.4827107 & -41.1917357 & 0.4659 & 3 & IMACS & \\
71 & 351.2136769 & -41.1890435 & 0.3605 & 2 & IMACS & \\
72 & 351.1140503 & -41.1849005 & 0.2999 & 2 & IMACS & \\
73 & 351.3939791 & -41.1787078 & 0.3777 & 3 & IMACS & \\
74 & 350.9263901 & -41.1795695 & 0.4174 & 3 & IMACS & \\
75 & 351.4277158 & -41.1742006 & 0.3748 & 3 & IMACS & \\
76 & 351.3794157 & -41.1741300 & 0.4108 & 3 & IMACS & \\
77 & 350.9935886 & -41.1697505 & 0.4160 & 3 & IMACS & \\
78 & 350.9680695 & -41.1679272 & 0.3124 & 3 & IMACS & \\
79 & 351.1282096 & -41.1689561 & 0.3681 & 2 & IMACS & \\
80 & 351.0685270 & -41.1668891 & 0.5927 & 3 & IMACS & \\
81 & 351.4073441 & -41.1579148 & 0.3597 & 3 & IMACS & \\
82 & 351.1513607 & -41.1565862 & 0.3667 & 3 & IMACS & \\
82 & 351.1517969 & -41.1563888 & 0.4063 & 1 & IMACS & \\
83 & 351.2608321 & -41.1583439 & 0.3625 & 3 & IMACS & \\
84 & 351.4254882 & -41.1566662 & 0.3553 & 3 & IMACS & \\
85 & 351.1652313 & -41.1546266 & 0.2826 & 1 & IMACS & \\
86 & 350.9918061 & -41.1505843 & 0.3572 & 3 & IMACS & \\
87 & 351.1599352 & -41.1516310 & 0.3515 & 1 & IMACS & \\
88 & 351.1360050 & -41.1470098 & 0.3708 & 3 & IMACS & \\
89 & 351.4205442 & -41.1382882 & 0.3729 & 3 & IMACS & \\
90 & 351.3324299 & -41.1383155 & 0.3583 & 3 & IMACS & \\
91 & 351.1183336 & -41.1359333 & 0.3021 & 3 & IMACS & \\
92 & 351.3531781 & -41.1347803 & 0.4128 & 3 & IMACS & \\
93 & 351.4219075 & -41.1241488 & 0.3744 & 1 & IMACS & \\
94 & 351.2055221 & -41.1237165 & 0.3499 & 2 & IMACS & \\
95 & 351.3152562 & -41.1132695 & 0.3563 & 3 & IMACS & \\
96 & 351.3813927 & -41.0953665 & 0.4896 & 3 & IMACS & \\
97 & 351.3453414 & -41.0850716 & 0.4351 & 3 & IMACS & \\
98 & 351.4456225 & -41.0799043 & 0.2407 & 1 & IMACS & \\
99 & 351.3581700 & -41.0769170 & 0.3112 & 3 & IMACS & \\
100 & 351.4312135 & -41.0736623 & 0.4108 & 3 & IMACS & \\
101 & 351.4091380 & -41.0565935 & 0.4109 & 3 & IMACS & \\
\enddata 
\tablecomments{Spectroscopic redshifts measured in \SPTXXIII\ from Gemini, Magellan/LDSS3, and Magellan/IMACS. See \autoref{sec:specdata} for details of the spectroscopy analysis. Strongly lensed galaxies are tabulated in \autoref{tab:arcstable}. }
\end{deluxetable*}

\startlongtable
\begin{deluxetable*}{lllllll} 
\tablecolumns{7} 
\tablecaption{List of spectroscopically identified objects in \SPTZERO\ field of view \label{tab:spt0049allz}  } 
\tablehead{\colhead{ID} &
            \colhead{R.A.}    & 
            \colhead{Decl.}    & 
            \colhead{$z_{spec}$}     & 
            \colhead{conf.}       & 
            \colhead{Instruments}      &
            \colhead{}       \\[-8pt]
            \colhead{} &
            \colhead{J2000}     & 
            \colhead{J2000}    & 
            \colhead{}       & 
            \colhead{}       & 
            \colhead{}       & 
            \colhead{}             }
\startdata 
030 & 12.2958676 & -24.6396331 & 0.3445 & 1 & LDSS3  & \\
025 & 12.2976676 & -24.6445438 & 0.2752 & 2 & LDSS3  & \\
024 & 12.2947339 & -24.6482300 & 0.5279 & 3 & LDSS3  & \\
027 & 12.3155564 & -24.6437909 & 0.5318 & 3 & LDSS3  & \\
023 & 12.2940359 & -24.6542052 & 0.2492 & 3 & LDSS3  & \\
026 & 12.3073222 & -24.6526637 & 0.3426 & 3 & LDSS3  & \\
028 & 12.3072829 & -24.6551361 & 0.3448 & 3 & LDSS3  & \\
022 & 12.2909294 & -24.6640596 & 0.4341 & 3 & LDSS3  & \\
006 & 12.2876854 & -24.6674970 & 0.4440 & 1 & LDSS3  & \\
010 & 12.2893727 & -24.6700378 & 0.0 & 1 & LDSS3  & \\
011 & 12.2751750 & -24.6766800 & 0.9398 & 3 & LDSS3  & \\
004 & 12.3026499 & -24.6699485 & 0.746 & 1 & LDSS3  & \\
002 & 12.2849216 & -24.6888735 & 1.0302 & 2 & LDSS3  & \\
001 & 12.2888840 & -24.6913042 & 0.4299 & 1 & LDSS3  & \\
009 & 12.2985897 & -24.6906095 & 0.5355 & 3 & LDSS3  & \\
005 & 12.3041248 & -24.6907131 & 1.72 & 1 & LDSS3  & \\
012 & 12.3042900 & -24.6931631 & 0.5293 & 3 & LDSS3  & \\
013 & 12.2957763 & -24.6986553 & 0.5321 & 3 & LDSS3  & \\
014 & 12.2821687 & -24.7062031 & 0.5265 & 3 & LDSS3  & \\
015 & 12.2867523 & -24.7068513 & 0.5185 & 2 & LDSS3  & \\
034 & 12.2776163 & -24.7128064 & 0.3572 & 3 & LDSS3  & \\
017-A & 12.3229906 & -24.7002583 & 0.3141 & 3 & LDSS3  & \\
017-B & 12.3227230 & -24.6997445 & 0.2482 & 1 & LDSS3  & \\
016 & 12.3205370 & -24.7033231 & 0.4047 & 3 & LDSS3  & \\
018 & 12.2758188 & -24.7225908 & 0.1910 & 2 & LDSS3  & \\
020 & 12.3018904 & -24.7172519 & 0.9065 & 3 & LDSS3  & \\
019 & 12.2965644 & -24.7212467 & 0.437 & 1 & LDSS3  & \\
021 & 12.3004454 & -24.7249998 & 0.5247 & 3 & LDSS3  & \\
033 & 12.3159975 & -24.7222300 & 0.5234 & 2 & LDSS3  & \\
032 & 12.3054420 & -24.7283286 & 0.4038 & 1 & LDSS3  & \\
\enddata
\tablecomments{Spectroscopic redshifts measured in \SPTZERO\ from Magellan/LDSS3. See \autoref{sec:specdata} for details of the spectroscopy analysis. Strongly lensed galaxies are tabulated in \autoref{tab:arcstable}.  017-A and 017-B are two different galaxy detection from the same slit, ID017. } %obj003 and obj007 are the ID of the two multiple images with redshift from this
%obj007 is system 6 but my confidence for the redshift z=4.058 is 1 

\end{deluxetable*}

\section{\SPTV}\label{sec:appendixB}
\begin{figure*}
    \centering
    \includegraphics[width=\linewidth]{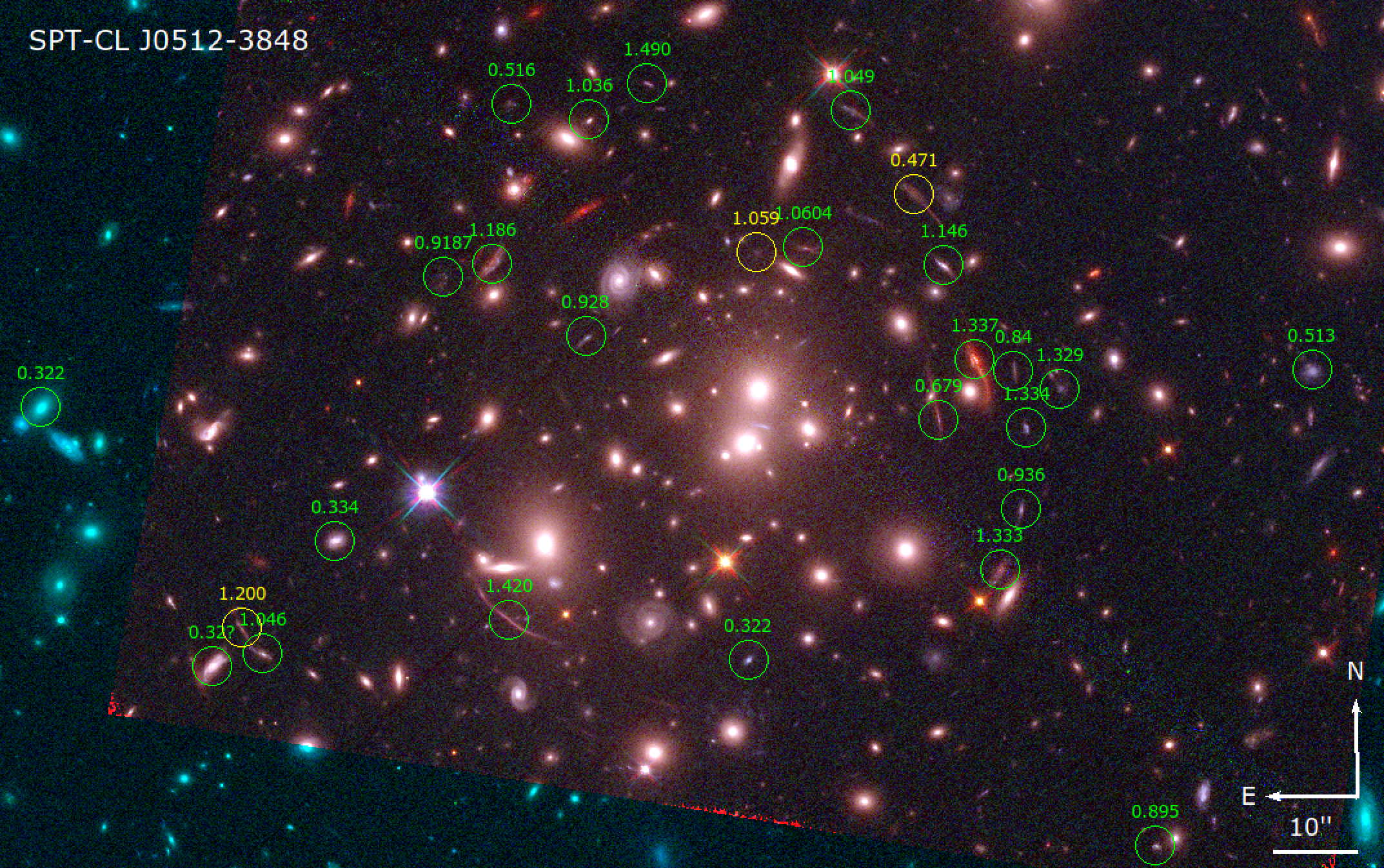}
    \caption{Composite color images of the field of \SPTV, from HST imaging in WFC3IR/F140W (red),  ACS/F814W (green), and ACS/F606W (blue). Circles labeled the spectroscopic redshift identified. Green circles correspond to high confidence identification while yellow circles mark tentative redshift identification. The full list of identification is listed in \autoref{tab:spt0512allz}. \label{fig:spt0512}}
\end{figure*}

A third cluster, \SPTV, was flagged as a promising strong lens candidate as part of our HST program to identify the next generation of extraordinary cluster lenses (GO-15937; PI: G. Mahler). The cluster was cataloged in \cite{Bleem2020} with significance $\xi=7.54$, $z=0.326$, and $M_{500c}=6.07^{+0.87}_{-0.88}\times10^{14} h_{70}^{-1}$\msun. Ground based optical imaging with Magellan/PISCO revealed promising arc-like features \citep[see Figure 8 in ][]{Bleem2020}. 
Despite the preliminary evidence, the combined analysis of HST morphology and extensive LDSS3 spectroscopy indicated that most of the arc-like features are high-flexure singly imaged lensed galaxies at $z<2$. While the cluster is likely a strong lens, there is not sufficient evidence for a robust lens model. In particular, we do not identify secure cluster-wide multiply-imaged systems. 
The field was observed with multi-object slit masks on 2017 Jan 1, 2017 Jan 30, 2017 Sep 2, 2018 Jan 9-10, and 2020 Jan 23 \citep[][Table D.1]{remolina_thesis}.  Data reduction and spectroscopic analysis were as described in \autoref{sec:specdata}. 
In this appendix, we provide the LDSS3 spectroscopic redshifts we obtained in this field, to facilitate possible future investigations of this line of sight by the community.
Galaxies with spectroscopic redshifts at the core of \SPTV\ are labeled in \autoref{fig:spt0512}, overplotted on the HST imaging from GO-15937. \autoref{tab:spt0512allz} presents the spectroscopic redshifts of all the objects in the field observed using Magellan/LDSS3 instruments as part of the follow-up of SPT clusters, out to $\sim 2.5$\arcmin\ from the BCG.

\startlongtable
\begin{deluxetable*}{lllllll} 
\tablecolumns{7} 
\tablecaption{List of spectroscopically identified objects in \SPTV\ field of view \label{tab:spt0512allz}  } 
\tablehead{\colhead{ID} &
            \colhead{R.A.}    & 
            \colhead{Decl.}    & 
            \colhead{$z_{spec}$}     & 
            \colhead{conf.}       & 
            \colhead{Instruments}      &
            \colhead{}       \\[-8pt]
            \colhead{} &
            \colhead{J2000}     & 
            \colhead{J2000}    & 
            \colhead{}       & 
            \colhead{}       & 
            \colhead{}       & 
            \colhead{}             }
\startdata
1 & 78.2523417 & -38.7977722 & 1.420 & 3 & LDSS3 & \\
2 & 78.2634083 & -38.7980306 & 1.200 & 2 & LDSS3 & \\
3 & 78.2550625 & -38.7866944 & 0.9187 & 3 & LDSS3 & \\
4 & 78.2330458 & -38.7893667 & 1.337 & 3 & LDSS3 & \\
5 & 78.2295042 & -38.7903139 & 1.329 & 3 & LDSS3 & \\
6 & 78.2595542 & -38.7952194 & 0.334 & 3 & LDSS3 & \\
7 & 78.2625458 & -38.7988722 & 1.046 & 3 & LDSS3 & \\
8 & 78.2646458 & -38.7992667 & 0.32 & 2 & LDSS3 & \\
9 & 78.2530417 & -38.7862778 & 1.186 & 3 & LDSS3 & \\
10 & 78.2343458 & -38.7863167 & 1.146 & 3 & LDSS3 & \\
11 & 78.2421125 & -38.7858972 & 1.059 & 2 & LDSS3 & \\
12 & 78.2355625 & -38.7840167 & 0.471 & 2 & LDSS3 & \\
13 & 78.2381708 & -38.7813083 & 1.049 & 3 & LDSS3 & \\
14 & 78.2490208 & -38.7816333 & 1.036 & 3 & LDSS3 & \\
15 & 78.2522458 & -38.7811111 & 0.516 & 3 & LDSS3 & \\
16 & 78.2466083 & -38.7804389 & 1.490 & 3 & LDSS3 & \\
17 & 78.2636375 & -38.7742222 & 0.872 & 3 & LDSS3 & \\
18 & 78.2345424 & -38.7913338 & 0.679 & 3 & LDSS3 & \\
19 & 78.2309167 & -38.7915861 & 1.334 & 3 & LDSS3 & \\
20 & 78.2320000 & -38.7961528 & 1.333 & 3 & LDSS3 & \\
21 & 78.2424292 & -38.7990694 & 0.322 & 3 & LDSS3 & \\
22 & 78.2190708 & -38.7897111 & 0.513 & 3 & LDSS3 & \\
23 & 78.2145583 & -38.7762639 & 0.324 & 3 & LDSS3 & \\
24 & 78.2255500 & -38.8050750 & 0.895 & 3 & LDSS3 & \\
25 & 78.2218708 & -38.8072361 & 0.444 & 3 & LDSS3 & \\
26 & 78.2558125 & -38.8102667 & 1.607 & 2 & LDSS3 & \\
27 & 78.2717333 & -38.7909111 & 0.322 & 3 & LDSS3 & \\
28 & 78.2491579 & -38.7886275 & 0.928 & 3 & LDSS3 & \\
29 & 78.2401483 & -38.7857283 & 1.0604 & 3 & LDSS3 & \\
30 & 78.2311376 & -38.7942000 & 0.936 & 3 & LDSS3 & \\
31 & 78.2314275 & -38.7897544 & 0.84 & 3 & LDSS3 & \\
\enddata 
\tablecomments{Spectroscopic redshifts measured in \SPTV\ from Magellan/LDSS3. See \autoref{sec:specdata} for details of the spectroscopy analysis. }
\end{deluxetable*}

\section{Lensing Candidates}\label{sec:appendixC} 
\begin{figure*}
    \centering
    \includegraphics[width=0.5\linewidth]{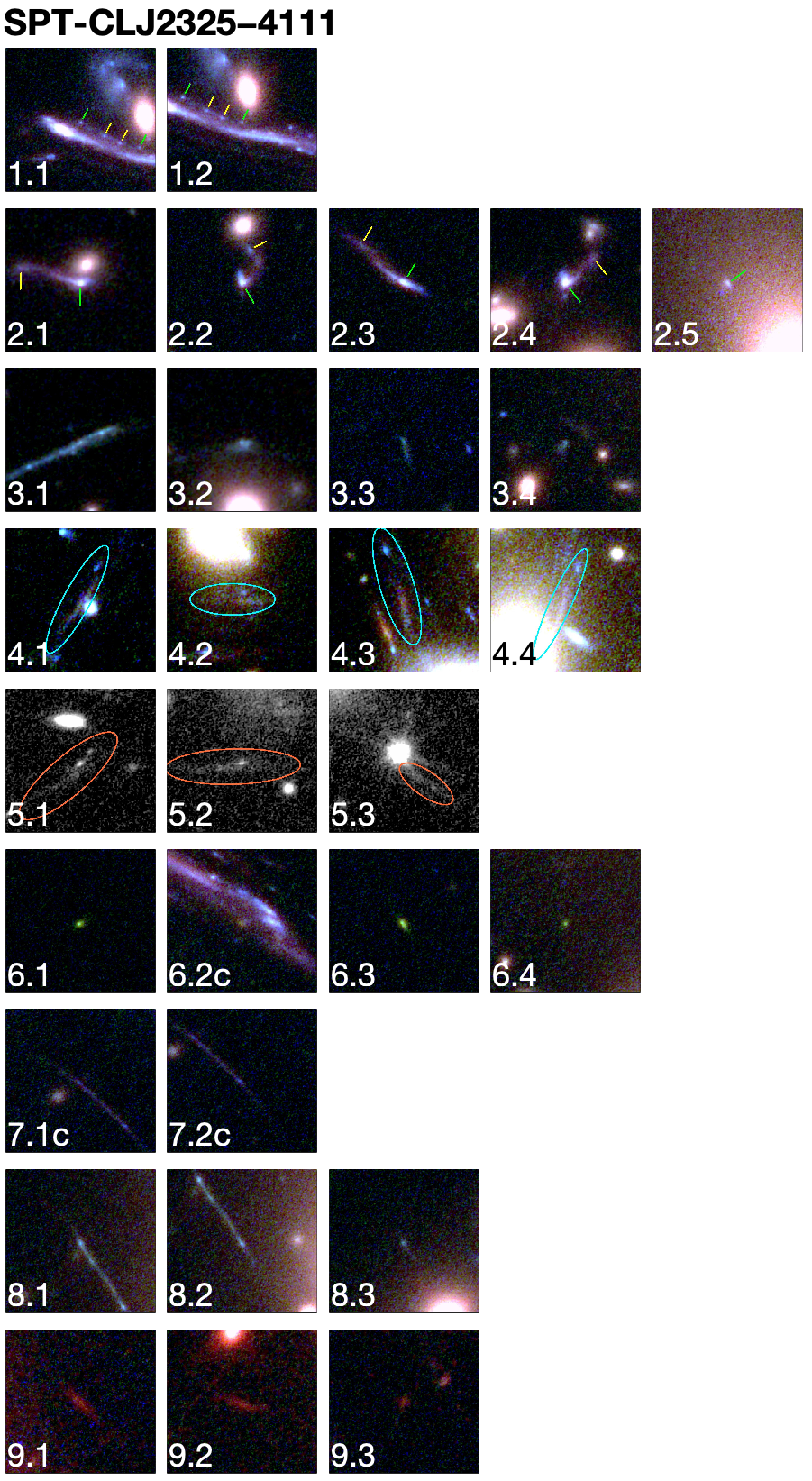}
    \includegraphics[width=0.4\linewidth]{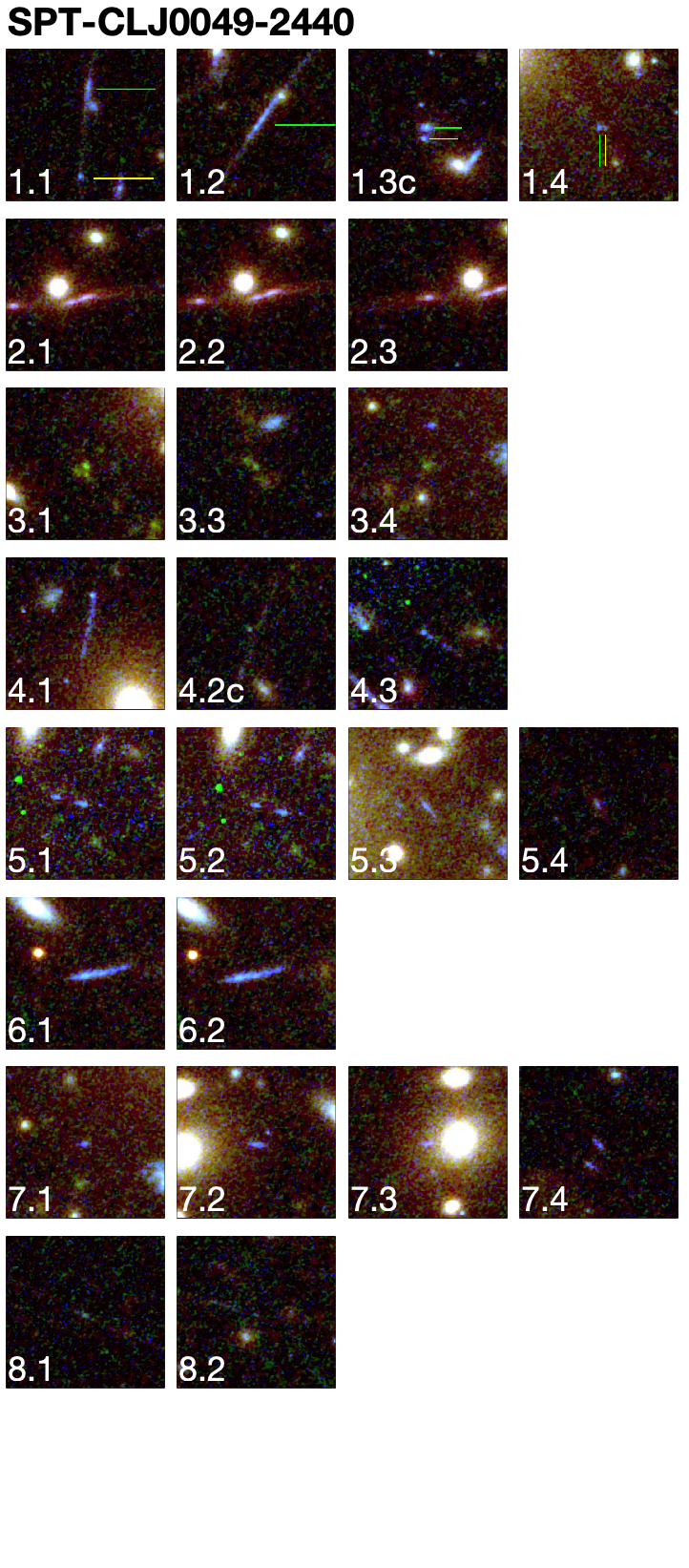}
    \caption{Zoom-in on the multiple images of lensed galaxies in \SPTXXIII\ (left) and \SPTZERO\ (right). Each square is $3\farcs0\times3\farcs0$ field of view, with the exception of images of sources 4 and 5 in \SPTXXIII, which are $4\farcs0\times4\farcs0$. Most images are centered on the clump that was identified as constraint. In \SPTXXIII~source~4, the blue emission knot at the north of each arc was used as the constraint. The color rendition is the same as \autoref{fig:hst2325} and \autoref{fig:hst0049}. For \SPTXXIII~source~5, we show only F814W in grayscale, since image 5.3 of the source is affected by diffraction spikes of a nearby star in the other bands. Green and yellow ticks mark the first and second clumps, respectively, in systems where multiple clumps were used as constraints. }
    \label{fig:stamps}
\end{figure*}

In this section we provide a zoom-in view of the multiple images identified in this work and used as lens modeling constraints, in \autoref{fig:stamps}. See \autoref{tab:arcstable} for coordinates, \autoref{fig:hst2325} and \autoref{fig:hst0049} for the full field of view of each cluster with arcs labeled. \autoref{sec:lensing-cstr} describes the identification.

%% Appendix material should be preceded with a single \appendix command.
%% There should be a \section command for each appendix. Mark appendix
%% subsections with the same markup you use in the main body of the paper.
%% Each Appendix (indicated with \section) will be lettered A, B, C, etc.
%% The equation counter will reset when it encounters the \appendix
%% command and will number appendix equations (A1), (A2), etc. The
%% Figure and Table counter will not reset.

\end{document}